\documentclass[10pt]{article}

\usepackage{tabularx}
\usepackage{booktabs}
\usepackage{pifont} 
\usepackage{xfrac} 
\usepackage{fancyhdr}
\usepackage{amsmath}
\usepackage[a4paper,margin=2cm]{geometry}
\usepackage{lipsum,multicol,float,graphicx}
\usepackage[font=footnotesize,labelfont=bf,labelsep=period]{caption}
\usepackage{hyperref}
\usepackage[scaled=.90]{helvet}
\usepackage{titlesec}
\usepackage{siunitx}
\usepackage[symbol]{footmisc}
\hypersetup{colorlinks=false,linkcolor=blue,filecolor=blue,urlcolor=blue}

\usepackage[natbib=false,backend=biber,bibencoding=utf8,sorting=none,url=false,doi=false,isbn=false,eprint=false,style=numeric-comp,giveninits=true]{biblatex}
\bibliography{tutorialbib.bib}

\DefineBibliographyStrings{english}{
 page             = {\ifbibliography{}{\adddot}},
 pages            = {\ifbibliography{}{\adddot}},
}

\DeclareFieldFormat[inbook,article,inproceedings,incollection]{citetitle}{#1}
\DeclareFieldFormat[inbook,article,inproceedings,incollection]{title}{#1}
\DeclareFieldFormat[article]{volume}{\textbf{#1},\space} 
\AtEveryBibitem{\clearfield{title}}
 
\renewcommand*{\newunitpunct}{\addcomma\space}

\renewbibmacro*{volume+number+eid}{%
  \setunit*{\space}
  \printfield{volume}%
  \setunit*{\addcomma\space}
  \printfield{number}%
  \setunit{\addcomma\space}%
  \printfield{eid}}
  
\renewbibmacro*{date}{%
   \printtext[parens]{\printfield{year}}  
}

\renewbibmacro*{issue+date}{%
  \setunit*{\space}
    \iffieldundef{issue}
      {\usebibmacro{date}}
      {
       \setunit*{\addspace}
       \usebibmacro{date}
       }
  \newunit}
  
\renewbibmacro*{journal+issuetitle}{%
  \usebibmacro{journal}%
  \setunit*{\addspace}%
  \iffieldundef{series}
    {}
    {\newunit
     \printfield{series}%
     \setunit{\addspace}}%
  \usebibmacro{volume+number+eid}%
  \newunit}

  \DeclareBibliographyDriver{article}{%
  \usebibmacro{bibindex}%
  \usebibmacro{begentry}%
  \usebibmacro{author/translator+others}%
  \setunit{\labelnamepunct}\newblock
  \usebibmacro{title}%
  \newunit
  \printlist{language}%
  \newunit\newblock
  \usebibmacro{byauthor}%
  \newunit\newblock
  \usebibmacro{bytranslator+others}%
  \newunit\newblock
  \printfield{version}%
  \newunit\newblock
  \usebibmacro{journal+issuetitle}%
  \newunit
  \usebibmacro{byeditor+others}%
  \newunit
  \usebibmacro{note+pages}%
  \setunit{\addspace}
  \usebibmacro{issue+date}
  \setunit{\addcolon\space}
  \usebibmacro{issue}
  \newunit\newblock
  \iftoggle{bbx:isbn}
    {\printfield{issn}}
    {}%
  \newunit\newblock
  \usebibmacro{doi+eprint+url}%
  \newunit\newblock
  \usebibmacro{addendum+pubstate}%
  \setunit{\bibpagerefpunct}\newblock
  \usebibmacro{pageref}%
  \usebibmacro{finentry}}

\titleformat{\section}
  {\normalfont\sffamily\bfseries}
  {\thesection .}{5pt}{}
\titleformat{\subsection}
  {\normalfont\sffamily\itshape}
  {\thesubsection .}{5pt}{}

\makeatletter
\renewenvironment{abstract}
{\noindent\bfseries\sffamily\abstractname\vspace{1mm}\\
\normalfont\small}
\makeatother

\begin{document}

\thispagestyle{empty}
{

\vspace{5mm}
\noindent\sffamily\Huge\bfseries Dynamics and correlations of a Bose-Einstein condensate of photons

\vspace{5mm}
\noindent\bfseries\sffamily\large Julian Schmitt\normalfont

\vspace{4mm}
\noindent\footnotesize Institut f\"ur Angewandte Physik, Universit{\"{a}}t Bonn, Wegelerstra\ss e 8, D-53115 Bonn, Germany\\
Present address: Cavendish Laboratory, 19 JJ Thomson Avenue, Cambridge CB3 0HE, UK


\vspace{2mm}
\noindent\footnotesize 19 July 2018

\vspace{5mm}
}

\begin{abstract}
The Tutorial reports recent experimental advances in studies of the dynamics as well as the number and phase correlations of a Bose-Einstein condensed photon gas confined in a high-finesse dye-filled microcavity. Repeated absorption-emission-processes of photons on dye molecules here establish a thermal coupling of the photonic quantum gas to both a heat bath and a particle reservoir comprised of dye molecules. In this way, for the first time Bose-Einstein condensation under grand-canonical statistical ensemble conditions becomes experimentally accessible.
\end{abstract}
\vspace{1pc}

\section{Grand-canonical Bose-Einstein condensation}

Large statistical number fluctuations are a fundamental property known from the thermal behaviour of bosons, as has been strikingly revealed in experiments with light and material particles~\cite{Hanbury,Baym,Schellekens,Jeltes,Dall,Hodgman}. For low temperatures or high densities, however, when a Bose gas undergoes Bose-Einstein condensation (BEC), the persistence of large particle number fluctuations can conflict with the conservation of the total particle number. Accordingly, fluctuations are damped out and second-order coherence emerges ~\cite{Glauber1,Glauber2,Kocharovsky}. This notion grounds on the microcanonical or canonical statistical description of the system, which applies for systems well-isolated from their environment suppressing both energy and particle exchange with the environment as e.g. realised in ultra-cold atomic gases ~\cite{Kocharovsky,Holthaus2}. Following the first observation of BEC in dilute atomic vapour~\cite{Anderson2,Davis}, evidence for the the emergence of first-order coherence~\cite{Anderson1,Saba,Bloch2,Andrews} and the suppression of density fluctuations~\cite{Oettl,Burt,Perrin,Ketterle1,Schellekens} have provided hallmarks for the phase transition. More recently, BECs have also been observed in two-dimensional (2D) gases of exciton-polaritons~\cite{Deng1,Balili,Kasprzak1,YamamotoRev}, magnons~\cite{Demokritov} and photons~\cite{Klaers1,Klaers2,Marelic}. Quintessentially, these systems are open due to their coupling to the environment for e.g. particle injection or thermalisation, which reinforces the relevance of reservoirs for their description, as for example provided by grand-canonical statistics.

In the grand-canonical ensemble, the system is subject to particle (and energy) exchange with a reservoir~\cite{Huang}. For bosons, the population $n$ in each quantum state suffers large number fluctuations $\delta n\simeq \bar n$, while the fixed chemical potential (and temperature) accounts for a complete thermodynamic description of the gas. In the thermodynamic limit, all three statistical ensembles are generally expected to become equivalent due to vanishing relative fluctuations of the total particle number, i.e. $\delta N/\bar N\rightarrow 0$. Applied e.g. to the macroscopically occupied ground state in the Bose-Einstein condensed phase ($\bar n\simeq \bar N$), however, in the grand-canonical ensemble large fluctuations of the total particle number, $\delta N\simeq\bar N$ occur. Surprisingly, the statistical fluctuations here become enhanced as the system temperature approaches absolute zero instead of being frozen out. This so-called {\itshape grand-canonical fluctuation catastrophe} has been a long-standing issue in theoretical physics~\cite{Fierz,Fujiwara,Ziff,Holthaus1,Holthaus2,Kocharovsky,Weiss,Navez,Haar,Yukalov} and its observation has long remained elusive. Most notably, Ziff, Uhlenbeck and Kac altogether questioned the physical significance of the grand-canonical ensemble in the condensed phase~\cite{Ziff}; their arguments, however, apply only for diffusive contact between a spatially separated BEC and particle reservoir.

In contrast, for a BEC of photons in a dye-filled optical microcavity genuine grand-canonical statistical conditions in the condensed phase can become relevant. Here, the coupling of the condensed particles to an effective reservoir is realised by interparticle conversion between photons, ground and excited state dye molecules~\cite{Klaers5,Sobyanin}. In this system, we have for the first time observed grand-canonical number statistics in a BEC by demonstrating its coupling to both a heat bath and a particle reservoir~\cite{Schmitt1,Schmitt3,Ciuti2}. These results provided a first experimental hint at the fluctuation catastrophe. Moreover, our work revealed phase fluctuations of the condensate wave function in the wake of grand-canonical statistical number fluctuations~\cite{Schmitt1,Schmitt3,Schmitt4}.

The present Tutorial contains a theoretical and experimental study of the thermalisation dynamics and first and second-order temporal correlations of a Bose-Einstein condensed photon gas under canonical and grand-canonical ensemble conditions. The Tutorial is organised as follows: Section~\ref{BECofphotons} introduces the concept of photon BEC, Sections~\ref{thermalisierungsdynamik_theor}-\ref{phasenkohaerenz} give a theoretical description of the photon thermalisation process, along with the BEC number and phase correlations, while Sections~\ref{thermalisierung}-\ref{phasenkohaerenzexp} describe our corresponding experiments. Finally, Section~\ref{conclusion} concludes and gives an outlook.

\section{Bose-Einstein condensation of photons}
\label{BECofphotons}
Photons depict a prime example among the Bose gases known today and yet, it has taken almost a century to find ways to condense them -- {\itshape Why?} Thermal photons usually do not become quantum degenerate: in blackbody radiation, for example, the coupling of temperature and total photon number prohibits BEC at low temperatures as photons at $T\rightarrow 0$ vanish instead of forming a condensate\footnote{In other words: chemical potential $\mu=0$}. In optical gases with a conserved particle number as e.g. in nonlinear microcavities, photon-photon interactions are usually too small to achieve efficient thermalisation of the light~\cite{Chiao1,Chiao2}.

Quantum fluids of light have nevertheless emerged in recent years by synthesising dressed light-matter-states~\cite{Ciuti}, such as exciton-polaritons in microcavities~\cite{YamamotoRev} or suface-plasmon-polaritons~\cite{Toermae2}. These platforms have provided long sought-after evidence for condensation, coherence~\cite{Kasprzak1,Deng1,Balili,Toermae3,Kasprzak3,Plumhof} and thermalisation~\cite{Kasprzak2,Deng2,Snoke,Pfeiffer} in optical quantum gases.

More recently, BEC of {\itshape pure} photons has also become tractable by implementing a photon thermalisation mechanism with an incoherent molecular medium that realises a non-zero chemical potential for the light~\cite{Klaers3}; see Refs.~\cite{Wang,Hafezi} for similar concepts. The first observation of photon BEC in 2010 by Klaers \itshape{et al.}\normalfont~\cite{Klaers1,Klaers2,Klaers4} has been confirmed in more recent work by Marelic \itshape{et al.}\normalfont~\cite{Marelic} and Greveling \itshape{et al.}\normalfont~\cite{Greveling2}. In the meantime, a number of experiments have elaborated on the thermalisation~\cite{Schmitt2,Schmitt3,Marelic3}, the calorimetry~\cite{Damm}, the first-order spatial coherence~\cite{Marelic2,Damm2}, the first- and second-order temporal correlations~\cite{Schmitt1,Schmitt4}, the polarisation properties\cite{Greveling3}, non-local interactions~\cite{Greveling} and the generation of lattices and micropotentials for photon condensates~\cite{Dung,Walker}. Key aspects that are related to the topics discussed in this Tutorial have been studied in a (non-exhaustive) series of theoretical work on photon condensation and its dynamics~\cite{Kirton1,Kirton2,Keeling1,Nyman2}, on grand-canonical particle number correlations~\cite{Klaers4,Sobyanin,Weiss2}, on phase diffusion~\cite{Leeuw} and on the relation of photon condensation and lasing~\cite{Chiocchetta,Scully2}.

In this Section, we introduce the scheme for BEC of photons in a dye-filled optical microcavity with a focus on thermal and chemical equilibrium, the microcavity dispersion and the statistical physics of the photons.

\subsection{Photons in a dye-filled microcavity}
\begin{figure}[t]
\centering\includegraphics{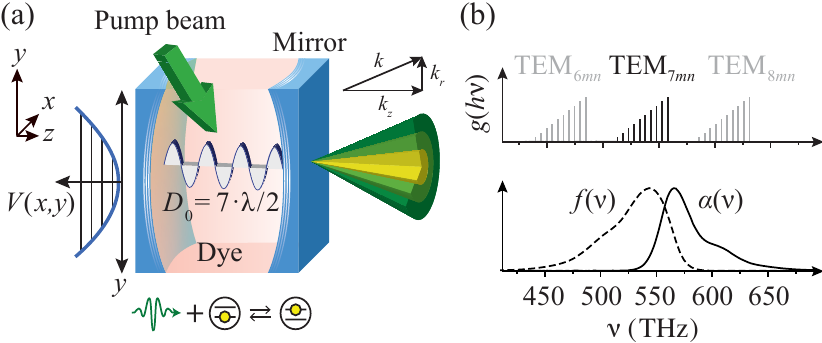}
\caption{Scheme of the experiment. (a) The dye-filled optical cavity consists of two highly-reflecting mirrors separated by $q=7$ half-wavelengths. The spherical mirror curvature introduces a harmonic potential $V(x,y)$ for the transverse motion of the photons. A pump laser excites the dye molecules and subsequent emission-absorption-cycles lead to a thermalisation of the photon gas at $T=300~\si{\kelvin}$. The cavity emission is monitored in a spatially, spectrally and time-resolving way. (b) Cavity mode spectrum of the photons (top), and spectral profiles of dye fluorescence $f(\nu)$ and absorption $\alpha(\nu)$ (bottom). The height of the bars indicates the degeneracy of the cavity eigenmodes. Fluorescence photons are emitted into transverse modes with fixed $q=7$ (black bars), making the photon gas effectively 2D. Reproduced with permission from \cite{Klaers1,Damm2}. Copyright 2010 \& 2017 managed by the Nature Publishing Group.}
\label{fig5}
\end{figure}

Figure~\ref{fig5} shows our microcavity experiment, which consists of two curved mirrors spaced by $D_0\simeq1.4~\si{\micro m}$ (or $1.6~\si{\micro m}$) and filled with a liquid dye solution. At a mirror separation $D_0=q \lambda/2\tilde n_0$ the resonator encloses $q=7$ (or 8) half waves, which corresponds to a free spectral range of adjacent longitudinal cavity modes $\Delta\lambda=\lambda^2/2\tilde n_0 D_0\simeq80~\si{\nano m}$ ($\Delta\nu\simeq 75~\si{\THz}$) comparable to the spectral width of the dye fluorescent emission (Fig.~\ref{fig5}(b)). Here, $\lambda$ denotes the optical wavelength in vacuum and $\tilde n_0$ is the refractive index of the dye solution. Accordingly, photons associated with a fixed longitudinal wave number $q$ are absorbed and emitted into the resonator differing only in their transverse quantum numbers $m$ and $n$. Effectively, this reduced dimensionality introduces a low-energy ground state ("cutoff") for the photon gas $\hbar\omega_\textrm{\tiny c}$, which is given by the TEM$_{q00}$ cavity mode. In other words, the photons can be ascribed an effective mass $m_\textrm{\tiny ph}=\hbar\omega_\textrm{\tiny c}/(c/\tilde n_0)^2$ and the photon kinetics is reduced to the transverse plane of the resonator. Additionally, the curvature of the cavity mirrors imposes an in-plane confinement (Fig.~\ref{fig5}(a), left). The photon gas behaves formally equivalent to a 2D, harmonically trapped ideal Bose gas, for which in thermal equilibrium BEC is expected below a finite critical temperature $T_\textrm{\tiny c}$ or above a critical particle number $N_\textrm{\tiny c}$~\cite{Bagnato,Griffin}. 

The solved dye molecules are optically pumped by a laser beam and the electronically excited molecules decay via emission of fluorescence photons in the cavity modes, as sketched in Fig.~\ref{fig5}(b, top). Inside the high-finesse cavity, frequent absorption-emission-cycles of photons by dye molecules establish a thermal contact between both subsystems in the sense of the grand-canonical ensemble: the photon gas acquires a temperature $T$ (room temperature) and chemical potential $\mu$, as determined by the much larger molecular reservoir. Firstly, for the (energy) thermalisation the spectral distributions of fluorescence $f(\omega)$ and absorption $\alpha(\omega)$, see Fig.~\ref{fig5}(b), are required to scale with a Boltzmann-factor $f(\omega)/\alpha(\omega)\propto \omega^3 \exp(-\hbar\omega/k_\textrm{\tiny B}T)$\setcounter{footnote}{2}\footnote{For many dye solutions at room-temperature the scaling is based on the Kennard-Stepanov relation~\cite{Kennard1,Kennard2,Stepanov1,Stepanov2}.}. The fluorescence-induced energy exchange between the photons and the (thermal) molecular bath then translates to the spectrum of the photon gas. Secondly, the chemical (particle number) equilibration rests on the fact that the interaction between photons and molecules can be considered as a photochemical reaction, see Fig.~\ref{fig5}(a, bottom). The required energy to electronically excite a dye molecule $\hbar \omega\simeq2.3~\si{\electronvolt}$ exceeds thermal energy $k_\textrm{\tiny B}T\simeq 0.025~\si{\electronvolt}$ by far, which suppresses fluctuation-driven dye excitations by a factor of order $\exp(-\hbar\omega/k_\textrm{\tiny B}T)\approx 10^{-37}$. Similarly, the thermally excited emission of photons into the cavity modes ($\hbar\omega\simeq 2~\si{\electronvolt}$) is very unlikely. An optical photon (of energy $\hbar\omega$) is emitted only, if another optical photon ($\hbar\omega'$) has been previously absorbed. If this condition is maintained throughout the experiments, the photon number does not decrease as the gas is cooled down, i.e. $\mu\neq 0$, in contrast to blackbody radiation.

The thermalisation process equilibrates photons over the set of TEM$_{mn}$ modes, leading to an average internal energy of the photon gas ${\sim}k_\textrm{\tiny B} T$ above $\hbar\omega_\textrm{\tiny c}$. "Cold" photons propagate near the optical axis, while "hot" photons exhibit large angles with respect to the optical axis. By heating up the dye solution, an enhanced population of highly excited transverse states is observed~\cite{Klaers2,Damm2}. The equilibrium Bose-Einstein distribution has been experimentally confirmed in the dye-microcavity experiment~\cite{Klaers1,Klaers2,Klaers3,Marelic,Damm,Schmitt1,Schmitt2,Schmitt3,Schmitt4}: For small total particle numbers, $N\leq N_\textrm{\tiny c}$, the photon energies are Boltzmann-distributed, while for $N>N_\textrm{\tiny c}$ adding more photons results in the accumulation of a BEC in the transverse ground state accompanied by a saturation of excited transverse modes.

\begin{figure}[t]
\flushright
\centering\includegraphics{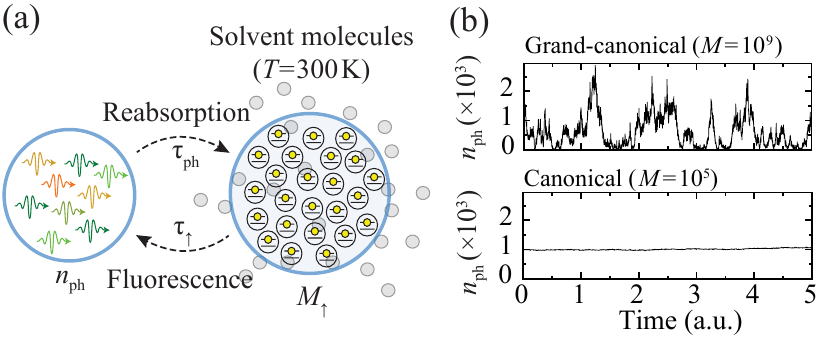}
\caption{Statistical ensemble and particle number fluctuations. (a)~The photon gas couples to the reservoir of electronically excited molecules by reabsorption after a photon lifetime $\tau_\textrm{\tiny ph}$. The molecular excitations decay within $\tau_\uparrow$, establishing chemical equilibrium between the photon gas and the particle reservoir. Simultaneously, multiple collisions of dye and solvent molecules lead to a thermalisation of the rovibronic dye states at room temperature. The light-matter-interaction imprints the thermal equilibrium state onto the photon gas and the molecules can be regarded as a heat bath. (b)~Temporal fluctuations of the photon number under grand-canonical ensemble conditions (large reservoir $M=10^9$, top), and damped fluctuations in the canonical ensemble (small reservoir $M=10^5$, bottom).}
\label{fig6}
\end{figure}

The light-matter-interaction between photons and molecules at room temperature is incoherent, due to many dephasing collisions between dye and solvent molecules during the dye excited state lifetime~\cite{Lakowicz,Yokoyama}. Consequently, the dynamics of photons and molecules can be modelled by rate equations, which also determine the mean population of photons $\bar n$ and excited dye molecules $\bar M_\uparrow$ (see Fig.~\ref{fig6}~(a)). As for typical experimental parameters $\bar M_\uparrow\gg \bar n$, the ensemble of excited dye molecules can be viewed as an effective particle reservoir for the photon gas. The heat energy and particle exchange with the dye reservoir paves the way for studies of the transition from canonical to grand-canonical ensemble conditions, as illustrated in Fig.~\ref{fig6}(b).

In the grand-canonical ensemble, where each eigenstate suffers strong number fluctuations $\delta n_i/ \bar n_i\simeq 1$, the second-order coherence of a BEC is expected to be substantially reduced, i.e. $g^{(2)}(0) \simeq ({\delta n_0}/{\bar n_0})^2 +1 = 2$. Experimentally, we find evidence for large statistical intensity fluctuations in BECs, which persist up condensate fractions of $\bar n_0/\bar N\simeq70\%$ as long as the particle reservoir complies with grand-canonical conditions~\cite{Klaers5,Schmitt1,Schmitt4}. This is in contrast to experiments with ultra-cold atoms, where a reduction of density fluctuations in the Bose-Einstein condensed phase has been observed~\cite{Dall,Oettl,Hodgman,Schellekens}. In this case, the emergence of second-order coherence is related to the isolation of the atomic ensemble from its environment, which necessitates a statistical description in the microcanonical ensemble with fixed particle number and Poissonian fluctuations $\delta n_0/\bar n_0 = 1/\sqrt{\bar n_0}\stackrel{\bar n_0\gg 1}{\longrightarrow}0$, i.e. $g^{(2)}(0)=1$. Interestingly, also the photon statistics in a laser follows a Poissonian distribution~\cite{Siegman,Fox,Loudon}. The fluctuation properties of photon BECs under grand-canonical conditions differ strikingly from those of both lasers and BECs in the microcanonical or canonical ensemble.

\subsection{Thermal equilibrium}
We outline the fluorescence-induced thermalisation process of the microcavity photon gas, which is based on radiative energy exchange between photons and dye molecules by absorption-emission-processes. The latter establish a thermal contact between the system (photon gas) and a heat bath at room temperature (dye solution) by dissipating excess energy of "hot" photons and providing energy for "colder" photons. For this, the spectral absorption and emission profiles of the dye molecules are required to fulfil the so-called Kennard-Stepanov relation, as will be discussed in the following. A more refined derivation of the thermalisation process can be found in refs.~\cite{Klaers2,Klaers3,Klaers4}.

\begin{figure}[t]
\centering\includegraphics{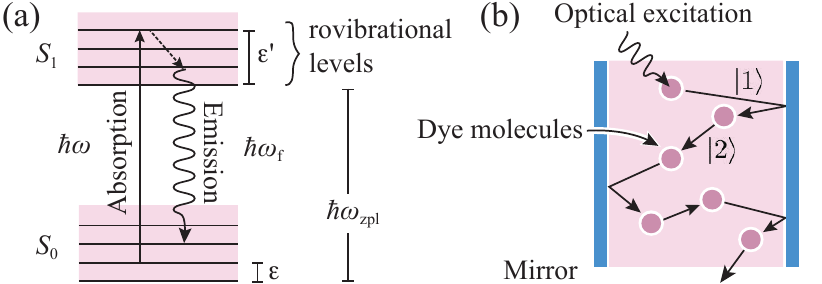}
\caption{(a) Electronic and rovibrational energy levels in a dye molecule. (b) In the "photon box" absorption-emission-cycles by dye molecules frequently change the configuration of the light field ($|1\rangle\rightarrow|2\rangle\rightarrow...$) to thermalise the photon gas.}
\label{fig13}
\end{figure}

The relevant, underlying molecular processes are sketched in the simplified energy diagram of a dye molecule in Fig.~\ref{fig13}(a). The electronic ground and excited singlet states $S_{0,1}$ exhibit a (quasi-)continuous subset of rotational and vibrational modes (shaded areas), and the energy difference between the ground states in $S_0$ and $S_1$ is on the order of $\hbar\omega_\textrm{\tiny zpl}\simeq 2~\si{\electronvolt}$ (zero-phonon line). After a photon absorption ($\hbar\omega_\textrm{\tiny a}$), frequent collisions between dye and solvent molecules ($10^{-15}~\si{\second}$ time scale at room temperature) rapidly alter the rovibrational molecular state, resulting in a thermal distribution in the electronically excited manifold. During the relaxation, any excess energy is dissipated by the solvent bath on a $10^{-12}~\si{\second}$ timescale. To this end, the subsequent fluorescence emission ($\tau=10^{-9}~\si{\second}$, $\hbar\omega_\textrm{\tiny f}$) occurs from a thermally equilibrated state $S_1$ to the ground state $S_0$, which is subject to the same relaxation mechanism.

This insight allows us to derive a Boltzmann-type law relating the spectral absorption and emission profiles of the dye molecules, known as the Kennard-Stepanov relation~\cite{Kennard1,Kennard2,Stepanov1,Stepanov2,Sawicki,McCumber}. We obtain the ratio of fluorescence $f(\omega)$ and absorption $\alpha(\omega)$ by integrating over the rovibrational energy levels
\begin{equation}
\frac{f(\omega)}{\alpha(\omega)}\propto \frac{\int D_\uparrow(\epsilon') p(\epsilon') A(\epsilon',\omega)d\epsilon'}{\int D_\downarrow(\epsilon) p(\epsilon) B(\epsilon,\omega)d\epsilon},
\label{2_0_2}
\end{equation}
where $\epsilon, \epsilon'$ denote energies and $D_\downarrow(\epsilon),D_\uparrow(\epsilon')$ the rovibrational density of states in ground ($\downarrow$) and excited ($\uparrow$) state. Due to the collisional relaxation, $p(\epsilon^{(\prime)} )=\exp( - \epsilon^{(\prime)} /k_\textrm{\tiny B} T)$ in both states. Considering energy conservation $\hbar\omega+\epsilon = \hbar\omega_\textrm{\tiny zpl} + \epsilon'$, the Einstein coefficients $A(\epsilon',\omega)$ and $B(\epsilon,\omega)$ are related by $D_\uparrow(\epsilon')A(\epsilon',\omega)d\epsilon' = \frac{2\hbar\omega^3}{\pi c^2}D_\downarrow(\epsilon)B(\epsilon,\omega)d\epsilon$~\cite{Sawicki}. Assuming identical rovibrational substructures, $D_\downarrow(\epsilon)=D_\uparrow(\epsilon')$,~(\ref{2_0_2}) yields the Kennard-Stepanov relation
\begin{equation}
\frac{f(\omega)}{\alpha(\omega)}\propto \frac{2\hbar\omega^3}{\pi c^2} \exp\left(-\frac{\hbar(\omega-\omega_\textrm{\tiny zpl})}{k_\textrm{\tiny B} T}\right).
\label{KennardStepanov}
\end{equation}
Experimentally, the scaling has been verified in e.g. liquid dye solutions~\cite{Bjourn,Szalay,Singhal,Knox,Klaers4}, dye-doped polymers~\cite{Schmitt2}, semiconductors~\cite{Ihara} or ultra-dense gases~\cite{Moroshkin}.

The Kennard-Stepanov relation is the key ingredient for the photon thermalisation mechanism to work. In our high-finesse "photon box", see Fig.~\ref{fig13}(b), the fluorescence photons undergo many absorption-emission-cycles, corresponding to a random walk of the light field configuration~\cite{Klaers4}. The ratio of the transition rates between two configurations $|1\rangle\rightarrow |2\rangle$, which differ from each other by the absorption of one photon with frequency $\omega_i$ and the emission of one photon with frequency $\omega_j$, is
\begin{equation}
\frac{R_{12}}{R_{21}}= \frac{\alpha(\omega_i)f(\omega_j)\omega_j^3}{\alpha(\omega_j)f(\omega_i)\omega_i^3} = e^{ - {\hbar(\omega_j-\omega_i)}/{k_\textrm{\tiny B} T}}\ \ \forall i,j.
\label{2_0_9}
\end{equation}
From the theory of Markov processes it is known that exactly such a Boltzmann-scaling of the transition rates evokes a thermal state of the master equation (detailed balance)~\cite{Metropolis,LandauBinder,Klaers4}. The Kennard-Stepanov relation ensures that the photon gas for sufficiently long times acquires a thermal equilibrium state.

\subsection{Chemical equilibrium}
Besides energy exchange with the heat bath (temperature $T$), the effective particle exchange between the photon gas and the dye reservoir assigns the photons a chemical potential $\mu_\gamma$. In the following, we will see that it is determined by the excitation level of the dye medium.

Due to $\hbar\omega_\textrm{\tiny zpl}\gg k_\textrm{\tiny B} T$, purely thermal excitation of molecules from their ground ($\downarrow$) to excited electronic ($\uparrow$) states is strongly suppressed, and optical photons ($\gamma$) are required to drive the transition. Vice versa, the decay of a molecule results (with an efficiency of $\simeq95\%$) in the emission of a photon. Altogether, such a behavior resembles a photochemical reaction:
\begin{equation}
\gamma\   + \downarrow\ \  \rightleftharpoons \ \ \uparrow
\label{2_0_10}
\end{equation}
In chemical equilibrium (i.e. zero net particle flux between different species) the particular chemical potentials balance, $\mu_\gamma + \mu_\downarrow = \mu_\uparrow$. Thus, the fugacity of the photons reads
 \begin{equation}
z=e^{\frac{\mu_\gamma}{k_\textrm{\tiny B} T}}=\left. {e^{\frac{\mu_\uparrow}{k_\textrm{\tiny B} T}}}\middle/{e^{\frac{\mu_\downarrow}{k_\textrm{\tiny B} T}}}. \right.
\label{eq_fugacity}
\end{equation}
The partition function of a dye molecule $\mathcal{F}=w_\downarrow \exp({\mu_\downarrow}/{k_\textrm{\tiny B} T})  + w_\uparrow \exp[{(\mu_\uparrow-\hbar\omega_\textrm{\tiny zpl}})/{k_\textrm{\tiny B} T}]$, with the statistical weights $w_{\downarrow,\uparrow}=\int_{\epsilon\geq 0}{D_{\downarrow,\uparrow}(\epsilon) \exp[-\epsilon/k_\textrm{\tiny B} T] \textrm{d}\epsilon}$, allows one to associate the molecular chemical potential with the probability of finding a molecule in the ground or excited electronic state, respectively:
\begin{equation}
w_\downarrow \frac{e^{\frac{\mu_\downarrow}{k_\textrm{\tiny B} T}}}{\mathcal{F}}= \frac{M_\downarrow}{M},\ \
w_\uparrow \frac{e^{\frac{\mu_\uparrow-\hbar\omega_\textrm{\tiny zpl}}{k_\textrm{\tiny B} T}}}{ \mathcal{F}}= \frac{M_\uparrow}{M}
\end{equation}
This probability is determined by the ratio of the number of excited and relaxed dye molecules $M_{\uparrow,\downarrow}$ and the total number of molecules $M$. By renormalising the chemical potential with respect to the cavity ground state energy, $\mu=\mu_\gamma - \hbar \omega_\textrm{\tiny c}$,~(\ref{eq_fugacity}) yields
\begin{equation}
e^{\frac{\mu}{k_\textrm{\tiny B} T}} =\frac{w_\downarrow}{w_\uparrow}  \frac{M_\uparrow}{M_\downarrow} e^{-\frac{\hbar\Delta}{k_\textrm{\tiny B} T}}.
\label{chemeq}
\end{equation}
In equilibrium, the chemical potential $\mu$ is thus determined by the dye molecular excitation level $M_\uparrow/M_\downarrow$ and the detuning between the condensate frequency and dye resonance $\Delta=\omega_\textrm{\tiny c}-\omega_\textrm{\setcounter{footnote}{2}\footnotesize zpl}$.

\subsection{Microcavity dispersion relation}
The microcavity photons can be formally described as a 2D harmonically trapped Bose gas. In the resonator filled with a dye medium with index of refraction $\tilde n$, the energy-momentum-relation of a photon in free space $E={\hbar c}/{\tilde n}\sqrt{k_r^2+k_z^2}$ with $c$ the speed of light, $k_{r}$ the radial and $k_z=\pi q/D(r)$ the longitudinal wave vector ($q$ is an integer longitudinal wave number) component gets modified by the cavity boundary conditions. These are determined by the mirror spacing $D(r)\simeq D_0 - r^2/R$ at a radial distance $r$ from the optical axis, where $D_0$ denotes the mirror separation on the optical axis. These parameters are illustrated in Fig.~\ref{fig8}(a). Using the paraxial approximation ($k_z\gg k_r$), the dispersion relation of the microcavity photons becomes
\begin{equation}
E  \simeq \frac{\pi \hbar c q }{\tilde n D_0} + \frac{\pi \hbar c q }{\tilde n R D_0^2} r^2 + \frac{\hbar c D_0}{2\pi q \tilde n} k_r^2. 
\label{2_1_6}
\end{equation}
In addition, we extend~(\ref{2_1_6}) by accounting for a nonlinear response of the refractive index subject to changes of the 2D photon density, i.e. the intensity of the light field. The total index of refraction $\tilde n=\tilde n_0 + \Delta \tilde n_r = \tilde n_0 +\tilde n_2 I(\textrm{\bfseries r\normalfont})$ can be written as a sum of the linear refractive index in the absence of photons $\tilde n_0$ and a nonlinear contribution $\tilde n_2$. The nonlinear term results from physical effects that lead to intensity-dependent energy shifts, as e.g. the optical Kerr effect~\cite{Boyd}, or temporally slow thermal lensing~\cite{Dung}. Assuming $\tilde n_0\gg \Delta \tilde n_r$, we obtain
\begin{equation}
E=m_\textrm{\tiny ph} \frac{c^2}{\tilde n_0^2} + \frac{\hbar^2 k_r^2}{2m_\textrm{\tiny ph}} + \frac{1}{2}m_\textrm{\tiny ph}\Omega^2 r^2 - \frac{m_\textrm{\tiny ph}c^2}{\tilde n_0^3}\tilde n_2 I(\textrm{\bfseries r\normalfont}).
\label{2_1_8}
\end{equation}
Here, the effective photon mass $m_\textrm{\tiny ph}={\pi\hbar q \tilde n_0}/(D_0 c) $ and trapping frequency $\Omega = c/\tilde n_0 \sqrt{D_0 R/2}$ have been introduced, revealing the formal equivalence of~(\ref{2_1_8}) with the dispersion of a massive, harmonically trapped particle moving non-relativistically in a 2D plane, see Fig.~\ref{fig8}(a, right). The first term in~(\ref{2_1_8}) determines the effective rest energy of the photons, a global energy shift determined by the cavity boundary conditions. It corresponds to the energy of the $q$-th longitudinal mode without any transverse excitations $m_\textrm{\tiny ph} ({\tiny c}/{\tilde n_0})^2 = E_{q00}=\hbar\omega_\textrm{\tiny c}$ with the cutoff frequency $\omega_\textrm{\tiny c}$. The eigenenergies in the cavity are given by 2D (isotropic) harmonic oscillator states $E_{n_x,n_y} =m_\textrm{\tiny ph} {c^2}/{n_0^2} + \hbar\Omega (n_x + n_y + 1)$ with quantum numbers $n_x$ und $n_y$. The eigenfunctions $\psi_{n_x,n_y}(x,y)= \psi_{n_x}(x)\cdot\psi_{n_y}(y)$ are given by the 1D solutions $\psi_{n}(x) = (\sqrt{2^n n!} \sqrt{\pi}b)^{-1} H_n({x}/{b}) \exp[{-{ x^2}/(2 b^2)}]$, where $b=\sqrt{{\hbar}/{m_\textrm{\tiny ph}\Omega}}$ denotes the oscillator length and $H_n(x)$ the Hermite polynomials.

\subsection{Statistical physics of microcavity photons}

\begin{figure*}[t]
\centering\includegraphics{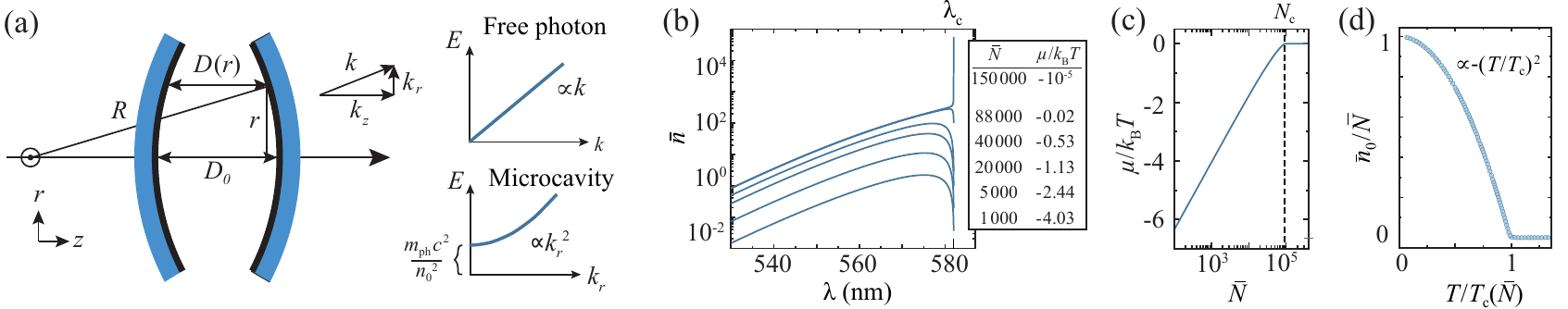}
\caption{(a) Microcavity geometry showing mirrors (radius of curvature $R$) separated by $D(r)$ at a transverse position $r$. In the paraxial approximation ($k_z\gg k_r$), one finds a modification of the photon dispersion relation from linear scaling $E=\hbar kc/\tilde n_0 $ in 3D free space (top) to a quadratic scaling with small transverse momenta $k_r$ in the microcavity (bottom), similar to the dispersion relation of a 2D massive particle. (b) Spectral occupation versus wavelength. The rest energy of the photons is determined by the cutoff wavelength $\lambda_\textrm{\tiny c}=h\tilde n_0 /(m_\textrm{\tiny ph} c)$. Below the critical photon number $N_\textrm{\tiny c}\approx 90\thinspace000$, the spectra show a Boltzmann scaling. For $\bar N>N_\textrm{\tiny c}$, the ground state becomes macroscopically occupied. (c) The chemical potential grows in the classical region with increasing particle number until it saturates at $\mu= 0$ around $N_\textrm{\tiny c}$. (d) The condensate fraction exhibits an quadratic scaling as a function of the reduced temperature. Experimentally, we adjust the reduced temperature by varying $T_\textrm{\tiny c}\propto \sqrt{N}$ to match room temperature $T=300~\si{\kelvin}$ when operating at the phase transition.}
\label{fig8}
\end{figure*}

In the following, we will describe the temperature behaviour of the (ideal) 2D photon gas in a harmonic trap~\cite{Pethick,Bagnato,Dalfovo,Hadzibabic2,Petrov}. We derive the critical particle number and temperature, respectively, as well as the spectral and spatial distributions for the experimentally studied photon gas. We can specify the transversal excitation energies in the harmonic trap
\begin{equation}
u_{n_x,n_y}=E_{n_x,n_y}-m_\textrm{\tiny ph}\frac{c^2}{\tilde n_0^2}-\hbar\Omega = \hbar\Omega \left(n_x+n_y\right)
\label{2_2_1}
\end{equation} 
with a degeneracy of the eigenstates $g(u) = 2 \left(u/(\hbar\Omega)+1\right)$, where the factor $2$ accounts for the two-fold polarisation degeneracy of the photons. At temperature $T$, the average occupation of an excited state with energy $u_{n_x,n_y}$ is given by the Bose-Einstein distribution
\begin{equation}
\bar n_{T,\mu}(u) = \frac{g(u)}{\exp[(u-\mu)/k_\textrm{\tiny B} T]-1}.
\label{2_2_3}
\end{equation} 
Here, we have implicitly assumed that the system is grand-canonical with a chemical potential $\mu$ adjusting the average total particle number $\bar N$ under the constraint $\bar N = \sum_{u=0,\hbar\Omega,2\hbar\Omega,...}{\bar n_{T,\mu}(u)}$. At high temperatures or low total photon numbers, the chemical potential obeys $\mu/k_\textrm{\tiny B} T\ll0$, and~(\ref{2_2_3}) equals the classical Boltzmann distribution. In the opposite limit ($T\rightarrow 0$ or $N\rightarrow \infty$), the chemical potential converges asymptotically to the ground state energy $\mu\rightarrow 0^-$ (Fig.~\ref{fig8}~(c)) and the ground state becomes macroscopically occupied. The phase transition to a BEC occurs at the critical photon number or temperature, respectively,
\begin{equation}
N_\textrm{\tiny c} = \frac{\pi^2}{3}\left(\frac{k_\textrm{\tiny B} T}{\hbar \Omega}\right)^2\ \ \ \ \textrm{,}\ \ \ \ \ T_\textrm{\tiny c}=\frac{\sqrt{3}}{\pi}\frac{\hbar\Omega}{k_\textrm{\tiny B}} \sqrt{\bar N},
\label{2_2_8}
\end{equation}
and as a function of the cavity parameters $T_\textrm{\tiny c}\propto (\bar N/R)^{1/2}$. Notably, an equilibrium phase transition requires its critical temperature to remain finite in the thermodynamic limit ($\bar N,V\rightarrow \infty$). It can be achieved by increasing the particle number $\bar N$ and volume $V\propto R^2$ in a way that conserves $\bar N/R$, i.e. by gradually switching off the trapping potential $R\rightarrow\infty$.

The expected spectral photon distributions for increasing chemical potentials are shown in Fig.~\ref{fig8}(b). The condensation fraction scales quadratically with the reduced temperature, ${\bar n_0}/{\bar N}=1-\left(T/T_\textrm{\tiny c}\right)^2$, see Fig.~\ref{fig8}(d), as expected for a 2D harmonically trapped ideal Bose gas~\cite{Pethick,Petrov}. In this confinement, BEC occurs not only in momentum space but also in position space. The spatial intensity distribution of the condensed photon gas is the sum over all oscillator eigenfunctions weighted with the Bose-Einstein factor:
\begin{eqnarray}
I_{T,\mu}(x,y) \simeq \frac{2m_\textrm{\tiny ph}c^2}{\tilde n_0^2 \tau_\textrm{\setcounter{footnote}{2}\footnotesize rt}} \sum_{n_x,n_y}{   \frac{|\psi_{n_x,n_y}(x,y)|^2}{ \exp\left({\frac{\hbar\Omega(n_x+n_y)-\mu}{k_\textrm{\tiny B}T}}\right) - 1 }  }
\label{2_2_14}
\end{eqnarray}
The power per photon is accounted for by $m_\textrm{\tiny ph} (c/\tilde n_0)^2/\tau_\textrm{\setcounter{footnote}{2}\footnotesize rt}$, where $\tau_\textrm{\setcounter{footnote}{2}\footnotesize rt}=2 D_0\tilde n_0/c$ denotes the photon round trip time of in the resonator. This approximation is valid due to $\hbar \Omega\sim 0.1~\si{\meV}$ being much smaller than the rest energy $m_\textrm{\tiny ph}(c/\tilde n_0)^2\sim 1~\si{\electronvolt}$. 

\begin{figure}[t]
\flushright
\centering\includegraphics{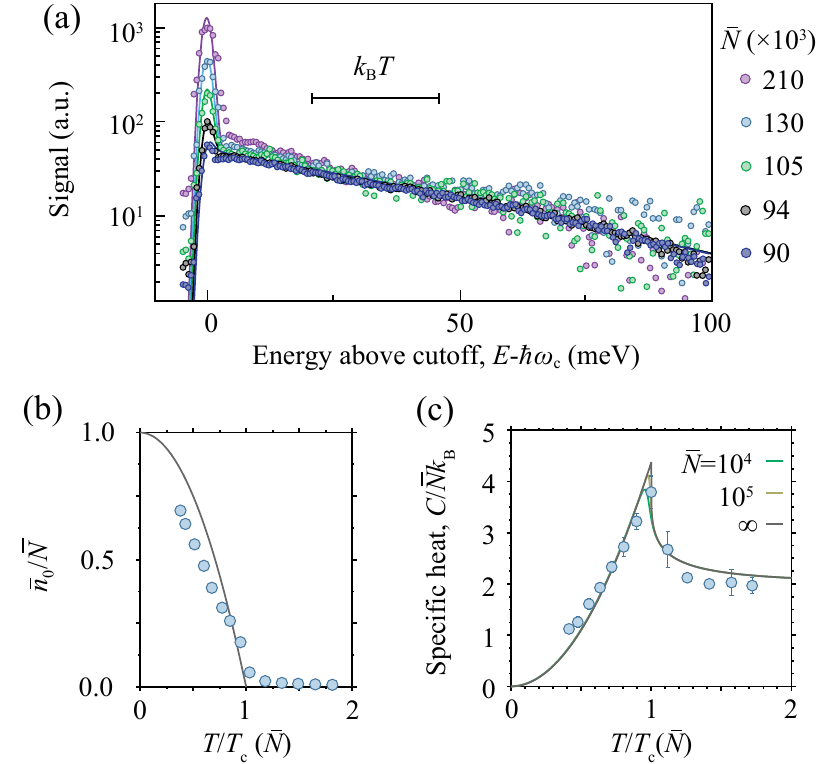}
\caption{Evidence for BEC of light. (a) Photon spectra for increasing total photon numbers $\bar N$. For $\bar N\geq N_\textrm{\tiny c}$, the excited mode population saturates and the ground state becomes macroscopically occupied. (b) The condensate fraction and (c) the specific heat of the photon gas is derived from the spectral distributions (not shown) at various particle numbers above and below $N_\textrm{\tiny c}$. The discontinuity close to $T=T_\textrm{\tiny c}(\bar N)$ reveals the phase transition. Reproduced with permission from \cite{Schmitt4,Damm}. Copyright 2014 \& 2016 by the American Physical Society \& Nature Publishing Group.}
\label{fig12}
\end{figure}

\begin{table}[t]
\centering
\begin{tabularx}{0.5\columnwidth}{c|XXXXc}
\hline
$q$ & $7$ & $7$ & 8 & 7  &  \\ \hline
$R$ & 1 & 6 & 1 & 1  & $\si{m}$\\ \hline
$\lambda_\textrm{\tiny c}$ & 580 & 580 & 580 & 560  & $\si{\nano m}$ \\ \hline
$D_0$ & 1.42 & 1.42 & 1.62 & 1.37 & $\si{\micro m}$\\ \hline
$m_\textrm{\tiny ph}$ & 7.79 & 7.79 & 7.80 & 8.07  & $10^{-36}~\si{\kilogram}$ \\ \hline
$\Omega/2\pi$ &  39.6 & 16.2 & 37.1 & 40.3  & $\si{\GHz}$ \\ \hline
$N_\textrm{\tiny c}$ &  81\thinspace700 & 490\thinspace200 & 93\thinspace200 & 78\thinspace800  &  \\ \hline
\end{tabularx}
\caption{Photon gas parameters for some different microcavity geometries. (Refractive index $\tilde n_0=1.43$ for ethylene glycol, temperature $T=300~\si{\kelvin}$)}
\label{tab1}
\end{table}

The spectral and spatial distributions of the photon gas have been experimentally verified for the first time for both the classical and Bose-Einstein condensed phase in pioneering work by Klaers \itshape{et al.}\normalfont~\cite{Klaers1,Klaers2,Klaers3,Klaers4}. Subsequent studies have provided further insight into the phase transition, and revealed e.g. thermodynamic properties such as condensate fraction or specific heat~\cite{Damm,Damm2,Schmitt1,Schmitt2,Schmitt3,Schmitt4,Marelic,Marelic2,Greveling}; see Fig.~\ref{fig12} for an overview of some experimental signatures of photon BEC.

The purpose of the present Tutorial is to elucidate the coherence properties of BECs of light. For this study, typically realised experimental parameters for the microcavity setup are $q=8$ at a cavity cutoff wavelength of $\lambda_\textrm{\tiny c}=580~\si{\nano m}$, which is associated with a mirror separation of $D_0=1.62~\si{\micro m}$. The refractive index of the dye medium (Rhodamine 6G solved in ethylene glycol) amounts to $\tilde n_0=1.43$ at room temperature $T=300~\si{\kelvin}$. Most of the experiments are conducted using mirrors with a radius of curvature of $R=1~\si{m}$. Therefore, the effective photon mass is $m_\textrm{\tiny ph}=7.8\times 10^{-36}~\si{\kilogram}$ and the frequency of the harmonic trap is $\Omega/2\pi = 37~\si{\GHz}$. With this one expects a critical particle number $N_\textrm{\tiny c}=93\thinspace000$. Due to the short resonator round trip time $\tau_\textrm{\footnotesize rt}=15~\si{\fs}$, the average circulating optical power in the resonator at threshold becomes $P_\textrm{\tiny c}\approx 2.1~\si{\watt}$. The highly-reflecting mirrors transmit a fraction $\tilde T \simeq 2.5\times 10^{-5}$ of the optical power. At criticality, the continuous power of the cavity emission is approximately $5~\si{\nano\watt}$. Table~\ref{tab1} summarises parameter sets, which are used in the course of the discussed experimental sequences.

\section{Multimode photon dynamics}
\label{thermalisierungsdynamik_theor}

Bose-Einstein condensation is a phase transition of the Bose gas in thermal equilibrium. The thermalisation of a nonequilibrium system can occur via different mechanisms and with characteristic dynamics. Atomic gases e.g. relax into equilibrium by contact interactions, while microcavity photons inherit their temperature solely from the thermal contact to a molecular heat bath. The atomic equilibration requires several interatomic collisions~\cite{Monroe,Snoke2}, whereas the photons can be thermalised after only a single absorption-emission-cycle. In this Section, we will theoretically investigate the photon dynamics.

\subsection{Rate equation model}
We start our discussion by analysing the rate equations for absorption and emission of photons in the microcavity modes. As the dye solution is embedded into the cavity volume, fluorescence emission occurs directly into the reabsorbing medium. After an absorption process, the high collision rate between solvent and dye molecules at room temperature leads to decoherence of the molecular dipoles~\cite{Lakowicz}. The photon-dye-system here correspondingly is in the weak coupling regime~\cite{Yokoyama,Angelis}. In first-order perturbation theory, the photon dynamics can thus be adequately modelled by semiclassical rate equations, which are determined by the time evolution of the diagonal elements of the density matrix.

To begin with, we consider a configuration of the light field $\{n_1,n_2,...,n_i,...\}$ with $n_i$ photons in the $i$-th cavity mode. The transition rates (per volume)
\begin{equation}
\begin{aligned}
R_{12}^i(\textrm{\bfseries r\normalfont}) & = B_{12}(\omega_i) u^i(\textrm{\bfseries r\normalfont}) \rho_\downarrow n_i
\\
R_{21}^i(\textrm{\bfseries r\normalfont}) & = B_{21}(\omega_i) u^i(\textrm{\bfseries r\normalfont}) \rho_\uparrow (n_i + 1),
\label{3_2}
\end{aligned}
\end{equation}
give the probability (per time) to absorb or emit a photon in mode $i$ at position \bfseries r \normalfont with the frequency-dependent Einstein coefficients for absorption and emission $B_{12,21}(\omega_i)$, the spectral energy density per photon $u^i(\textrm{\bfseries r\normalfont})$, and the densities of ground and excited state molecules $\rho_{\downarrow,\uparrow}$. Due to the densities on the right-hand-side in~(\ref{3_2}) one obtains rates per volume, which yield absolute rates after integrating out the resonator volume. For the transverse ground state ($i=q00$) with $\omega_i=E_{q00}/\hbar$ and $n_i=n$, this gives
\begin{equation}
R_n^{12} =  B_{12}\left(\frac{E_{q00}}{\hbar}\right) u^{q00}(0)  \frac{n \rho_\downarrow}{|f^{q00}(0)|^2}.
\label{3_3}
\end{equation}
Here, we have expressed the energy density $u^{q00}(\textrm{\bfseries r\normalfont})=u^{q00}(0)  {|f^{q00}(\textrm{\bfseries r\normalfont})|^2}/{|f^{q00}(0)|^2}$ by the normalised mode function $f^{q00}(\textrm{\bfseries r\normalfont})$. Using the effective mode volume $\tilde V^{q00}_\textrm{\tiny eff} = \int{{|f^{q00}(\textrm{\bfseries r\normalfont})|^2}/{\max \left\{|f^{q00}(\textrm{\bfseries r\normalfont})|^2\right\}} dV} = 1/|f^{q00}(0)|^2 $~\cite{Andreani,Kristensen}, the modified 
Einstein coefficients $\hat B_{12,21}= B_{12,21}(E_{q00}/\hbar) u^{q00}(0)$ and the number of ground and excited state molecules $M_{\downarrow,\uparrow}= \rho_{\downarrow,\uparrow} \tilde V^{q00}_\textrm{\tiny eff}$, one obtains the rate equations for the ground mode populated with $n$ photons
\begin{equation}
\begin{aligned}
R_n^{12}  & = \hat B_{12} M_\downarrow n   =  \hat B_{12} (M-X+n) n \\
R_n^{21}  & = \hat B_{21} M_\uparrow (n+1)  =  \hat B_{21} (X-n) (n+1).
\label{3_6}
\end{aligned}
\end{equation}
According to the photochemical reaction in~(\ref{2_0_10}), we have expressed the rates as a function of the sum of all molecular and photonic excitations $X=M_\uparrow + n$ and the total molecule number in the ground mode volume $M = M_\downarrow + M_\uparrow=M_\downarrow + X - n$, which we assume to be constant reservoir parameters.

The rate equations readily provide the temporal evolution of the photon number
\begin{equation}
\frac{\partial}{\partial t} n_i =  \hat B_{21} M_\uparrow (n_i+1) - (\hat B_{12} M_\downarrow +\gamma_{\textrm{\tiny ph},i})n_i
\label{3_7}
\end{equation}
with a photon loss rate $\gamma_{\textrm{\tiny ph},i}$ due to mirror transmission. To conserve the excitation number $X$, any loss must be compensated for by a net gain $P$ in the molecule rate equations, which is experimentally realised by pumping with a laser beam: 
\begin{equation}
-\frac{\partial}{\partial t} M_\downarrow = \frac{\partial}{\partial t} M_\uparrow = P - \sum_i{\frac{\partial}{\partial t} n_i} -\gamma_\textrm{\tiny M}M_\uparrow
\label{3_8}
\end{equation}
Additionally, $P$ must balance the molecular loss rate $\gamma_\textrm{\tiny M}$, which results from non-radiative decay and fluorescence into unconfined leakage modes. 

\subsection{Steady-state photon number}
The rate equation model enables a quantitative description of the photon thermalisation dynamics in the microcavity. For this, we consider a simplified model for the multimode photon gas in the uncondensed phase without spatial photon transport, losses or pumping ($P=\gamma_\textrm{\tiny M}=\gamma_{\textrm{\tiny ph},i}=0$). Here, the molecule number ($10^8$) exceeds the average photon number per mode ($10^1$), such that the rate equations of the dye medium in~(\ref{3_8}) can be considered as quasi-stationary with a fixed molecular excitation level $M_\uparrow/M_\downarrow$. A more refined model including dissipative spatial dynamics has been theoretically reported by Kirton and Keeling~\cite{Kirton1,Kirton2}. Our own detailed numerical simulations of the spatial photon dynamics are discussed in Section \ref{numerischesimulationdynamik}~\cite{Schmitt3}.

For a single cavity mode (angular frequency $\omega_i$), the Kennard-Stepanov relation reads ${\hat B^i_{21}}/{\hat B^i_{12}}={w_\downarrow}/{w_\uparrow}\exp[{-{\hbar(\omega_i-\omega_\textrm{\tiny zpl})}/{k_\textrm{\tiny B} T}}]$. Together with~(\ref{chemeq}), we obtain the average photon number in thermal and chemical equilibrium from~(\ref{3_7}):
\begin{equation}
\bar n_i  =  \left({\frac{\hat B^i_{12}}{\hat B^i_{21}}\frac{M_\downarrow}{M_\uparrow}-1}\right)^{-1} = \left(e^{\frac{\hbar(\omega_i-\omega_\textrm{\tiny c})-\mu}{k_\textrm{\tiny B} T}}-1\right)^{-1}
\label{3_11}
\end{equation}
By summing~(\ref{3_7}) over all degenerate cavity modes with the energy $\epsilon_i=\hbar \omega$, we obtain the rate equation for the photon number $n\equiv n(\omega,t)=\sum_{\epsilon_i=\hbar \omega}{n_i(t)}$ in the multimode cavity:
\begin{equation}
\begin{aligned}
\frac{\partial n}{\partial  t}& =\hat B_{21}\left[n+\sum_{\epsilon_i=\hbar\omega}{1}\right]M_\uparrow-\hat B_{12} n M_\downarrow\\
& = \hat B_{21} n M_\uparrow + \hat A_{21} M_\uparrow - \hat B_{12} n M_\downarrow,
\label{3_11_1}
\end{aligned}
\end{equation}
The term $\sum_{\epsilon_i=\hbar\omega}{1}$ gives the energy-dependent mode density $g(\omega)=2\left[(\omega-\omega_\textrm{\tiny c})/\Omega + 1\right]$. In the second step, we have identified the Einstein coefficient for spontaneous emission $\hat A_{21}=g(\omega) \hat B_{21}$. The steady-state photon number is $\bar n(\omega) = {g(\omega)}\left\{\exp\left[{\hbar(\omega-\omega_\textrm{\tiny c})-\mu}/{k_\textrm{\tiny B} T}\right] - 1\right\}^{-1}$.

\subsection{Spectral photon number evolution}
To determine the thermalisation time, we rephrase the single mode~(\ref{3_7}) as $\dot n_i + \alpha n_i + \beta = 0$, with the coefficients $\alpha  =  \hat B^i_{12} M_\downarrow - \hat B^i_{21} M_\uparrow$ and $\beta  =  - B^i_{21}M_\uparrow$. For the initial condition $n_i(0)=0$, this differential equation is solved by
\begin{equation}
n_i(t)=-\frac{\beta}{\alpha} \left[1-e^{-\alpha t}\right] = \bar n_i \left[1-e^{-t/\tau_i}\right],
\label{3_13}
\end{equation}
with the time constant $\tau_i=(\bar n_i + 1)/(\hat B^i_{12}  M_\downarrow)= {\bar n_i}/({\hat B^i_{21} M_\uparrow})$. We expand~(\ref{3_13}):
\begin{equation}
n_i(t)  = \hat B_{21} M_\uparrow t\ \left[1- \frac{t}{2}\left(\hat B_{12} M_\downarrow - \hat B_{21} M_\uparrow \right)\right].
\label{3_16_1_2}
\end{equation}
For early times, we can neglect the second-order term $\propto t^2$, so that the spectrum in this limit will be determined by the emission profile $\hat B_{21}(\omega)$. If we approximate~(\ref{3_16_1_2}) for the uncondensed regime with $M_\uparrow\ll M_\downarrow\hat B_{12}/\hat B_{21}$, we obtain the characteristic time after which the initial spectral redistribution of the photon gas occurs
\begin{equation}
\tau_{\textrm{\tiny th}}\simeq \frac{1}{\hat B_{12} M_\downarrow}.
\label{3_16_2}
\end{equation}
This equals the mean reabsorption time of a photon in the dye medium. The relative occupation of $n_i$ and $n_{i+1}$ of two neighbouring resonator modes (with frequencies $\omega_i,\omega_i+\Omega$) demonstrates, that the spectral slope is nearly thermal after $\tau_\textrm{\tiny th}$. Without loss of generality, we assume that the fluorescence strength into the modes is equal, $\hat B^i_{21}=\hat B^{i+1}_{21}$, as is indeed fulfilled for the used dyes (Section~\ref{thermalisierung}). Using the Kennard-Stepanov relation, the absorption coefficients $\hat B_{12}^i  = \hat B_{21}^i \exp[  {\hbar(\omega_i-\omega_\textrm{\tiny zpl})}/{k_\textrm{\tiny B} T } ]$ and $\hat B_{12}^{i+1} = \hat B_{12}^i \exp [{\hbar\Omega}/{k_\textrm{\tiny B} T } ]$ yield the photon dynamics
\begin{equation}
n_{i,i+1}(t) \simeq M_\uparrow \hat B_{21}^{i,i+1} t \left[1-\frac{t}{2} M_\downarrow \hat B_{12}^{i,i+1}\right].
\label{3_16_4}
\end{equation}
An expansion in $\hbar\Omega/k_\textrm{\tiny B}T$ determines the spectral population difference at the thermalisation time
\begin{equation}
\frac{n_{i+1}-n_i}{\Omega}(\tau_\textrm{\tiny th}) = - \frac12 \frac{\hbar}{k_\textrm{\tiny B} T}\frac{M_\uparrow \hat B_{21}^i}{M_\downarrow \hat B_{12}^i} = - \frac12 \frac{\hbar}{k_\textrm{\tiny B} T}\bar n_i
\label{3_16_6}
\end{equation}
in the limit of a Boltzmann distribution with $\bar n_i=M_\uparrow \hat B_{21}^i (M_\downarrow \hat B_{12}^i)^{-1}$. For the equilibrium distribution in~(\ref{3_11}), the similar scaling $({\bar n_{i+1} - \bar n_i})/{\Omega} = -{\hbar}/({k_\textrm{\tiny B} T}) \bar n_i$ demonstrates that the photon spectrum agrees except for a factor $1/2$ with the spectral shape of a Boltzmann distribution after the reabsorption time $\tau_\textrm{\tiny th}$. Accordingly, the microcavity photons have relaxed to a thermal-like equilibrium after completing approximately one emission-absorption-cycle.

\begin{figure}[t]
\centering\includegraphics{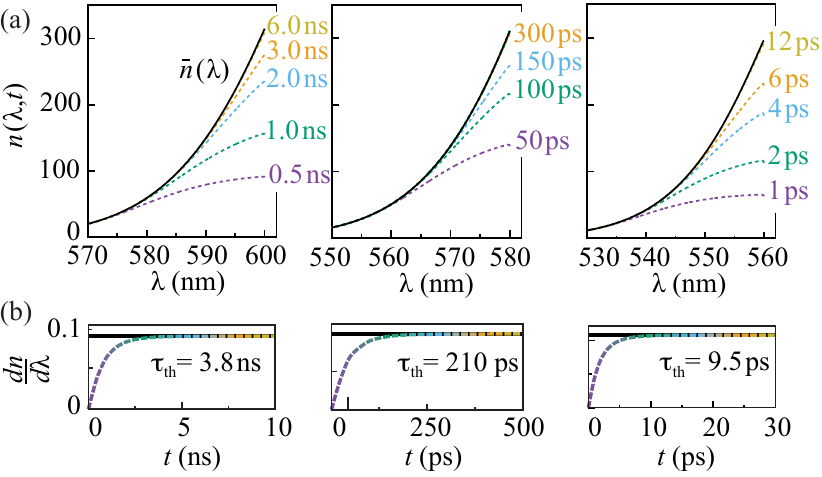}
\caption{(a) Spectral thermalisation dynamics for $\lambda_\textrm{\tiny c}=\{600;580;560\}~\si{\nano m}$. All shown equilibrium spectra $\bar n(\lambda)$ (solid lines) are close to the condensation threshold. The wavelength-dependent dye reabsorption, $\sigma(\lambda)\propto \exp(-\lambda)$, leads to a slower photon number evolution (dashed) in the red ($10^{-9}~\si{\second}$) than in the green spectral region ($10^{-11}~\si{\second}$). (b) The low-wavelength spectral slope (dashed) quantifies the degree of thermalisation, which approaches the equilibrium value (solid) with advancing times. After $\tau_\textrm{\tiny th}$ the relative difference between both curves is less than $1\%$. (Dye concentration $0.1~\si{\milli \mole/\litre}$, $\mu=-0.007k_\textrm{\tiny B}T$)}
\label{fig14}
\end{figure}

In general, the relaxation rates of individual modes depend on their frequencies. It is therefore helpful to express the Einstein coefficient for absorption $\hat B_{12}(\omega)$ as a function of the experimentally accessible cross section $\sigma(\omega)$. By comparing the coefficients in Beer's law ${\partial n}/{\partial t} = -M_\downarrow \tilde V_\textrm{\tiny eff}^{-1}\sigma(\omega) c\ n$ and the rate equation for absorption ${\partial n}/{\partial t} = -M_\downarrow \hat B_{12}(\omega) n$, we find the useful relation
\begin{equation}
\hat B_{12}(\omega) = \frac{\sigma(\omega)c}{\tilde V_\textrm{\tiny eff}}.
\label{3_15}
\end{equation}
Figure~\ref{fig14}(a) shows the calculated temporal evolution of the spectra in~(\ref{3_13}) for Rhodamine 6G dye (Section~\ref{thermalisierung}). Owing to the wavelength-dependence of the absorption cross section (maximum near $530~\si{\nano m}$), the time $\tau_\textrm{\tiny th}$ after which the spectral distribution has relaxed to a thermal equilibrium distribution $\bar n(\lambda)$ varies for different spectral regions. While the photon gas relaxation takes several nanoseconds in the red spectral region (Fig.~\ref{fig14}(b), left), for the yellow-green spectral region a thermalisation time of a few picoseconds is predicted (right). For example, Rhodamine 6G absorbs photons at $580~\si{\nano m}$ wavelength with cross section $\sigma(2\pi c/\lambda)\simeq 10^{-22}~\si{m}^2$. With the mirror separation $D_0\simeq 1.6~\si{\micro m}$ and the diameter of the TEM$_{00}$ mode $d_{0}\simeq 12~\si{\micro m}$ we can estimate the effective ground mode volume $\tilde V_\textrm{\tiny eff,00}=\pi (d_{0}/2)^2 D_0\simeq 1.8\times 10^{-16}~\si{m}^3$ and the rate coefficient $\hat B_{12}(\omega) \simeq 166~\si{\second}^{-1}$. For typical dye concentrations near $1~\si{\milli \mole/\litre}$, around $M_\downarrow\approx 10^8$ dye molecules reside in the mode volume. Therefore, the expected thermalisation time is $\tau_\textrm{\tiny th}\approx 50~\si{\ps}$.

\subsection{Chemical equilibration time}
The spectral thermalisation time of the photon gas $\tau_\textrm{\tiny th}$ in the uncondensed phase is approximately given by the photon reabsorption time in the dye solution, $1/\hat B_{21} M_\downarrow$, see~\ref{3_16_2}. In general, this value differs from the chemical equilibration time $\tau_\textrm{\tiny ch}$, which is the time after which the system has acquired its steady-state-population $\bar n(\omega)$. To see this, we extend the single mode description in~(\ref{3_13}) to the multimode system:
\begin{equation}
n(\omega,t)=\bar n(\omega) \left[1-e^{- t/\tau(\omega)}\right],
\end{equation}
with $\tau(\omega)={\bar n(\omega)}/[g(\omega)\hat B_{21}(\omega) M_\uparrow]$. For a Boltzmann distribution $\bar n(\omega) = g(\omega) \hat B_{21}(\omega)M_\uparrow/[\hat B_{12}(\omega)M_\downarrow]$, the chemical equilibration time is the weighted average over all frequency-dependent relaxation times
\begin{equation}
\tau_\textrm{\tiny ch}  =  \frac{\int_{\omega_\textrm{\tiny c}}^{\infty}{ \tau(\omega)\bar n(\omega) d\omega }}{ \int_{\omega_\textrm{\tiny c}}^{\infty}{ \bar n(\omega) d\omega } }  \stackrel{k_\textrm{\tiny B}T\gg\hbar\Omega}{\simeq} \frac{\tau_\textrm{\tiny th}}{4}e^{-\frac{\hbar \Delta}{k_\textrm{\tiny B}T}},
\label{3_16_2_2}
\end{equation}
where we have assumed $\hat B_{21}(\omega)$ to be independent of $\omega$, as is roughly fulfilled for Rhodamine 6G dye within the relevant wavelength range ($540$ to $600\si{\nano m}$). In our experiments, the dye-cavity detuning $\Delta=\omega_\textrm{\tiny c}-\omega_\textrm{\tiny zpl}$ takes values between $\Delta_{560\si{\nano m}} = -2.4k_\textrm{\tiny B} T/\hbar$ and $\Delta_{605\si{\nano m}} =-8.7k_\textrm{\tiny B} T/\hbar$, implying the chemical equilibration time to exceed the spectral relaxation time by $\tau_\textrm{\tiny ch} / \tau_\textrm{\tiny th} \approx 3$ ($560\si{\nano m}$) up to $1500$ ($605\si{\nano m}$). This prediction is experimentally verified (Section~\ref{thermalisierung}). Below the critical photon number, our simplified analytical model provides an adequate description of the photon number thermalisation dynamics. It should be noted, that this model is insufficient to predict the dynamics in the Bose-Einstein condensed phase where the optical feedback onto the dye requires using the molecular rate equations. In this regime, the large photon number speeds up the dynamics by stimulated emission events and the chemical equilibration can become much faster than the spectral thermalisation, as will be discussed Section~\ref{numerischesimulationdynamik} on the basis of numerical simulations.

\section{Grand-canonical photon statistics}
\label{photonenzahlstatistik}

For BEC in the grand-canonical statistical ensemble regime, i.e. in the presence of a large particle reservoir, large statistical number fluctuations on the order of the total particle number $N$ have been predicted~\cite{Fierz,Fujiwara,Ziff,Holthaus1,Holthaus2,Kocharovsky,Weiss,Navez,Haar,Yukalov}. In contrast to this, the (micro-)canonical statistical ensemble features Poissonian number fluctuations, i.e. a scaling with $\sqrt{N}$; a situation realised in most atomic BECs~\cite{Ketterle1,Oettl,Dall,Hodgman,Schellekens}. In the dye-cavity system, Bose-Einstein condensed photons couple to electronic transitions of a specific number of dye molecules, which realises the repeated exchange of photon- and molecule-like excitations. The latter can be interpreted as an effective particle reservoir for the photons, with a size that depends on the molecule number and the dye-cavity detuning. We find that the photon number statistics of the ground state resembles a (nearly) Bose-Einstein distributed thermal light source, in contrast to both atomic BECs and the laser~\cite{Oettl,Freed,Milonni}. Under these conditions, the phase transition can be regarded as a BEC in the grand-canonical ensemble regime.

\subsection{Photon number distribution}
\noindent
We start by considering the master equation for the probability $p_n\equiv p_n(t)$ to find $n$ photons in the ground state at time $t$\setcounter{footnote}{2}\footnote{We will denote $n$ (and $\bar n$) as the (average) photon number in the BEC in Sections~\ref{photonenzahlstatistik} und~\ref{phasenkohaerenz}.}. The flow of probability between photons in the condensate and the reservoir is
\begin{equation}
\dot p_n = R^{21}_{n-1} p_{n-1} - (R^{12}_n + R_n^{21})p_n + R_{n+1}^{12} p_{n+1},
\label{3_17}
\end{equation}
with the rates given by~(\ref{3_6})~\cite{Abraham,Klaers5}. According to the experiment, we assume $M=M_\uparrow+M_\downarrow$ and $X=M_\uparrow+n$ to be constant. For $t\rightarrow\infty$ the probability flow $p_n(t)$ is expected to become stationary, $\dot p_n(\infty)=0$, and the photon number distribution converges to its equilibrium value $\mathcal{P}_n:=p_n(\infty)$. In this limit,~(\ref{3_17}) is solved by the recursive ansatz $\mathcal{P}_n = \mathcal{P}_0\prod_{k=0}^{n-1}{R_k^{21}/R_{k+1}^{12}}$, and one obtains the photon number statistics 
\begin{equation}
\frac{\mathcal{P}_n}{\mathcal{P}_0} = \frac{(M-X)!X!}{(M-X+n)! (X-n)!}\left(\frac{\hat B_{21}}{\hat B_{12}}\right)^n,
\label{3_18}
\end{equation}
which is used to calculate the average condensate number and its fluctuations. Similarly, the statistics can been derived by a entropy maximisation principle~\cite{Sobyanin}.

In general,~(\ref{3_18}) has to be evaluated numerically. At constant temperature $T$, we induce the phase transition by increasing the particle number $\bar N$, which effectively lowers the reduced temperature $T/T_\textrm{\tiny c}(\bar N)$. For each $\bar N$, the following numerical method then computes the  excitation number $X$ that recovers the given particle number $\bar N$: For a starting value $X$, the average photon number in the condensate
\begin{equation}
\bar n = \sum_{n\geq 0}{n \mathcal{P}_n}
\end{equation}
and the molecular excitation level of the medium in the ground mode volume
\begin{equation}
\frac{M_\uparrow}{M_\downarrow} = \frac{X-\bar n }{M-X+\bar n}
\end{equation}
are computed. As the density of excited molecules (and thus the excitation level) is required to be spatially homogeneous in chemical equilibrium, the ratio ${M_\uparrow}/{M_\downarrow}$ controls the chemical potential for the photon gas, see~(\ref{chemeq}). Accordingly, the number of photon in excited states is $\bar n_\textrm{exc}= \sum_{u>0}{{g(u)}/({\exp[(u-\mu)/(k_\textrm{\tiny B} T)]-1}})$. If there are residual deviations between $\bar n + \bar n_\textrm{\tiny exc}$ and the target photon number $\bar N$, the numerical method is iterated with an adjusted excitation number $X$ until a certain level of precision is reached.

\begin{figure}[t]
\centering\includegraphics{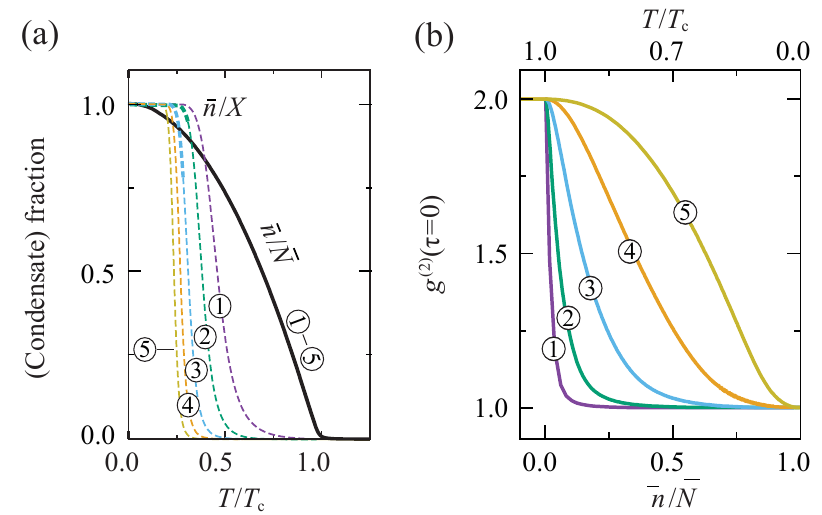}
\caption{(a) Condensate fraction $\bar n / \bar N$ and photonic fraction of the excitation number $\bar n /X$ versus reduced temperature $T/T_\textrm{\tiny c}(\bar N)$ for reservoir sizes $M_\textrm{\tiny \ding{172}-\ding{176}}=\{10^8;...;10^{12}\}$ and dye-cavity detuning $\hbar\Delta = -4.67k_\textrm{\tiny B} T$ (adapted from~\cite{Klaers5}). (b) The autocorrelation function $g^{(2)}(0)$ of the condensate (reservoirs as in (a)) predicts large photon number fluctuations even deep in the condensed phase. Reproduced with permission from \cite{Klaers5}. Copyright 2012 by the American Physical Society.}
\label{fig15}
\end{figure}

Figure~\ref{fig15}(a) shows the calculated condensate fraction $\bar n/\bar N$ and the photonic fraction of the excitation number $\bar n  / X$ for five different-sized molecular reservoirs as a function of the reduced temperature. The constant dye-cavity detuning $\hbar\Delta= -4.67k_\textrm{\tiny B} T$ controls the Kennard-Stepanov relation $\hat B_{21}/\hat B_{12}$ and hence the photon statistics in~(\ref{3_18}). For all studied reservoirs, the condensate fraction follows the analytic solution $\bar n / \bar N = 1- \left(T/T_\textrm{\tiny c}\right)^2$ and the curves for $\bar n / X$ reveal that a large number of excitations are present as molecular excitations down to very low temperatures. Furthermore, Fig.~\ref{fig15}(b) shows the zero-delay autocorrelation function
\begin{equation}
g^{(2)}(\tau = 0)= \frac{\langle n(n-1)\rangle }{\bar n^2} = \frac{\sum_{n\geq 0}{n(n-1)\mathcal{P}_n}}{\left(\sum_{n\geq 0}{n\mathcal{P}_n}\right)^2}
\label{3_22}
\end{equation}
for the same reservoir parameters as a function of the condensate fraction and the reduced temperature. For $T\geq T_\textrm{\tiny c}$, the ground state occupation exhibits the usual, strong intensity fluctuations in a single mode of the thermal Bose gas, $g^{(2)}(0)=2$, and the photon number statistics is Bose-Einstein-distributed. In the presence of large reservoirs, the intensity correlations maintain when the temperature is lowered deep into the condensed phase, as attributed to the grand-canonical particle exchange with the dye reservoir. For $T/T_\textrm{\tiny c}\ll1$, the statistical number fluctuations are damped out and our calculations demonstrate the emergence of second-order coherence, $g^{(2)}(0)=1$, with Poissonian statistics. We do not find indications that the transition between both statistical regimes is accompanied by a discontinuity in the thermodynamic quantities, excluding a further phase transition scenario within the Bose-Einstein condensed phase. The crossover of the photon statistics in the condensed phase remains valid also in the thermodynamic limit, as will be discussed later.

\subsection{Asymptotic photon number distributions}
We show that the photon number statistics interpolates between a Bose-Einstein- and Poissonian distribution. To analytically derive the limiting cases, we rewrite~(\ref{3_18}) in a recursion form:
\begin{equation}
\frac{\mathcal{P}_{n+1}}{\mathcal{P}_{n}} = \frac{X-n}{M-X+n+1} \frac{\hat B_{21}}{\hat B_{12}}
\label{3_23}
\end{equation}

For \itshape Bose-Einstein statistics \normalfont to apply, $\mathcal{P}_n$ must follow a geometric series with a ratio $\mathcal{P}_{n+1}/\mathcal{P}_n$ being independent of $n$. This is fulfilled if and only if the reservoir dimensions $M$ and $X$ are very large, so that the photon number on the right-hand-side of~(\ref{3_23}) can be safely neglected, i.e $X\gg n$ and $M-X\gg n$ ("grand-canonical limit"). With $X\simeq M_\uparrow$ and $M-X \simeq M_\downarrow$,
\begin{equation}
\frac{\mathcal{P}_{n+1}}{\mathcal{P}_{n}} \stackrel{\textrm{g.c.}}{=} \frac{M_\uparrow}{M_\downarrow} \frac{\hat B_{21}}{\hat B_{12}} \ \ \Rightarrow \ \ \frac{\mathcal{P}_{n}}{\mathcal{P}_{0}} \stackrel{\textrm{g.c.}}{=} \left(\frac{M_\uparrow}{M_\downarrow} \frac{\hat B_{21}}{\hat B_{12}}\right)^{n}.
\label{3_24}
\end{equation}
Here, $\mathcal{P}_n$ decays exponentially from its maximum at $n=0$. Normalisation of~(\ref{3_24}) gives
\begin{equation}
\mathcal{P}_{n}  =  \left(1-\frac{M_\uparrow}{M_\downarrow}\frac{\hat B_{21}}{\hat B_{12}}\right) \left(\frac{M_\uparrow}{M_\downarrow} \frac{\hat B_{21}}{\hat B_{12}}\right)^{n} =  \frac{\left(\frac{\bar n}{\bar n + 1}\right)^n}{\bar n + 1}.
\label{3_25}
\end{equation}
In the last step, we have identified the average condensate number from~(\ref{3_11}). This result remains valid also for increased $M$, as long as the excitation level $M_\uparrow/M_\downarrow\simeq X/(M-X)$ (and thus $\mu,\bar n,\bar N$) are kept constant. Equation~(\ref{3_25}) is the well-known Bose-Einstein statistics, see Fig.~\ref{fig16}, which also applies for example for chaotic, thermal light or blackbody radiation.

In the case of \itshape Poisson statistics\normalfont, the most probable photon number is finite, ${n_\textrm{\tiny max}}>0$. Under the assumption $\mathcal{P}_{n_\textrm{\tiny max}+1} = \mathcal{P}_{n_\textrm{\tiny max}}$,~(\ref{3_23}) yields
$n_\textrm{\tiny max} = X - ({M + 1})/({1 + {\hat B_{21}}/{\hat B_{12}}})$. Expanding for $\Delta n=n - n_\textrm{\tiny max}$,
\begin{equation}
\frac{\mathcal{P}_{n+1}}{\mathcal{P}_{n}} = 1 - \frac{\Delta n}{\lambda} + \frac{1}{1 + {\hat B_{21}}/{\hat B_{12}}} \left( \frac{\Delta n}{\lambda}  \right)^2 - ...
\label{3_27}
\end{equation}
with $\lambda ={\hat B_{21}}/{\hat B_{12}} {M+1}/{(\hat B_{21}/\hat B_{12}+1)^2}$. In the low temperature limit, the ratio of the Einstein coefficients scales with the dye-cavity detuning $\Delta$. For a negative detuning, as in our experiments, it diverges:
\begin{eqnarray}
{\frac{\hat B_{21}(\omega)}{\hat B_{12}(\omega)}} = {\frac{w_\downarrow}{w_\uparrow} e^{-\frac{\hbar\Delta}{k_\textrm{\tiny B} T}}}\ \stackrel{{T\rightarrow 0}}{\simeq}\left\{
\begin{array}{@{}ll@{}}
        0, & \Delta>0\\
        \infty, & \Delta<0
\end{array}\right.
\label{3_29}
\end{eqnarray}
Hence,~(\ref{3_27}) simplifies to ${\mathcal{P}_{n+1}}/{\mathcal{P}_{n}}{\simeq}{\lambda}/({\lambda + \Delta n})$ (or $\simeq({\lambda - \Delta n})/{\lambda}$) for $\Delta>0$ (or $\Delta<0$). This recursion formula implies the relative probability near $\Delta n$ around the maximum $n_\textrm{\tiny max}$:
\begin{equation}
\frac{\mathcal{P}_{n_\textrm{\tiny max}+\Delta n}}{\mathcal{P}_{n_\textrm{\tiny max}}} \simeq 
\left\{
\begin{array}{@{}ll@{}}
\frac{(\lambda-1)!}{(\lambda-1+\Delta n)!} \lambda^{\Delta n},	& \Delta>0\\
\frac{\lambda!}{(\lambda-\Delta n)!}  \lambda^{-\Delta n}, & \Delta<0
\end{array}\right.
\label{3_31}
\end{equation}
Upon transforming $\Delta n\rightarrow -\Delta n$, both distributions are the same and their relative scaling is analogous to a Poisson distribution
\begin{equation}
\mathcal{P}^\textrm{\tiny p}_n=e^{-\lambda}\frac{\lambda^n}{n!}\ \ \Rightarrow\ \  \frac{\mathcal{P}^\textrm{\tiny p}_{n_\textrm{\tiny max}+\Delta n}}{\mathcal{P}^\textrm{\tiny p}_{n_\textrm{\tiny max}}} = \frac{\lambda!}{(\lambda+\Delta n)!}\lambda^{\Delta n}
\label{3_32}
\end{equation}
which contains only one parameter $\lambda$ for mean and variance. The solutions in~(\ref{3_31}) however are only Poissonian with respect to the relative photon number $\Delta n$, as the an additional parameter $n_\textrm{\tiny max}$ tunes the most probable photon number. For example, in the limit $T\rightarrow 0$ ($\Delta\neq 0$) the statistics peaks at $n_\textrm{\tiny max}=\bar n=\bar N$ with $\lambda = 0$, where all photons of the systems have condensed into the ground state and the photon number is precisely known.

\subsection{Statistics crossover}
\begin{figure}[t]
\centering\includegraphics{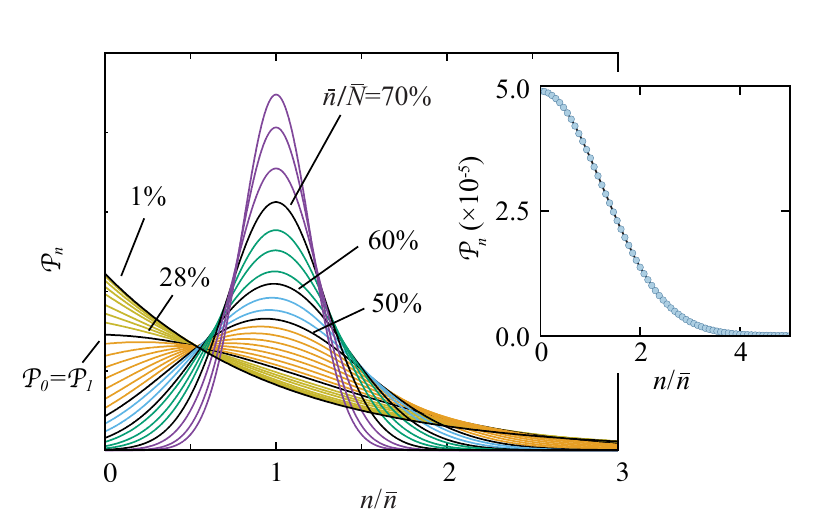}
\caption{Crossover from Bose-Einstein to Poissonian statistics for increasing $\bar n/\bar N$ at fixed reservoir size. The probability $\mathcal{P}_n$ is plotted versus the photon number normalised with the mean condensate population $\bar n$. Up to $\bar n /\bar N= 28.3\%$ the maximum of the distribution occurs at $n_\textrm{\tiny max}=0$. For increasing quantum degeneracy, the distribution shifts to $n_\textrm{\tiny max}>0$ and the variance is reduced, demonstrating a damping of particle number fluctuations and the emergence of second-order coherence. \itshape Inset\normalfont : Photon number statistics at the crossover $T_\textrm{\tiny x}=0.847~T_\textrm{\tiny c}$ ($\bar n/\bar N\simeq28.3\%$). The probability distribution corresponds to a gaussian with $\bar n \approx 27\thinspace000$ photons. ($M=10^{10}, \hbar\Delta=-2.4k_\textrm{\tiny B}T$, polarisation degeneracy neglected; the curves normalised in the $n/\bar n$-representation for clarity)}
\label{fig16}
\end{figure}

Figure~\ref{fig16} shows numerically calculated photon number distributions to find $n$ photons in the BEC for a fixed reservoir size ($M=10^{10}$, $\Delta=\omega_\textrm{\tiny c}-\omega_\textrm{\tiny zpl}=-2.4k_\textrm{\tiny B} T/\hbar$). By decreasing the reduced temperature from $T/T_\textrm{\tiny c}=1.0$ to $0.4$, or vice versa increasing the condensate fraction from $\bar n /\bar N \simeq 1\%$ to $80\%$, one observes a continuous crossover from Bose-Einstein to Poissonian statistics.

As the crossover point between both statistical regimes, we define the reduced temperature $T_\textrm{\tiny x}/T_\textrm{\tiny c}(\bar N)$ when the most probable photon number ceases to be $n_\textrm{\tiny max}~=~0$, or in other words when the condition $\mathcal{P}_0=\mathcal{P}_1$ is fulfilled (Fig.~\ref{fig16}, inset). Inserting into~(\ref{3_23}) yields:
\begin{equation}
\frac{M+1}{X} = 1 + \frac{\hat B_{21}}{\hat B_{12}}
\label{3_33}
\end{equation}
Due to the large number of molecules, we assume $M+1\simeq M$. With regard to the experimental conditions, we derive $T_\textrm{\tiny x}$ for fixed total numbers of molecules $M$ and photons $\bar N$. With $X=\bar n + M_\uparrow$ the average condensate population follows
 \begin{equation}
\bar n  =  M \frac{\frac{\hat B_{12}}{\hat B_{21}}\frac{M_\downarrow}{M_\uparrow}  - 1}{\left( 1+ \frac{\hat B_{12}}{\hat B_{21}} \right)\left( 1 + \frac{M_\downarrow}{M_\uparrow} \right)}.
\label{3_34}
\end{equation}
In the grand-canonical limit ($M_\downarrow, M_\uparrow \gg \bar n$), see~(\ref{3_25}), the nominator corresponds to the inverse of the average photon number $\bar n = \sum_{n=0}^{\infty} {n\mathcal{P}_n} = [ ({\hat B_{12}}/{\hat B_{21}}) ({M_\downarrow}/{M_\uparrow}) - 1]^{-1}$. Although grand-canonical conditions do not strictly apply in the crossover region, we use this to estimate $T_\textrm{\tiny x}$. We find
\begin{equation}
\bar n = \sqrt{\frac{M}{\left[ 1+ \frac{w_\uparrow}{w_\downarrow} e^{ \frac{\hbar\Delta}{ k_\textrm{\tiny B} T}} \right] \left[ 1 + \frac{w_\downarrow}{w_\uparrow} e^{\frac{\hbar\omega_\textrm{\tiny zpl}-\mu_\gamma}{k_\textrm{\tiny B} T} } \right]}},
\label{3_36}
\end{equation}
where both the Kennard-Stepanov relation and chemical equilibrium have been applied. Moreover, it is safe to assume $w_\downarrow =w_\uparrow$ for the statistical weights of ground and excited molecular states~\cite{Lakowicz}. In the condensed phase, the chemical potential of the photons $\hbar(\omega_\textrm{\tiny c} + \Omega)<\mu_\gamma<\hbar\omega_\textrm{\tiny c}$, and consequently we can use $\mu_\gamma \simeq \hbar\omega_\textrm{\tiny c} = \hbar(\Delta + \omega_\textrm{\tiny zpl})$ to simplify the second bracket term in the denominator.

Equation~(\ref{3_36}) resembles a boundary for the average number of condensed photons, up to which the particle number statistics can be considered Bose-Einstein-like. With $\bar n =\bar N [ 1-({T_\textrm{\tiny x}}/{T_\textrm{\tiny c}})^2 ]$, this implicitly determines the temperature $T_\textrm{\tiny x}$ for the crossover. To investigate the scaling of the reduced crossover temperature $t=T_\textrm{\tiny x}/T_\textrm{\tiny c}$ with the system parameters, we rewrite~(\ref{3_36}):
\begin{equation}
1-t^2 = {\frac{\sqrt{M/2}}{\bar N}}\left[1 + \cosh \left(\frac{\hbar\Delta}{k_\textrm{\tiny B}  T_\textrm{\tiny c}}\frac{1}{t} \right) \right]^{-{1}/{2}}
\label{3_38}
\end{equation}
The temperature depends only on relative size of the subsystems $\sqrt{M}/\bar N$ and the reduced detuning $\hbar\Delta/k_\textrm{\tiny B} T_\textrm{\tiny c}$, which plays an important role for the thermodynamic limit: $\bar N, R, M \rightarrow \infty$ with ${R}/{\bar N}=\textrm{const.}$ and ${\sqrt M}/{\bar N }=\textrm{const.}$ The first condition conserves the critical temperature $T_\textrm{\tiny c}$ and therefore fixes $\hbar\Delta/k_\textrm{\tiny B} T_\textrm{\tiny c}$, see~(\ref{2_2_8}). The second requirement conserves $T_\textrm{\tiny x}$ (below $T_\textrm{\tiny c}$), which rules out that the temperature difference arises from finite size effects. Notably, both regimes, Bose-Einstein- and Poissonian statistics, exist \itshape within \normalfont the condensed phase. While the former relates to the grand-canonical ensemble ($M\gg \bar n^2$), the latter refers to a canonical ensemble scenario ($M\ll \bar n^2$). The crossover between both regimes is induced by changing $\bar n/\bar N$ or $\Delta$, respectively. To highlight this,Tab.~\ref{tab2} summarises numerically calculated values for $T_\textrm{\tiny x}/T_\textrm{\tiny c}$ for different reservoirs $M_\textrm{\setcounter{footnote}{2}\footnotesize\ding{172}-\ding{176}}=10^8-10^{12}$ at fixed photon gas sizes $\bar N =10^5$ and dye-cavity-detuning $\hbar \Delta=-2.4 k_\textrm{\tiny B} T_\textrm{\tiny c}$ ($\lambda_\textrm{\tiny c}=560~\si{\nano m}$ and $\lambda_\textrm{\tiny zpl}=545~\si{\nano m}$). For sufficiently large reservoirs, the Bose-Einstein-like grand-canonical statistics extends deep into the condensed phase.
\begin{table}[t]
\centering
\begin{tabularx}{0.5\columnwidth}{c|c|c|c|c|c}
\hline
$M_\textrm{\ding{172}-\ding{176}}$ & $ 10^8$ & $10^9$ & $10^{10}$ & $10^{11}$ & $10^{12}$  \\ \hline\hline
$T_\textrm{\tiny x}/T_\textrm{\tiny c}$ & 0.979 & 0.946 & 0.847 & 0.602 & 0.359  \\ \hline
$\bar n/\bar N $ & 3.0\% & 9.2\% & 28.3\% & 63.1\% & 86.7\%  \\ \hline
$g^{(2)}(0)$ & 1.5706 & 1.5707  & 1.5708  & 1.5708   & 1.5708 \\ \hline
\end{tabularx}
\caption{Numerically calculated reduced temperatures and condensate fractions, at which the crossover between Bose-Einstein and Poissonian statistics occurs for different-sized molecular particle reservoirs $M_\textrm{\ding{172}-\ding{176}}$. Here, the autocorrelation gives $g^{(2)}(0)\simeq \pi/2$, see~(\ref{3_41}). ($\hbar \Delta=-2.4 k_\textrm{\tiny B} T_\textrm{\tiny c}$, $\bar N =10^5$)}
\label{tab2}
\end{table}

The inset of Fig.~\ref{fig16} shows the Gaussian photon number distribution at $T_\textrm{\tiny x}/T_\textrm{\tiny c}=0.847$, given by $\mathcal{P}_n= {2}/{(\pi \bar n)} \exp[{-( {n}/{\bar n} )^2/\pi}]$. Accordingly, the zero-delay second-order correlation function reads
\begin{equation}
g^{(2)}(0)  =  \frac{\sum_{n\geq 0}{n(n-1)\mathcal{P}_n}}{\left[\sum_{n\geq 0}{n\mathcal{P}_n}\right]^2} = \frac{\pi}{2}-\frac{1}{\bar n } \stackrel{\bar n\gg 1}{\simeq} \frac{\pi}{2},
\label{3_41}
\end{equation}
which analytically reproduces the numerical results inTab.~\ref{tab2}. It corresponds to relative condensate number fluctuations of ${\delta n}/{\bar n}=\sqrt{g^{(2)}(0)-1} \simeq 75\%$.

\subsection{Second-order time correlations}

We extend our discussion of the photon statistics to the temporal dynamics of the statistical fluctuations~\cite{Castin}. The condensate photons are absorbed by $M_\downarrow$ molecules in the electronic ground state, and $M_\uparrow$ excited molecules decay by emission of photons into the condensate mode. Neglecting losses, the rate~(\ref{3_7}) becomes
\begin{equation}
\frac{\partial}{\partial t} n =\hat B_{21} \left(X-n\right)\left(1+n\right)-\hat B_{12} n \left(M-X+n\right)
\label{3_44}
\end{equation}
with $X=M_\uparrow+n$, $M=M_\downarrow + M_\uparrow$ and the steady-state solution $\bar n = 1/(\hat B_{12}\bar M_\downarrow/\hat B_{21}\bar M_\uparrow-1)$. To quantify the time evolution of deviations from $\bar n$, we define $\delta n(t) = n(t) - \bar n$ and obtain
\begin{eqnarray}
\frac{\partial}{\partial t} \delta n(t)  &=&  -(\hat B_{12} + \hat B_{21}) \delta n(t)^2 - \gamma \delta n(t),
\label{3_45}\\
\gamma & = & \frac{\hat B_{21} X}{\bar n}+(\hat B_{12} + \hat B_{21})\bar n\nonumber\\
& \simeq & \frac{\hat B_{12}\hat B_{21}}{\hat B_{12}+\hat B_{21}}\frac{M}{\bar n} + (\hat B_{12} + \hat B_{21})\bar n .
\label{3_46}
\end{eqnarray}
For typical experimental parameters, $(\hat B_{12}+\hat B_{21})\delta n\simeq 10^6~\si{\second}^{-1}$ and $\gamma\simeq 10^9~\si{\second}^{-1}$, the coefficients in~(\ref{3_45}) comply with $(\hat B_{12}+\hat B_{21})\bar n \sqrt{g^{(2)}(0)-1} \ll\gamma$, where $\delta n/\bar n =\sqrt{g^{(2)}(0)-1}$. The equation of motion can thus be linearised, ${\partial_t}\delta n(t)\simeq -\gamma  \delta n(t)$. The second-order autocorrelations at times $t$ and $t'=\tau+t$ reads
\begin{equation}
g^{(2)}(\tau)=\frac{\langle n(t+\tau) n (t)\rangle }{\bar n^2}= 1 + \frac{\langle \delta n(\tau)\delta n(0)\rangle}{\bar n^2},
\label{3_47}
\end{equation}
where $\langle \delta n(t+\tau)\rangle = \langle \delta n(t)\rangle =0$ has been used. Using the quantum regression theorem~\cite{Castin,Scully,Lax1} allows us to trace back the dynamics $\langle \delta n(\tau)\delta n(0)\rangle$ to the evolution of $\delta n(t)$: ${\partial_t}\langle \delta n (\tau) \delta n (0)\rangle \simeq  -\gamma \langle \delta n (\tau) \delta n (0)\rangle$. We find
\begin{equation}
g^{(2)}(\tau)-1\simeq  \left[g^{(2)}(0)-1\right] \exp\left(-\frac{\tau}{\tau^{(2)}_\textrm{\tiny c}}\right),
\label{3_51}
\end{equation}
where $\tau^{(2)}_\textrm{\tiny c}=\gamma^{-1}$ denotes the second-order correlation time. With the Kennard-Stepanov relation, the inverse correlation time can be recast as a function of experimental parameters:
\begin{equation}
\frac{1}{\tau^{(2)}_\textrm{\tiny c}} = \hat B_{21}\left[  \frac{M}{\bar n \left({1+e^{-\frac{\hbar \Delta}{k_\textrm{\tiny B} T}}}\right)} + \bar n \left({1+e^{\frac{\hbar \Delta}{k_\textrm{\tiny B} T}}}\right) \right]
\label{3_52}
\end{equation}
Around the crossover temperature $T_\textrm{\tiny x}$ from grand-canonical to canonical ensemble conditions, the second-oder correlation rate exhibits a piecewise scaling
\begin{figure}[t]
\centering\includegraphics{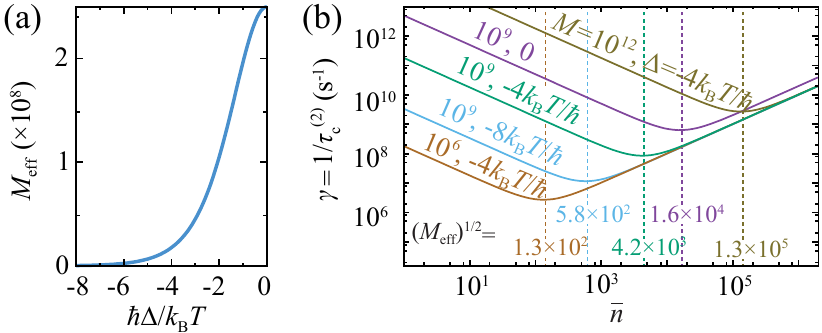}
\caption{(a) Effective reservoir size as a function of the dye-cavity-detuning. (b) Inverse second-order correlation time versus condensate number $\bar n$ for five effective reservoir sizes ${M_\textrm{\tiny eff}}$. The curves exhibits minima at $\bar n_\textrm{\tiny min} =({M_\textrm{\tiny eff}})^{1/2}$, which highlight the crossover point from grand-canonical ($\bar n <\bar n_\textrm{\tiny min}$) to canonical statistics ($\bar n > \bar n_\textrm{\tiny min}$). ($\hat B_{21}=10^4~\si{\second}^{-1}$, $M=10^9$).}
\label{fig18}
\end{figure}
\begin{equation}
\frac{1}{\tau^{(2)}_\textrm{\tiny c}} =
\left\{
\begin{array}{@{}ll@{}}
\frac{M}{\bar n }\frac{\hat B_{12}\hat B_{21}}{\hat B_{12} + \hat B_{21}},& T\gg T_\textrm{\tiny x}\ (\bar n^2\ll M_\textrm{\tiny eff})\\
2\bar n (\hat B_{12} + \hat B_{21}),& T=T_\textrm{\tiny x}\ \ (\bar n^2= M_\textrm{\tiny eff})\\
\bar n (\hat B_{12} + \hat B_{21}),&	T\ll T_\textrm{\tiny x}\ (\bar n^2\gg M_\textrm{\tiny eff}),\\
\end{array}\right.
\label{3_53}
\end{equation}
where we have introduced the effective reservoir size
\begin{equation}
{M_\textrm{\tiny eff}}=\frac{M}{{2}}{\left[1 + \cosh \left(\frac{\hbar\Delta}{k_\textrm{\tiny B} T}\right) \right]^{-1}}.
\label{3_53a}
\end{equation}
Figure~\ref{fig18}(a) illustrates the variation of the effective reservoir size as a function of the dye-cavity-detuning, and Fig.~\ref{fig18}(b) gives a plot of $\gamma$ versus $\bar n$ for various reservoir sizes. For a specific $M_\textrm{\tiny eff}$ the inverse correlation time decays in the grand-canonical regime ($g^{(2)}(0)\rightarrow 2$) with increasing condensate number, until it reaches a minimum at $\bar n_\textrm{\tiny min}=\sqrt{M_\textrm{\tiny eff}}$. In the canonical regime ($g^{(2)}(0)\rightarrow 1$), the fluctuation rate exhibits the opposite behaviour growing linearly with increasing photon numbers. This analytic prediction is confirmed by numerical Monte Carlo simulations, see Fig.~\ref{fig21}(b) in Section~\ref{phasenkohaerenz}. To exemplify the order of magnitude of $\gamma$, we give an estimate based on the typical experimental parameters discussed near the statistics crossover. For a condensate wavelength $\lambda_\textrm{\tiny c}=580~\si{\nano m}$, corresponding to a dye-cavity-detuning $\hbar\Delta=-5.3k_\textrm{\tiny B} T$ (Rhodamine 6G) and Einstein coefficients  $\hat B_{12}\simeq 170~\si{\second}^{-1}$ and $\hat B_{21}\simeq 3.4\times 10^4~\si{\second}^{-1}$, one obtains in the presence of $M= 10^{10}$ molecules an average condensate number of $\bar n_\textrm{\tiny min} =M_\textrm{\tiny eff}\simeq 7000$ photons. From~(\ref{3_53}) a time scale for the intensity fluctuations $\tau_\textrm{\tiny c}^{(2)}\approx 2~\si{\ns}$ is expected, which is close to the experimental observation.

\begin{figure}[b]
\centering\includegraphics{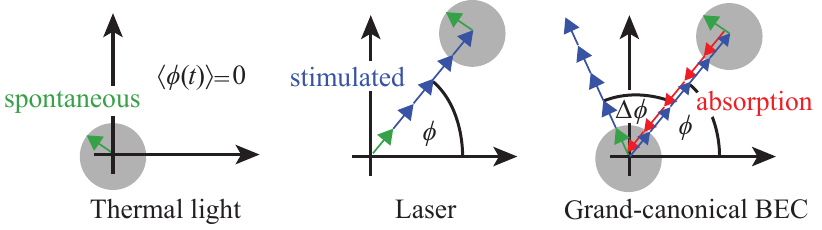}
\caption{(a) Phasor model for different light sources. While spontaneous emission on average does not develop a global phase, stimulated processes in a laser result in a macroscopic phase-stable light field. For the photon BEC, similarly a global phase $\phi$ emerges. In the presence of large reservoirs, however, the condensate vanishes due to strong dye-mediated reabsorption and subsequently emerges with a broken symmetry $\phi+\Delta \phi$.}
\label{fig19}
\end{figure}

\section{Phase coherence}
\label{phasenkohaerenz}

The phasor model allows a description of the temporal amplitude and phase evolution, $\sqrt{n(t)}$ and $\phi(t)$, of an optical single-mode field containing $n$ photons. As shown in Fig.~\ref{fig19}, it can be a valuable tool to consider qualitative differences between light sources:

(i) In a \itshape thermal light source\normalfont, the superposition of spontaneously emitted photons with arbitrary phases leads to a random walk of the total phase and destructive interference prohibits the emergence of a macroscopically occupied state with a stationary phase~\cite{Loudon}. Bose-Einstein photon statistics here gives rise to a most probable photon number $n_\textrm{\tiny max}=0$. 

(ii) Stimulated emission in a \itshape laser \normalfont results in a macroscopic occupation of a single optical mode with a nearly stable phase. The mode selection is induced by engineering losses in all undesired modes; making laser emission in general an out-of-equilibrium phenomenon. Residual spontaneous emission into the laser mode causes an amplitude and phase uncertainty~\cite{Schawlow}. Poissonian photon number statistics with $n_\textrm{\tiny max}>0$ lead to a vanishing probability to find zero photons $\mathcal{P}_0=0$.

(iii) Finally, the phasor diagram is also helpful to illustrate the phase dynamics of a \itshape BEC of photons\normalfont , where a reservoir induces large statistical, thermal-like (Bose-Einstein statistics) fluctuations of the condensate amplitude $\sqrt{n(t)}$. In this setting, the photon number eventually drops to $n=0$ and the subsequent spontaneous emission of a photon starts a cascade of stimulated processes forming a new macroscopically occupied ground mode. Due to the randomness of spontaneous emission we expect to observe the total phase of the wave function to change discretely in the course of time.

\subsection{Phase dynamics of the wave function}
Based on the rate~(\ref{3_6}) we perform Monte Carlo simulations of the photon number and phase evolution of the BEC based on the phasor model~\cite{LandauBinder}. While stimulated absorption and emission do not alter the phase of the wave function, spontaneously emitted photons cause a Heisenberg-type phase diffusion~\cite{Lewenstein,Imamoglu1,Leeuw}.

The phasor of the condensate with $n$ photons and phase $\phi$ is described by the complex number $\sqrt{n} e^{i\phi}$. Following a spontaneous emission event with random phase $\theta$ and amplitude $\sqrt{1}$ the phasor is modified to
\begin{equation}
\sqrt{n+\Delta n}\ e^{i(\phi+\Delta\phi)} = \sqrt{n} e^{i\phi}+(\sqrt{n+1}-\sqrt{n})e^{i\theta},
\label{3_55}
\end{equation}
which corresponds to a length change $\Delta n = 1 + 2\sqrt{n} \cos\theta$ and phase rotation $\Delta\phi = {\sin\theta}/{\sqrt{n}}$. The spontaneous phase becomes relevant for small photon numbers. Notably, the phase rotation describes only relative changes of the phase and does not apply to the case $n=0$, when the randomly selected phase $\theta$ breaks the symmetry to determine the overall phase of a re-emerging BEC.

\begin{figure}[t]
\centering\includegraphics{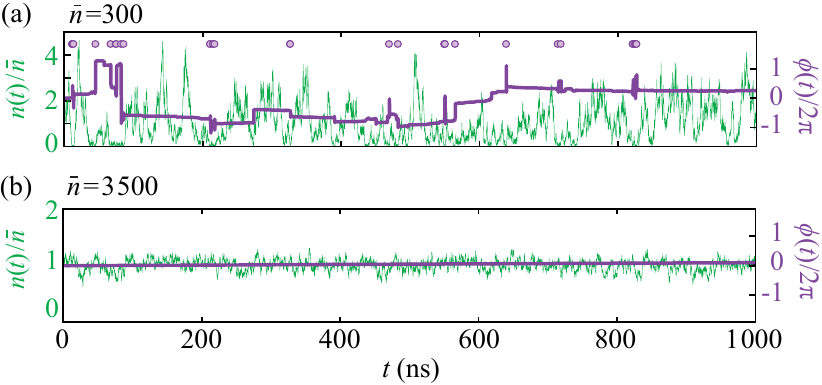}
\caption{Simulation of condensate number $n(t)/\bar n$ (green) and phase evolution $\phi(t)$ (purple) in the presence of a particle reservoir $M_\textrm{\tiny eff}=2.5\times 10^5$. (a) Under grand-canonical conditions, large number fluctuations occur accompanied by discrete phase jumps at points, when the photon number drops to zero (top circles). (b) For larger condensate sizes (canonical), the fluctuations are damped out. No phase jumps occur as attributed to the vanishing probability to find zero photons for Poissonian statistics. ($\Delta = 0k_\textrm{\tiny B} T$ and $M=10^6$)}
\label{fig20}
\end{figure}

Figure~\ref{fig20}(a) shows a Monte Carlo simulation of the time evolution of the (normalised) occupation number and corresponding phase for a fluctuating BEC coupled to a reservoir that is compatible with grand-canonical statistics, $\bar n < \sqrt{M_\textrm{\tiny eff}}=500$. The data reveal discrete phase jumps at points when no photons are present in the ground mode. For the same reservoir, Fig.~\ref{fig20}(b) gives the number and phase dynamics of a BEC in the canonical ensemble, with $\bar n \simeq 3\thinspace 500 >\sqrt{M_\textrm{\tiny eff}}$. Due to its finite size the reservoir starts to saturate and the number fluctuations get damped. Notably, the zero-photon-probability $\mathcal{P}_0$ vanishes, such that discrete phase jumps are suppressed. To quantify the temporal phase stability of the condensate, we introduce the phase jump rate $\Gamma_\textrm{\tiny PJ} = \#\textrm{phase jumps}/\textrm{time interval}$. 

Moreover, our simulations of the photon number evolution yield the second-order correlation function $g^{(2)}(\tau)$ and its associated timescale $\tau^{(2)}_\textrm{\tiny c}$. The phasor amplitude modulation that results from the fluctuating condensate population (time constant $\tau^{(2)}_\textrm{\tiny c}$) is expected to affect the degree of first-order coherence $g^{(1)}(\tau)$ via phase diffusion. Figure~\ref{fig20}(a) (bottom), however, suggests that this effect is negligible in comparison to the large phase jumps. Experimentally, continuous phase drifts cannot be resolved with the applied interferometric method described in Section~\ref{phasenkohaerenzexp}.

\begin{figure}[t]
\centering\includegraphics{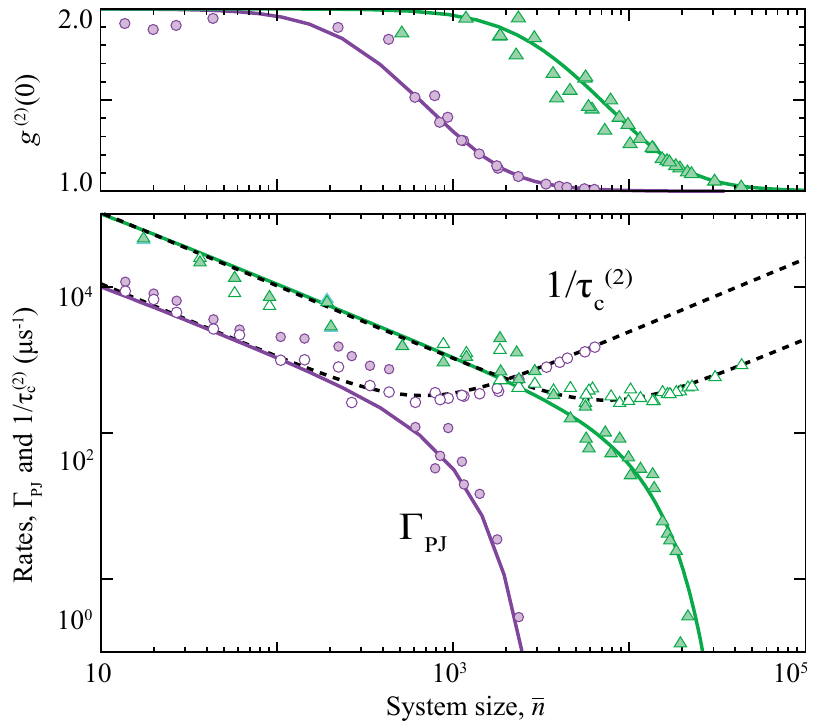}
\caption{Simulated autocorrelation $g^{(2)}(0)$ (top), phase jump rates (bottom, filled symbols) and second-order correlation rates (bottom, open) versus $\bar n$ for two effective reservoir sizes (violet circles and green triangles). The agreement of $\Gamma_{\textrm{\tiny PJ}}$ with $\hat B_{12}M \mathcal{P}_0$ (solid line) supports the assumption that phase jumps emerge when the condensate population vanishes. First-order coherence is enhanced as the photon statistics merges from being Bose-Einstein-like to Poissonian, which occurs at the $1/\tau^{(2)}_\textrm{\tiny c}$ minimum at $\bar n = \sqrt{M_\textrm{\tiny eff}}$, where $g^{(2)}(0)\simeq\pi/2$ (top) in good agreement with the analytical prediction (dashed line). ($\hat B_{12}=1\thinspace000~\si{\second}^{-1}$, $M=10^9$, $\hbar\Delta = \{-2.8;-7.7\} k_\textrm{\tiny B}T$)}
\label{fig21}
\end{figure}

Figure~\ref{fig21} shows $\Gamma_\textrm{\tiny PJ}$ and $1/\tau^{(2)}_\textrm{\tiny c}$ as a function of the average condensate number for two reservoirs, realised by varying $\Delta$. The data points are obtained from simulations similar to Fig.~\ref{fig20}. In the grand-canonical regime ($\bar n < \sqrt{M_\textrm{\tiny eff}}$), the phase jump and correlation rate decrease simultaneously with increasing system size. At the crossover to canonical statistics ($\bar n \geq \sqrt{M_\textrm{\tiny eff}}$), in Fig.~\ref{fig21} discernible by the autocorrelation value $g^{(2)}(0)\approx \pi/2$ as well as the minimum of $1/\tau_\textrm{\tiny c}^{(2)}$ (dashed line), both time scales separate. Beyond the minimum, we find the phase jumps to be more strongly suppressed, while an increase in the rate of second-order correlations is revealed in good agreement with the analytical prediction~(\ref{3_46}). This separation of coherence times is evident for both reservoirs in their respective crossover regions; the larger the reservoir, the further we find the phase jumps to persist in the condensed region. Strikingly, in this regime the photon condensate is expected to exhibit phase coherence despite large statistical number fluctuations characterised by $g^{(2)}(0)>1$.

\subsection{Phase jump rate}
Our numerical findings suggest that the phase jump rate of a fluctuating condensate $\Gamma_\textrm{\tiny PJ}$ correlates with the probability $\mathcal{P}_0$ to find zero photons in the ground state. To quantify this, we assume that discrete phase changes occur only in the absence of photons from the cavity ground state at a rate $\Gamma_\textrm{\tiny PJ}^0 = \mathcal{P}_0/{\tau_0}$. Here, $\tau_0$ labels a characteristic system time scale, i.e. the average time that a zero-photon-state exists in the cavity. Following a statistical fluctuation to a zero-photon-state, a certain time passes by until the condensate builds up with a new macroscopic phase, which is given by the inverse rate for spontaneous emission of a photon at the cutoff frequency. For $n=0$, the rate~(\ref{3_6}) depends only on the Einstein coefficient $\hat B_{21}$ and the number of excited dye molecules $M_\uparrow$, and thus $1/\tau_0 = \hat B_{21} M_\uparrow$. For the steady-state with $\bar n \gg 1$, see~(\ref{3_11}), one further obtains $\hat B_{21} M_\uparrow \simeq \hat B_{12} M_\downarrow$. With the typically fulfilled $M_\downarrow\simeq M$, we have
\begin{equation}
\Gamma_\textrm{\tiny PJ}^0 = B_{12} M \mathcal{P}_0.
\label{3_57_2}
\end{equation}
As $\mathcal{P}_0$ is determined by the photon number statistics, we consider the limiting cases: for a large reservoir (grand-canonical statistics), see~(\ref{3_25}), Bose-Einstein photon statistics gives $\mathcal{P}_0 = {1}/({\bar n + 1}) \simeq {1}/{\bar n}$. For small reservoirs (canonical statistics), the crossover to Poissonian statistics leads to a strong suppression of the zero-photon probability:
\begin{equation}
\Gamma_\textrm{\tiny PJ}^0 = \frac{\hat B_{12}M}{ \bar n^{\alpha}},\ \ \alpha = \left\{
\begin{array}{@{}ll@{}}
        1, & \textrm{ BE \& Gaussian}\\                              
        \infty, & \textrm{ Poisson}
\end{array}\right.
\label{3_58}
\end{equation}
To quantify the scaling of $\Gamma_\textrm{\tiny PJ}^0$ with $\bar n$, we have introduced the exponent $\alpha$, which interpolates between $1$ and $\infty$ when connecting grand-canonical and canonical ensemble conditions. In the Poissonian limit ($\alpha\rightarrow \infty$), the condensate thus exhibits the usual phase coherence. Using the above ansatz, the simulated phase jump rates can be reproduced as shown in Fig.~\ref{fig21} (solid line). The values for $\mathcal{P}_0$ were numerically calculated (Section~\ref{photonenzahlstatistik}).

\subsection{Thermodynamic limit: Correlation times}
Figure~\ref{fig21} illustrates the separation of time scales for phase and intensity fluctuations, which suggests that the coherence properties of a photon BEC differ fundamentally from those of a thermal light source, i.e. violating $g^{(2)}(\tau)=1+|g^{(1)}(\tau)|^2$~\cite{Loudon}. Therefore, the question arises whether the separation remains relevant in the thermodynamic limit. Combining~(\ref{3_46}) and~(\ref{3_58}) yields the ratio of the correlation times 
\begin{eqnarray}
\frac{\tau_\textrm{\tiny c}^{(1)}}{\tau_\textrm{\tiny c}^{(2)}} & = &\bar n^{\alpha-1} \ \left[\frac{\hat B_{12}}{\hat B_{12} + \hat B_{21}} + \frac{\hat B_{12}+\hat B_{21}}{\hat B_{21}}\frac{\bar n^2}{M}\right]\nonumber\\
& = & \bar N^{\alpha-1} \left[ K_1(T) + K_2(T)\left(\frac{\bar N}{\sqrt{M}}\right)^2 \right],
\label{3_59}
\end{eqnarray}
where we have used $\bar n =\bar N [1-(T/T_\textrm{\tiny c})^2] $ and included the temperature dependence in the constants $K_{1,2}(T)$. In the thermodynamic limit, $\bar N \rightarrow \infty$, $T_\textrm{\tiny c} =\textrm{const.} $ and $\bar N /\sqrt{M} = \textrm{const.}$, one expects the relative correlation times to scale with the parameter $\alpha$, which (like $\bar N/\sqrt{M}$) determines the photon number distribution $\mathcal{P}_n$ and $g^{(2)}(0)$. On the one hand, for a condensate in the grand-canonical regime ($\alpha=1$) at a temperature $T_\textrm{\tiny x}\leq T<T_\textrm{\tiny c}$, the dependence on the total particle number $\bar N$ in~(\ref{3_59}) vanishes and the correlation times coincide also in the thermodynamic limit. On the other hand, Poisson-like ($\alpha > 1$) and genuine Poisson statistics ($\alpha\rightarrow \infty$) are expected to cause a divergence of the first-order coherence time with respect to the second-order correlation time. Despite the here relatively  large condensate fluctuations with $g^{(2)}(0)\simeq 1.57$ ($1<\alpha <\infty$), a separation of time scales for first- and second-order is predicted for the thermodynamic limit. Similarly, the heuristic phase jump rate becomes
\begin{equation}
\Gamma_\textrm{\tiny PJ}^0 = K_3(T) \frac{M}{\bar N^2} \thinspace \frac{1}{\bar N^{\alpha-2}},
\label{3_60}
\end{equation}
where $K_3(T)$ denotes a temperature-dependent parameter, which does not change with the system size. We expect phase jumps to be fully suppressed in the thermodynamic limit only for Poissonian states with $\alpha > 2$.

Physically, the phase jumps originate from the persistence of fluctuations to zero-photon-states also in the thermodynamic limit caused by Bose-Einstein-like statistics. Provided the particle reservoir is sufficiently large, the time scales for number and phase fluctuations remain coupled even upon extrapolation of $\Gamma_\textrm{\tiny PJ}^0$ to the thermodynamic limit. On the one hand, the zero-photon-probability decays as $\mathcal{P}_0\propto \bar n^{-1}$ with increasing photon numbers ($\alpha=1$). On the other hand, this is counteracted by a quadratical increase of the molecule number required to conserve the photon statistics. Ultimately, this results in a larger phase jump rate.

\section{Thermalisation dynamics}
\label{thermalisierung}
In this section, we discuss experimental results of time-resolved measurements of the spectral photon kinetics, which shed light on the thermalisation dynamics to the molecular heat bath. Our measurements are performed for photon numbers near the critical particle number $N_\textrm{\tiny c}\simeq 90\thinspace000$ ($q=8$). Moreover, the experiment enables a spatially and spectrally-resolved observation of the transition dynamics from out-of-equilibrium, laser-like states to thermal equilibrium BECs for $\bar N\gg N_\textrm{\tiny c}$.

\begin{figure}[t]
\centering\includegraphics[width=0.5\columnwidth]{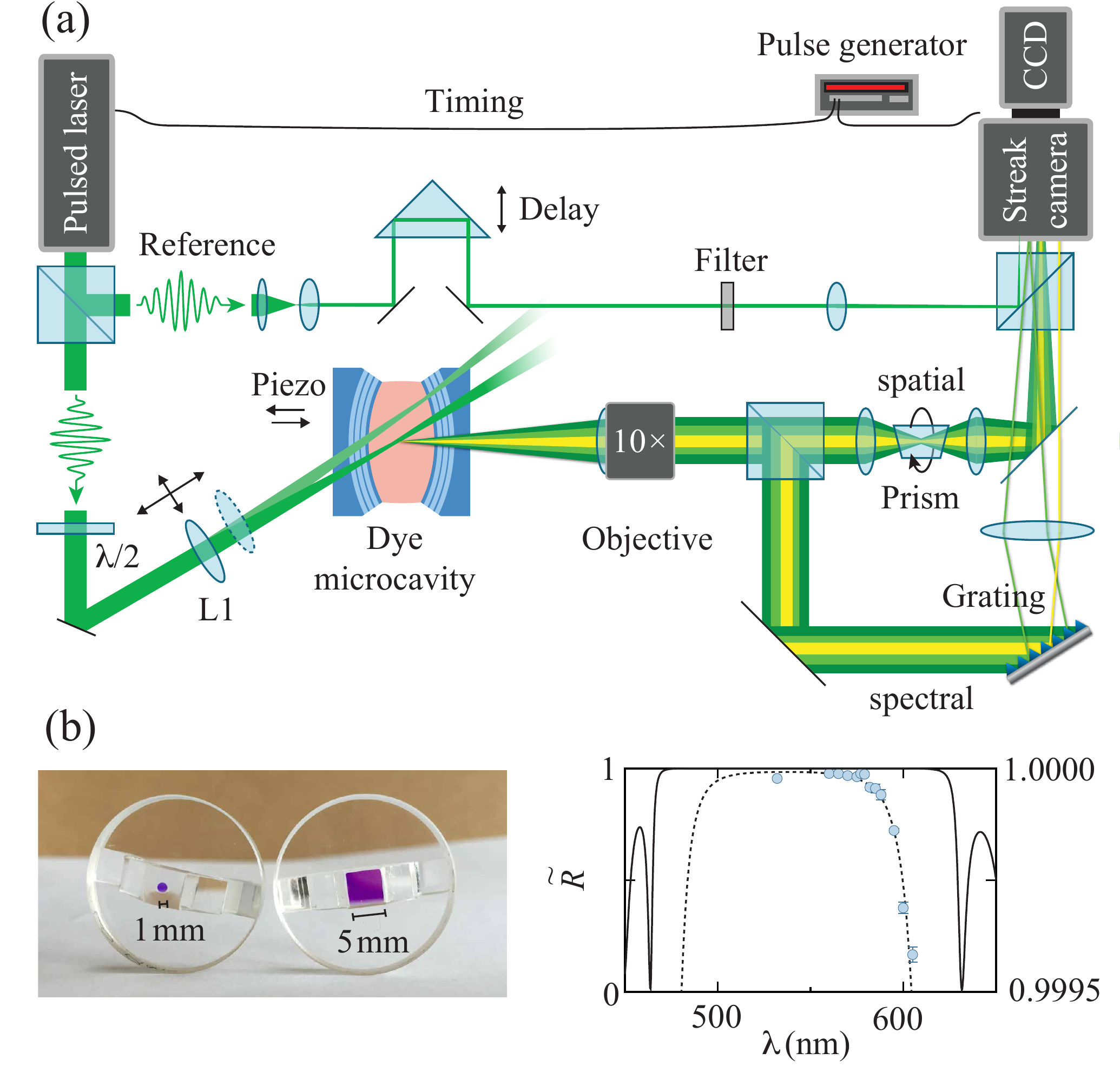}
\caption{(a) Experimental scheme for the time-resolved measurements of the spatial and spectral thermalisation dynamics. The microcavity is pumped with a pulsed laser beam and the emission imaged spatially and spectrally onto a streak-camera system. (b) Picture of the prepared cavity mirrors (left) and spectral mirror reflectivity $\tilde R$~\cite{Wahl}.}
\label{fig22}
\end{figure}

\subsection{Experimental scheme}
\begin{figure}[t]
\centering\includegraphics{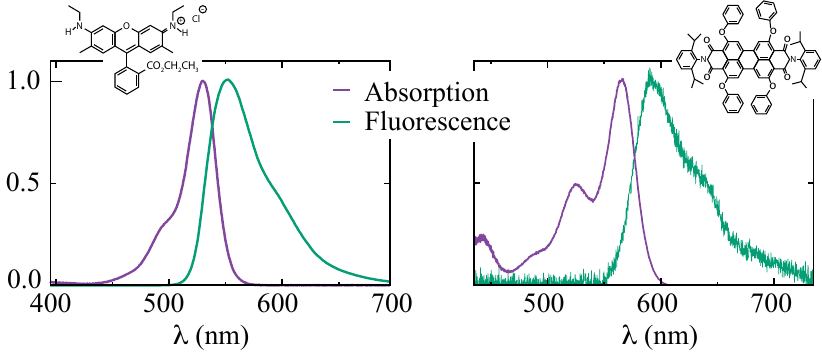}
\caption{Normalised absorption and emission spectra of Rhodamine 6G and Perylene red with structure formulae.}
\label{fig24}
\end{figure}

Figure~\ref{fig22}(a) shows a schematic of the experimental setup, which is comprised of the optical microcavity, the pump source and an analysis section~\cite{Schmitt3}. For this time-resolved study, the dye-cavity is pumped under an angle of approximately $42^\circ$ with respect to the optical axis using a picosecond pulsed laser. The cavity emission is detected by a streak camera in a spatially- and spectrally-resolved way. The microcavity is composed of highly-reflecting dielectric mirrors (CRD Optics, 901-0010-0550, radius of curvature $R=1~\si{m}$) with a maximum reflectivity $\tilde R=99.9988(2)\%$ around $550~\si{\nano m}$, while the bandwidth of a reflectivity beyond $99.98\%$ extends over a broad range $500{-}595~\si{\nano m}$, see Fig.~\ref{fig22}(b)\setcounter{footnote}{2}\footnote{Obtained from cavity-ring-down measurements using a dye laser tuned to $560{-}605~\si{\nano m}$~\cite{Wahl}.}. At the maximum, the cavity finesse amounts to $\mathcal{F}\approx 260\thinspace000$. To realise mirror separations in the micrometer range, the curved surface of one of the cavity mirrors is downsized in an in-house grinding process to ${\sim}1~\si{\milli m}$ diameter and equipped with prisms, see Fig.~\ref{fig22}(b). The latter enables optical pumping of the dye reservoir under the above mentioned angle, which together with the appropriate polarisation maximises mirror transmission to approximately $80\%$. By adjusting the lens L1 shown in Fig.~\ref{fig22} we control both pump spot position and diameter $d$ in the cavity plane, in order to initially excite the dye medium in a spatially homogeneous ($d\sim 500~\si{\micro m}$) or localised ($\sim 20~\si{\micro m}$) way. For comparison, the spatial extent of the ground mode is $d_0\approx 15~\si{\micro m}$, whereas the thermal cloud covers a region of a few hundred $\mu$m~\cite{Klaers3}. For a variation of the condensate wavelength, the cavity length can be piezo-tuned over a total length of $25~\si{\micro m}$. This allows us to actively stabilise the condensate wavelength with an accuracy $\delta\lambda\simeq 0.2~\si{\nano m}$ at $10~\si{\hertz}$ bandwidth, which compensates for long-term thermal or mechanical drifts. As dye materials we use Rhodamine 6G and Perylene red solutions of concentrations between $0.1$ and $5~\si{\milli \mole/\litre}$, see the spectra in Fig.~\ref{fig24} andTab.~\ref{tab3} for an overview of relevant properties. We expect the thermalisation time to be close to the reabsorption time and use dye concentrations of $0.1~\si{\milli \mole/\litre}$ to perform our time-resolved studies of the thermalisation dynamics.

\begin{table}
\centering
\begin{tabularx}{0.5\columnwidth}{c|cccc}
& Rh6G & PDI red &   & Ref.\\ \hline\hline
$m$ & $479.02$ & $1079.24$ & g/mol &\cite{Du,Fennel}\\\hline
$\Phi$ & $95$ & $96$ & \% &\cite{Magde,Seybold,Brouwer}\\\hline
$\lambda_\textrm{\tiny zpl}$ & $545$ & $585$ & nm & \cite{Drexhage} \\\hline
$\omega_\textrm{\tiny zpl}/2\pi$ & $ 550$ & $513$ & THz & \\\hline
$\varepsilon_\textrm{532nm}$ & $114\thinspace 000$ & $18\thinspace000$ &  & \cite{Du}\\\hline
$\sigma_\textrm{\tiny 532nm}$ & $43$ & $6.9$ & $10^{-17}\textrm{cm}^2$ &\\\hline
$\tilde n_0$ & 1.43 & 1.48 & &\\\hline
 $\tau_\textrm{\tiny 532nm}$  & 0.13 & 0.8 & ps &\\\hline
\end{tabularx}
\caption{Properties of the used dye media Rhodamine 6G (Rh6G) (\itshape Radiant Dyes\normalfont) solved in ethylene glycol and Perylene (PDI) red (\itshape Kremer Pigmente\normalfont) solved in inviscid paraffin oil. The absorption cross section follows from $\sigma= 3.82\times 10^{-21}\varepsilon$ (in units of $\textrm{cm}^2$)~\cite{Lakowicz} and the reabsorption time from $\tau_\textrm{\tiny 532nm}=(\rho \sigma_\textrm{532nm} c)^{-1}$, $\rho=1~\si{\milli \mole/\litre}$.}
\label{tab3}
\end{table}

To initialise the dye medium in a time-resolved way a mode-locked Nd:YAG pulse laser (EKSPLA PL2201) near $532~\si{\nano m}$ with $47~\mu\textrm{J}$ pulse energy and $15~\si{\ps}$ pulse length at $100~\si{\hertz}$ repetition rate is at our disposal. Both its spatial and temporal intensity profile of the pump beam pulse are gaussian. The laser system acts as the clock source for the experimental setup with electronic trigger noise around $100~\si{\ps}$. To obtain picosecond temporal resolution, we must therefore simultaneously detect the pump pulse and correct for its temporal jitter, which is achieved by directing part of the laser emission through a variable delay path onto the streak camera entrance slit. Subsequent to a pump pulse, the divergent microcavity emission is collimated by a $10\times$ long-working-distance objective (Mitutoyo M-Plan Apo $10\times$) and split into two beams. One part of the light is directed onto a diffraction grating (600 rules/mm), and the spectrally dispersed light is focussed on the streak camera entrance slit with a width of $1.5~\si{\cm}$ and $30~\si{\micro m}$ height. In the second optical path, a telescope images the photon gas onto a dove prism (Thorlabs PS992M-A), which rotates the spatial $(x,y)$ coordinates around the optical axis to align the emission with the entrance slit. The streak camera (Hamamatsu C10910) offers the time-resolved investigation in windows of $\{50;20;10;5;2;1;0.5;0.2;0.1\}\textrm{ns}$ with a temporal resolution of $1\%$ of the time range at $1\%$ detection efficiency. The data acquisition for all measurements is performed in a photon counting mode.

\subsection{Spectral thermalisation dynamics}
\label{spektralethermalisierungsdynamik}

\begin{figure}[t]
\centering\includegraphics{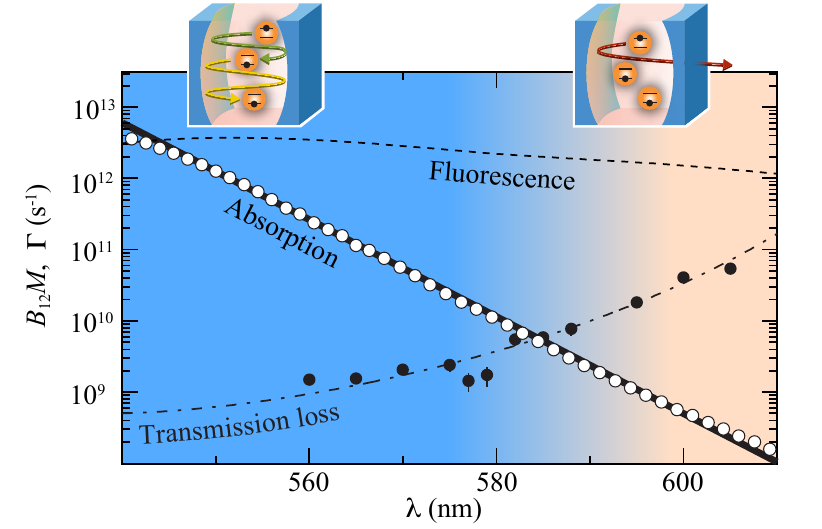}
\caption{Rates for absorption $\hat B_{12}M$, fluorescence and photon loss $\Gamma$ versus wavelength for Rhodamine 6G dye, $1~\si{\milli \mole/\litre}$, and CRD mirrors. The fluorescence (dashed line, normalised to absorption maximum) is approximately constant over the shown spectral range. For $\lambda<590~\si{\nano m}$, the photon dynamics is dominated by the exponential scaling of the reabsorption rate (open circles) that has been fitted with $\propto\exp[hc(\lambda-\lambda_\textrm{\tiny zpl})/k_\textrm{\tiny B} T]$ yielding $T=308(14)~\si{\kelvin}$ (solid line). Here, a thermal equilibrium state is expected to emerge (blue region). Increased losses from mirror transmission (filled circles, dash-dotted line) lead to dissipative dynamics for larger wavelengths (orange region).}
\label{fig26}
\end{figure}

First, we focus on the spectral thermalisation dynamics of the photon gas. For this, we realise different coupling strengths to the molecular heat bath and different loss rates due to mirror transmission by variation of the cutoff wavelength $\lambda_\textrm{\tiny c}=\{601;585;577;571\}~\si{\nano m}$. Additionally, we control the reabsorption by using different dye concentrations $\rho=\{0.1;1\}~\si{\milli \mole/\litre}$ (Rhodamine 6G). Figure~\ref{fig26} gives the measured spectral profiles of absorption, emission and loss rates. In the shown wavelength range, the fluorescence is approximately constant, whereas the absorption rate exhibits an exponential decay with increasing wavelength; their relative scaling confirms the validity of the Kennard-Stepanov ratio for the used dye, see~(\ref{KennardStepanov}). In contrast, the photon loss by mirror transmission increases with $\lambda$, suggesting incomplete thermalisation for $\lambda_\textrm{\tiny c}\geq 580~\si{\nano m}$. For $\lambda_\textrm{\tiny c}580~\si{\nano m}$, however, we expect the photon gas to acquire a thermal state within its cavity lifetime.

\begin{figure*}
\centering\includegraphics{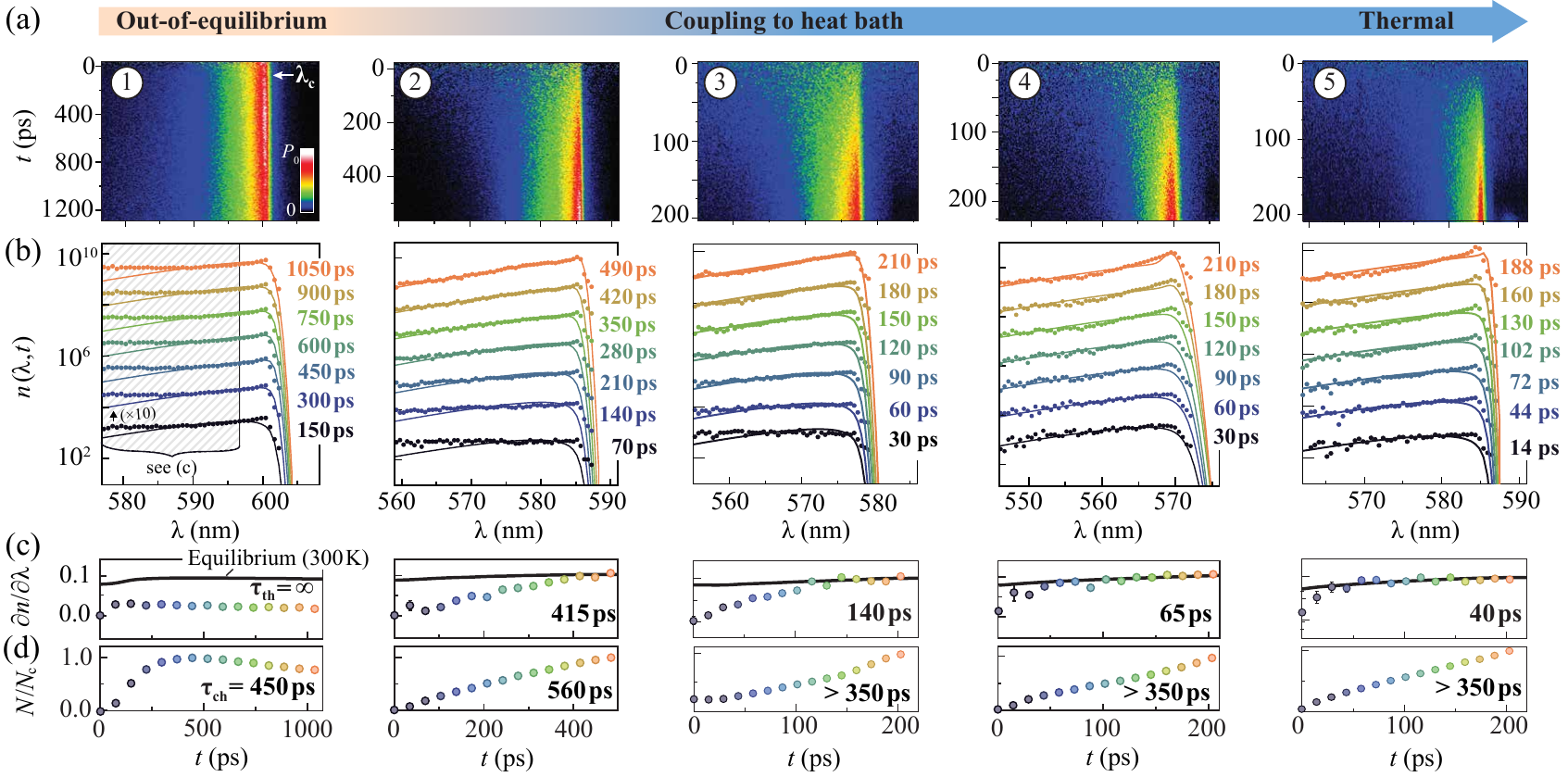}
\caption{Thermalisation dynamics of the photon gas for increased heat bath coupling. (a) Streak camera traces (line normalised) and (b) extracted spectra at different times along with $300~\si{\kelvin}$-Bose-Einstein distributions (solid lines). (c) By comparing the measured spectral slope (hatched area in (b)) with its equilibrium counterpart (line) the thermalisation time can be quantified. It reduces as the coupling to the heat bath is enhanced (from left to right). (d) The temporal evolution of the total power of the cavity emission (normalised to $N_\textrm{\tiny c}$) indicates the time scale for chemical equilibration between photons and dye molecules. ($\lambda_\textrm{\tiny c,\ding{172}-\ding{176}}=\{ 601;585;577;571;585 \}~\si{\nano m}$, Rhodamine 6G $\rho_\textrm{\tiny \ding{172}-\ding{175}}=0.1~\si{\milli \mole/\litre}$, $\rho_\textrm{\tiny \ding{176}}=1.0~\si{\milli \mole/\litre}$). Reproduced with permission from \cite{Schmitt3}. Copyright 2015 by the American Physical Society.}
\label{fig27}
\end{figure*}

The starting point for the measurement is a spatially homogeneous excitation of the dye medium using a broad pump beam ($2w_0=500~\si{\micro m}$), which minimises any gradients in the excitation level of the medium at $t=0$, realising well-defined initial conditions in chemical equilibrium. Figure~\ref{fig27}(a) gives line-normalised, false-colour streak camera traces showing the evolution of the spectral mode occupation. Here, we define $t=0$ as the time when the first fluorescence photons are detected, see $N/N_\textrm{\tiny c}(t)$ in Fig.~\ref{fig27}(d). From left to right, we successively increment $\lambda_\textrm{\tiny c}$ to gradually decouple the photon gas from the heat bath. All spectral distributions are weighted with the spectral mirror transmission coefficient. Individual excited modes, which are spaced by $42~\si{\pico m}$ ($\Omega/2\pi = 37~\si{\GHz}$), are not resolved due to limited spectral resolution of the diffraction grating of $1~\si{\nano m}$\setcounter{footnote}{2}\footnote{A high-resolution spectrum is shown in Fig.~\ref{fig34}.}. The recorded data span a total spectral range $\Delta\lambda=25~\si{\nano m}$ ($\Delta E= 3.5k_\textrm{\tiny B}T$), which is expected to contain the following fraction of photons:
\begin{equation}
\bar N_\textrm{\tiny exp.} = \int_{0}^{3.5k_\textrm{\tiny B} T}{\frac{2(u/\hbar\Omega + 1)}{e^{(u-\mu)/k_\textrm{\tiny B}T}-1}\textrm{d}u}\approx 0.93  \bar  N
\label{4_1}
\end{equation}
Our experimental data thus provides reliable information about the degree of thermalisation of the photon gas. In all measurements, we choose the laser power to be such that a macroscopic ground state occupation emerges at the end of the detection window, which allows us to calibrate the photon number at arbitrary times $N(t)$ with respect to its asymptotic value, i.e. the critical photon number $N(t\rightarrow\infty)\equiv N_\textrm{\tiny c}\approx 90\thinspace000$. Therefore, we compare our spectral data with thermal equilibrium Bose-Einstein distributions at $300~\si{\kelvin}$ (solid lines) with a chemical potential that satisfies the total photon number.

\begin{figure}[t]
\centering\includegraphics{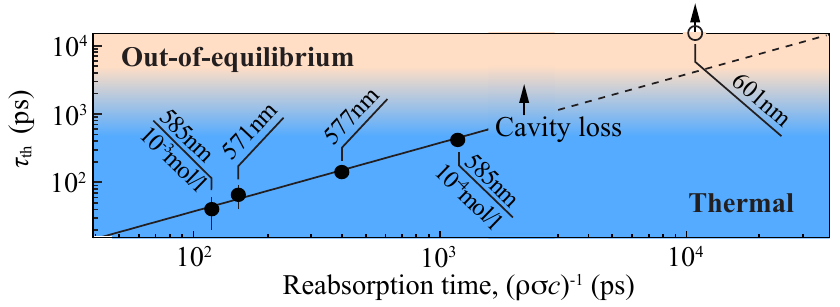}
\caption{Observed thermalisation time $\tau_\textrm{\tiny th}$ (circles) versus reabsorption time in the dye solution. For $\tau_\textrm{\tiny th}<\tau_\textrm{\tiny res}$ the photon gas acquires a thermal state (blue); otherwise it remains an out-of-equilbrium system. Reproduced with permission from \cite{Schmitt3}. Copyright 2015 by the American Physical Society.}
\label{fig28}
\end{figure}

For weak dye reabsorption and large cavity losses, $\lambda_\textrm{\tiny c}=601~\si{\nano m}$ (Fig.~\ref{fig27}, left), the spectral wing (hatched) deviates from its equivalent in equilibrium for all observed times. The photon gas fails to thermalise during its microcavity lifetime. However, as the absorptive coupling rate to the molecule bath is enhanced, $\lambda_\textrm{\tiny c}\leq585~\si{\nano m}$, (Fig.~\ref{fig27}, columns 2 to 5), we observe a thermalisation process that redistributes the photon energies, transforming the out-of-equilibrium distribution to a room temperature spectrum. The characteristic thermalisation time $\tau_\textrm{\tiny th}$ can be quantified by the spectral slope $\partial n(\lambda,t)/\partial\lambda$, as illustrated in Fig.~\ref{fig27}(c, circles), which in the presence of thermalisation converges to the equilibrium spectral slope (solid line). We observe $\tau_\textrm{\tiny th}=\{415;140;65;40\}~\si{\ps}$ defined as the time when the relative deviation between measured and equilibrium spectral slope is less then $1\%$. Figure~\ref{fig27}(d) shows the temporal increase of the total cavity emission, revealing the gradual establishment of chemical equilibrium between photons and dye molecules. Notably, the time scales for either thermal or chemical equilibration differ (Section~\ref{thermalisierungsdynamik_theor}). Even if the total photon is non-stationary, the spectral profile of the photon gas present in the microcavity can already be thermally distributed. In Fig.~\ref{fig28} we plot the measured thermalisation times as a function of the free absorption time in the medium $\left(\rho \sigma(\lambda) c\right)^{-1}$, which follows a linear scaling $\tau_\textrm{\tiny th}=0.37(5)\cdot \left(\rho \sigma(\lambda) c\right)^{-1}$. Indeed, the photon gas equilibrates due to an energy exchange with a heat bath at a rate that can be tuned via the reabsorption. If the thermalisation time exceeds the photon lifetime in the cavity, $\tau_\textrm{\tiny th}>\tau_\textrm{\tiny res}\approx 500~\si{\ps}$, the photons constitute an out-of-equilibrium ensemble.

\subsection{Bose-Einstein condensation dynamics}
\label{dynbec}
We turn our attention to the temporal photon dynamics subject to (i) \itshape spatially inhomogeneous  \normalfont and (ii) \itshape strongly inverting \normalfont pump excitation of the dye medium. As before, we investigate the dynamics for different coupling rates to the molecular heat bath $\rho \sigma(\lambda) c$ and resonator losses $\Gamma(\lambda)$, realised by varying the cutoff wavelength. In the out-of-equilibrium regime, $\Gamma(\lambda)>\rho \sigma(\lambda) c$, the nonequilibrium state of the dye medium manifests itself in transient multimode laser operation. When coupling the photons to a heat bath, $\Gamma(\lambda)<\rho \sigma(\lambda) c$ paves the way for the photon gas to thermal equilibrium and give rise to the emergence of BEC.

\begin{figure}[t]
\centering\includegraphics{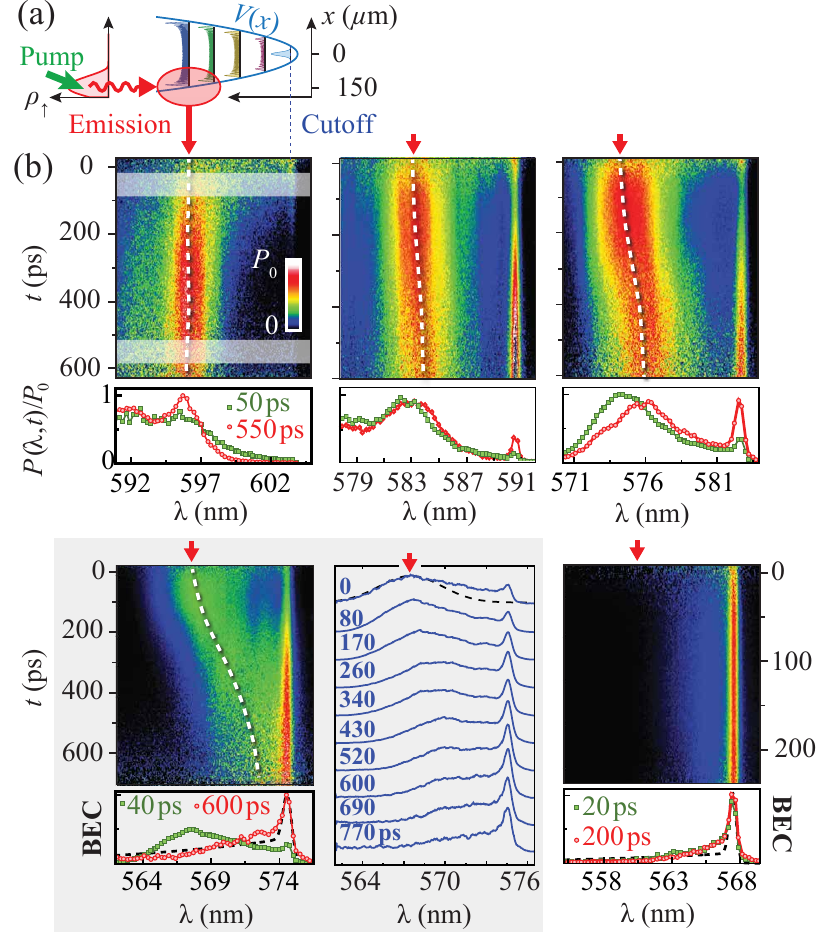}
\caption{Spectral condensation dynamics following an off-centre pump pulse for increasing thermal contact to the heat bath. (a) The pump pulse initialises an inhomogeneous excited molecule density $\rho_\uparrow$, which decays by stimulated emission into spatially overlapping modes. (b) Streak camera images (top) with spectral cuts (bottom) shortly after the photon injection (green squares) and at the end of the detection window (red circles). While for weak thermal contact (top, left) no thermalisation is observed, the enhanced reabsorption leads to a partial relaxation (top, middle \& right) of the photon emission towards larger wavelengths. Finally, the photon gas fully thermalises to a Bose-Einstein distribution (bottom row). The grey-shaded data set highlights the thermalisation dynamics with cuts in the dynamical transition region. (Rhodamine 6G $0.25~\si{\milli \mole/\litre},\ \lambda_\textrm{\tiny c}=\{603;590;582;574;567\}\si{\nano m}, \tau_\textrm{\tiny th}=\{2\thinspace300;388;124;38;13\}\textrm{ps}$, cavity lifetime $\tau_\textrm{\tiny res}=\{18;83;165;303;500\}\textrm{ps}$). Reproduced with permission from \cite{Schmitt3}. Copyright 2015 by the American Physical Society.}
\label{fig29}
\end{figure}

Figure~\ref{fig29}(a) indicates the experimental sequence to realise initial conditions far from equilibrium. We focus the pump beam to a diameter $2w_0=80~\si{\micro m}$ and position it at $(x=150~\si{\micro m},y=0)$ transversally displaced from the optical axis in the microcavity plane. As a result, the spatially inhomogeneous density $\rho_\uparrow(x)$ decays by emission of (initially) spontaneous photons into excited transverse cavity modes that overlap most with the pumped region. These eigenstates of the harmonic oscillator potential are at higher energies (lower wavelengths) than the transverse ground state $\hbar\omega_\textrm{\tiny c}$ ($\lambda_\textrm{\tiny c}$). Due to the sub-nanosecond time scales of the radiative processes, the comparatively slow effect of spatial diffusion of molecules can be safely neglected~\cite{Wellek}. Following the initial photon emission, the photon gas kinetics depends critically on both the dye reabsorption and the cavity loss rates.

The first data set in Fig.~\ref{fig29}(b) shows the spectral photon evolution in the weakly reabsorbing regime near $\lambda_\textrm{\tiny c}=603~\si{\nano m}$. The streak camera time traces have been line-normalised to clarify the spectral redistribution of the photons. Here, the reabsorption time $\left(\rho\sigma(\lambda_\textrm{\tiny c})c\right)^{-1}=5.8~\si{\ns}$ (corresponding to $\tau_\textrm{\tiny th}=2.3~\si{\ns}$) exceeds by far the average photon storage time in the cavity $\tau_\textrm{\tiny res}(\lambda_\textrm{\tiny c})=18~\si{\ps}$, such that no equilibrium distribution emerges. Instead, the optical feedback onto the inverted active medium causes stimulated amplification of the light field in the excited modes around $\lambda_\textrm{\tiny max}=595.7~\si{\nano m}$ after approximately $100~\si{\ps}$ and the maximum of the emission is maintained throughout the entire detection window. At $r=150~\si{\micro m}$ the resonant wavelength deviates from the cutoff wavelength by $\Delta \lambda (r)= \lambda_\textrm{\tiny c} - \lambda(r) = {2nr^2}/{qR}=8~\si{\nano m}$, which agrees with the observed value of $7.3~\si{\nano m}$. In further measurements, see Fig.~\ref{fig29}(b), we successively enhance the coupling to the heat bath by reducing $\lambda_\textrm{\tiny c}$. Accordingly, we observe a more and more accelerated spectral redistribution of the light towards an equilibrium distribution. This is a consequence of fast reabsorption processes, which chemically equilibrate any gradients in the density of the ground and excited state molecules. Due to the harmonic trapping potential, this light-induced diffusion is directed towards transverse modes with lower energies than the modes overlapping with the pump beam region. Strikingly, the ground state becomes macroscopically occupied for data with shorter cutoff wavelength, and for $\lambda_\textrm{\tiny c}=574~\si{\nano m}$ and $567~\si{\nano m}$ a BEC with thermally occupied excited states forms. In the case of $\lambda_\textrm{\tiny c}=567~\si{\nano m}$, the rapid thermalisation prevents the detection of any non-equilibrium emission at the given temporal resolution.

Using a spatially-selective photon injection technique to prepare a photon gas far from equilibrium, our measurements have demonstrated that a high-density (critical) photon gas thermalises to a Bose-Einstein condensate provided that the coupling to the heat bath is sufficiently strong. In the opposite limit, the high-density photon gas resembles an out-of-equilibrium state similar to a multimode laser. In contrast to the homogeneously pumped protocol (Section~\ref{spektralethermalisierungsdynamik}), the photon thermalisation dynamics is not universal but depends crucially on the initial conditions of the pumped dye medium, in excellent agreement with our numerical simulations, see Section~\ref{numerischesimulationdynamik}.

\subsection{Spatial photon kinetics}
\label{kinetik}
We focus on the \itshape spatial \normalfont condensation dynamics subsequent to an inhomogeneous inversion of the dye medium. For this, a tightly focussed pump beam ($2w_0\simeq27~\si{\micro m}$) irradiates the dye microcavity spatially displaced by $50~\si{\micro m}$ from the position of the trap minimum. To analyse the spatial intensity distribution, a real image of the cavity plane is projected onto the streak camera.

\begin{figure}[t]
\centering\includegraphics{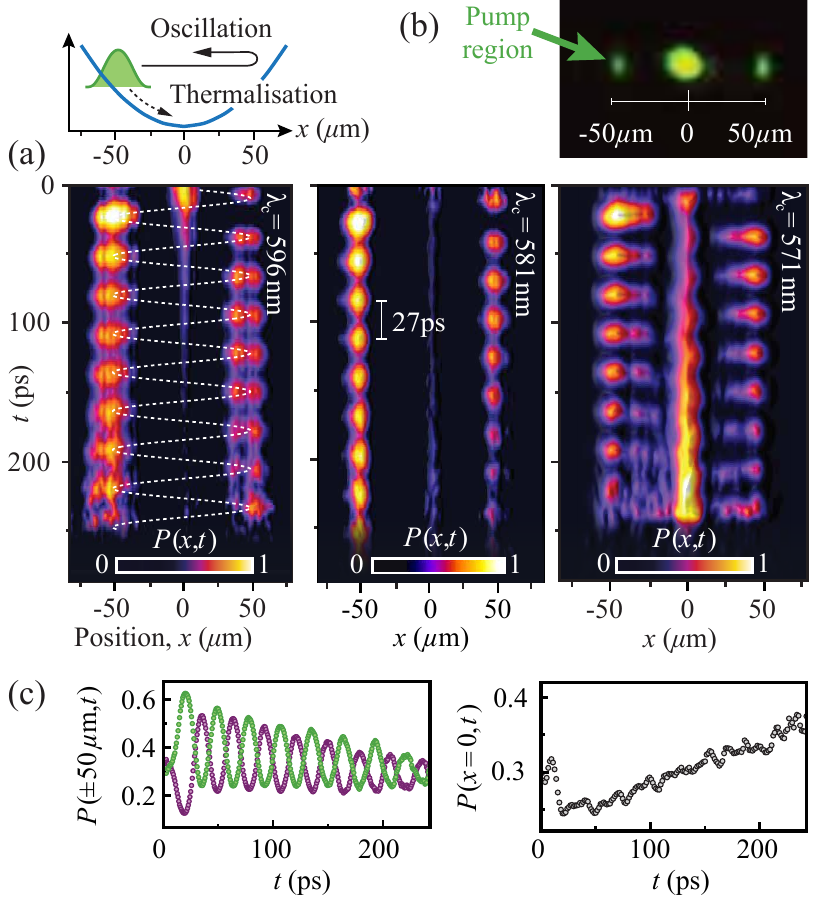}
\caption{Mode-locked laser operation and BEC of photons. (a) Line-normalised spatial evolution of the cavity emission for different cutoff wavelengths. The pump beam excites an oscillating wave packet in the harmonic potential (top). For weak reabsorption (left), a stable mode-locked laser oscillation occurs (dashed line), where most light leaves the cavity at the reversal points. For stronger reabsorption (middle and right) the photons thermalise and a condensate emerges in the trap centre. (b) CCD camera image of the (average) emission. (c) Temporal variation of the detected relative intensity from (a, right) at the reversal points (left) and in the condensate mode (right). (Rhodamine 6G, $\rho=0.1~\si{\milli \mole/\litre}$, $\lambda_\textrm{\tiny c}=\{596;581;571\}\si{\nano m}$). Reproduced with permission from \cite{Schmitt3}. Copyright 2015 by the American Physical Society.}
\label{fig30}
\end{figure}

Figure~\ref{fig30}(b) shows a typical CCD camera image of the average cavity emission. Besides the emission from the trap centre $(x=0,y=0)$, two bright spots are visible: the first one near the pumping region at $(-50\si{\micro m},0)$, the second one at $(50\si{\micro m},0)$, i.e. mirrored respectively to the trap centre. A time-resolved measurement (Fig.~\ref{fig30}, line-normalised) of the intensity distribution along the $x$-axis yields an explanation for the centro-symmetric emission: following the inhomogeneous dye excitation an optical wave packet forms, which oscillates in the harmonic potential with reversal points that determine the observed emission spots. The observed oscillation period $T=27~\si{\ps}$ shows excellent agreement with the expected inverse trap frequency $2\pi/\Omega\simeq (37.1~\si{\GHz})^{-1}$ (seeTab.~\ref{tab1}). Moreover, the wave packet emerges within only a few picoseconds. As this is considerably faster than the spontaneous decay time of the Rhodamine molecules ($4$~ns), the dynamics are driven by stimulated processes. The wave packet dynamics can be understood as a coherent superposition of adjacent transverse eigenstates spaced by $\hbar\Omega$, i.e. with a fixed relative phase, in close analogy to a mode-locked laser with an extremely high repetition rate. Oscillator modes that exhibit their maximum probability in proximity to the pumped region, experience maximum gain. Classically speaking, the velocity of the wave packet is minimised at the reversal points of the oscillation, maximising here the photon leakage rate out of the resonator. Quantum mechanically, this can be interpreted as constructive interference between multiple harmonic oscillator wave functions.

In the limit of weak reabsorption and large cavity losses, see Fig.~\ref{fig30}(a, left) for $\lambda_\textrm{\tiny c}=596~\si{\nano m}$, the photon kinetics is determined by the highly-excited oscillating out-of-equilibrium state throughout the measurement time of $250~\si{\ps}$. The visible residual initial population at small times is attributed to overlap of the pump beam with the ground mode at $x=0$, which however quickly decays. The situation drastically changes, as the thermal contact to the heat bath is established by lowering $\lambda_\textrm{\tiny c}=581~\si{\nano m}$ and $571~\si{\nano m}$ (Fig.~\ref{fig30}(a), middle \& right). During its oscillation, the wave packet traverses the enclosed dye volume, effectively equilibrating the initially inhomogeneous excitation level $\rho_\uparrow/\rho_\downarrow(x)$ by multiple photon reabsorption events. Figure~\ref{fig31} shows corresponding numerical simulations. With advancing times, this effects a dynamical redistribution from the laser-like wave packet to a BEC. The damping of the coherent oscillations and the emergence of the macroscopic ground state in the presence of a thermal bath is shown in Fig.~\ref{fig30}(c). Qualitatively, the measured photon kinetics is in good agreement with results from numerical simulations.

To conclude, our experimental study demonstrates that a thermal state of the photon gas in the dye-filled microcavity is imprinted by a molecular heat bath. In particular, we find that the efficiency of the thermal contact, i.e. the thermalisation rate, can be tuned systematically by the optical density of the dye solution. With regard to the canonical and grand-canonical statistical ensemble, the temperature of the dye solution actually becomes an external parameter for the photon gas. Thermalisation induced by absorption and emission is a necessary prerequisite for the emergence of a photon BEC.

\subsection{Numerical simulations}
\label{numerischesimulationdynamik}
To gain deeper insight into the crossover from transient laser operation ({\itshape critical out-of-equilibrium gas}) to BEC ({\itshape critical equilibrium gas}), we perform numerical simulations of the photon dynamics in the microcavity~\cite{Schmitt3}. Our phenomenological model relies on semi-classical one-dimensional rate equations and incorporates both the coupling of the photons to the optically active dye medium as well as their oscillatory movement in the harmonic trap. Independently, the results have been confirmed using a master equation model including coherences between photon modes~\cite{Keeling1}.

\begin{figure}[t]
\centering\includegraphics{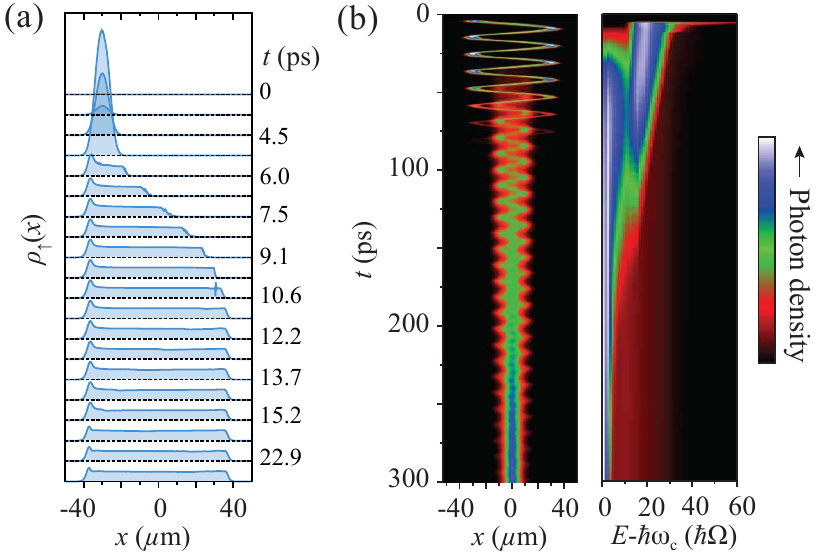}
\caption{Simulation of the thermalisation dynamics for off-centre pulsed excitation. (a) Density of excited molecules $\rho_\uparrow(x)$ for given times after the pump pulse. The locally excited medium near $x_0=-30~\si{\micro m}$ induces the formation of a photon wave packet (see (b)), which homogenises $\rho_\uparrow(x)$ in the course of its oscillation. (b) The temporal evolution of the spatial photon density (left) shows the oscillation of a mode-locked photon wave packet, which is damped out with time. After $250~\si{\ps}$, the photons have been redistributed to the cavity ground state at $x=0$. Photon evolution versus photon energy $E=(\hbar\omega-h\omega_c)/\hbar\Omega$ (right), showing the emergence of a BEC with $\bar n_0/\bar N\simeq95\%$. ($\Omega/2\pi=93~\si{\GHz}$, $\lambda_\textrm{\tiny c}=570~\si{\nano m}$,  $\hbar\Delta \simeq -4k_\textrm{\tiny B}T$ at $T=300~\si{\kelvin}$, $\hat B_{12}=1.3~\si{\kHz}$, $\rho_\uparrow(x)+\rho_\downarrow(x)=1.5\cdot 10^8~\si{\micro m}^{-1}$, $\Gamma_i=0$.)}
\label{fig31}
\end{figure}

Our approach is based on the equation of motion for the photon density:
\begin{equation}
\dot{\overline{n}}_i = \hat B^i_{21} \bar \rho_\uparrow \left(\bar n_i + \varepsilon_i \right) - \left( \hat B_{12}^i \bar \rho_\downarrow + \Gamma_i  \right) \bar n_i - v_i \frac{\partial}{\partial x} \bar n_i
\label{4_5}
\end{equation}
Here, $\bar n_i = \bar n_i(x,t)$ is the photon number density in the $i$-th mode at position $x$ and time $t$ (averaged over many realisations), the densities of molecules in ground and electronically excited state $\bar \rho_{\downarrow,\uparrow} = \bar \rho_{\downarrow,\uparrow}(x,t)$, and the rate coefficients for absorption and emission $\hat B^i_{12,21} = \hat B_{12,21}(\omega_i)$ at the photon angular frequency $\omega_i$. Furthermore, $\varepsilon_i = \varepsilon_i(x) $ denotes the density of a single photon in the $i$-the mode, $\Gamma_i=\Gamma(\omega_i)$ the cavity loss rate and $v_i=v_i(x)$ the photon velocity field, which will be discussed in the following. Assuming a conserved excitation number $X=n+M_\uparrow$, one finds
\begin{equation}
-\frac{\partial}{\partial t} \bar M_\downarrow = \frac{\partial}{\partial t} \bar M_\uparrow = P - \sum_i{\frac{\partial}{\partial t} \bar n_i},
\label{4_6}
\end{equation}
where $P=P(x,t)$ denotes the pump beam excitation. Heuristically, we consider a non-orthogonal set of optical modes consisting of coherent states $|\alpha_i\rangle$ with amplitudes $|\alpha_i|=\sqrt{u_i/\hbar\Omega}$, $u_i=i\cdot\hbar\Omega$ and the trap frequency $\Omega$. The mode energy spectrum corresponds to the eigenenergies of the harmonic oscillator potential. In contrast to stationary eigenstates, coherent states allow us to model the oscillation of particles or wave packets in the trap. The normalised photon density $\varepsilon_i(x)$ results from a temporal average
\begin{equation}
\varepsilon_i(x)=\frac{1}{T}\int_0^{T}{|\langle x|\alpha_i(t)\rangle|^2 dt}
\label{4_7}
\end{equation}
over an oscillation period $T=2\pi/\Omega$. By comparing the probability to find a particle within $dx$, $\varepsilon_i(x)dx$, with the temporal portion of a half-period that the particle is present in this interval, $dt/(T/2)$, we define the photon velocity field
\begin{equation}
v_i(x)={dx}/{dt}=\pm\Omega/\pi\varepsilon_i(x),
\label{4_8}
\end{equation}
where the sign changes after each half-period. Neglecting losses, $\Gamma_i=0$, the numerical results demonstrate that the model reproduces asymptotically the analytic Bose-Einstein distributions and the critical particle number, see Fig.~\ref{fig31}(b)). 

In analogy to the experiments described above, we simulate the photon thermalisation kinetics for initial out-of-equilibrium conditions realised by pumping the molecular medium with a Gaussian laser pulse $P(x,t)\propto \exp[{ - {(x-x_0)^2}/{2\sigma_x^2} - {(t-t_0)^2}/{2\sigma_t^2}}]$ with duration $\sigma_t=1.5~\si{\ps}$ and waist $\sigma_x=7.5~\si{\micro m}$. The pump pulse is positioned at $x_0=-30~\si{\micro m}$, where it locally excites molecules within a few picoseconds, as indicated in Fig.~\ref{fig31}(a). At this point, the chemical potentials of the photons and ground and excited state dye molecules exhibit strong gradients as visible in the spatially inhomogeneous dye excitation level. The cutoff wavelength $\lambda_\textrm{\tiny c}=570~\si{\nano m}$ is chosen such that the photon gas couples efficiently to the molecules\setcounter{footnote}{2}\footnote{The simulation parameters differ from experimental values for computational reasons. Cavity losses have been neglected.}.

Subsequent to the initialisation pulse, the simulations reveal the emergence of a high photon density in the pumped region, which reaches its maximum after only a few picoseconds. Owing to the trapping potential these photons are accelerated as a wave packet towards the trap minimum ($x=0$), see Fig.~\ref{fig31}(b). During their oscillation the photons are quickly reabsorbed by the enclosed dye medium, which results in a homogeneous density of excited molecules within the region traversed by the wave packet after nearly half an oscillation period ($T/2=5.4~\si{\ps}$) as visible in Fig.~\ref{fig31}(a)). The homogeneity of $(M_\uparrow/M_\downarrow)(x)$ is a prerequisite for chemical equilibrium among photons and molecules and the existence of a global chemical potential for the photon gas. Indeed, we find that the medium acquires a homogeneous state soon after the pump excitation, whereas the photon gas is still characterised by a non-thermal spectral distribution, see Fig.~\ref{fig31}(c). It takes additional $250~\si{\ps}$ until the laser-like wave packet has vanished and the photon energies are Bose-Einstein distributed with a macroscopic occupation of the ground mode.

\section{Number statistics of condensed light}
\label{secondordercorrelationsexp}

We describe measurements of the photon number statistics and second-order correlations of a photon BEC coupled to different-sized particle reservoirs. Our experiment gives access to canonical and grand-canonical statistical ensemble conditions, which are hallmarked by their particle number fluctuations: for small reservoirs (canonical), the photon statistics is Poissonian with small fluctuations, $\delta n/\bar n=1/\sqrt{\bar n}\approx 0$ (for $\bar n\gg 1$), whereas large reservoirs (grand-canonical) support unusually large fluctuations of the condensate population, $\delta n/\bar n = 1$. 

\subsection{Experimental scheme}

\begin{figure}[t]
\centering\includegraphics{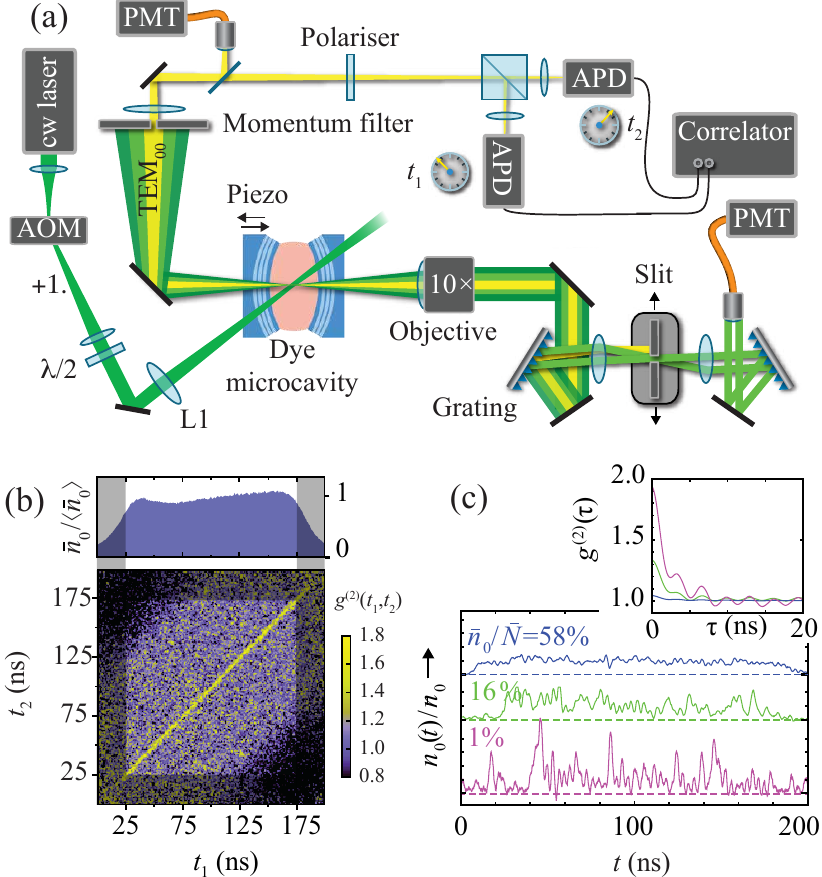}
\caption{(a) Time-resolved measurements of correlations, number statistics and fluctuations of the photon BEC. The microcavity is pumped (quasi-)continuously and part of the cavity emission is spectrally analysed. In the far field, the condensate emission passes several filtering stages and its correlations are detected in a Hanbury Brown-Twiss interferometer. (b) Typical average condensate emission (top) and second-order correlation function $g^{(2)}(t_1,t_2)$. (c) Time evolution of the (normalised) condensate population measured with a single photomultiplier (PMT) for different condensate fractions $\bar n_0/\bar N$. \itshape Inset: \normalfont corresponding autocorrelation functions with $\tau=t_2-t_1$. Reproduced with permission from \cite{Schmitt1}. Copyright 2014 by the American Physical Society.}
\label{fig32}
\end{figure}

Figure ~\ref{fig32} outlines the used experimental scheme. In contrast to the measurements of the thermalisation dynamics, a continuous pump and detection system is utilised. For all measurements, the microcavity is operated at $q=8$ and filled with either Rhodamine 6G (ethylene glycol) or Perylene red (inviscid paraffin oil) solutions at varying concentrations. The dye medium is pumped by a frequency-doubled Nd:YAG laser (Coherent Verdi V8) near $532~\si{\nano m}$, whose output power of up to $8~\si{\watt}$ is acousto-optically modulated (AOM) into $200~\si{\ns}$ pulses at $200~\si{\hertz}$ repetition rate, in order to reduce excitation of long-lived dye triplet states and to maintain condensate number constant throughout the pulse (Fig.~\ref{fig32}(b), top). For the latter, the rf-signal driving the AOM is mixed with a temporally increasing voltage from a function generator (Tektronix AFG3252). Additionally, a voltage-controlled attenuator actively stabilises the condensate power ($10~\si{\hertz}$ bandwidth), which is separately detected by a photomultipler. A $f_\textrm{\tiny L1}=400~\si{\milli m}$ focal length lens focuses the pump beam to a diameter of $2w_0\simeq150~\si{\micro m}$ into the microcavity plane to generate a photon gas. Here, the pump power controls the excitation level of the dye, as well as the chemical potential and the total number (and condensate fraction) of the photon gas. Any loss from the dye-microcavity-system is compensated by maintaining the pumping throughout the pulse.

To determine the condensate fraction $\bar n_0/\bar N$, we measure average photon spectra, see Fig.~\ref{fig34}(a), in a $4f$-spectrometer equipped with two diffraction gratings ($2400~\textrm{rules/mm}$) and two lenses with $f=100~\si{\milli m}$. A motion-controlled slit placed in the $2f$-Fourier-plane performs a wavelength selection of the multimode light, which is detected using a photomultiplier (Hamamatsu H10721-210). Although its spectral resolution $\Delta\lambda=0.5~\si{\nano m}$ precludes the measurement of individual transverse cavity modes spaced by $\Delta\lambda\simeq 41~\si{\pico m}$, we confirm the solitary macroscopic occupation of the ground state with a double monochromator (LTB Demon) with $6~\si{\pico m}$ resolution, see Figs.~\ref{fig34}(a) and~\ref{fig34}(c).

The photon correlations of the BEC are detected in a Hanbury Brown-Twiss interferometer, while for the direct observation of the time-resolved fluctuations and photon statistics a photomultiplier is at our disposal, see Fig.~\ref{fig32}. To measure the second-order correlations only of the condensate mode, the divergent cavity emission is first Fourier-filtered with a $5~\si{\milli m}$ iris in the far field approximately $850~\si{\milli m}$ behind the cavity\setcounter{footnote}{2}\footnote{The free propagation of the photon gas is equivalent to a free expansion of a harmonically trapped gas, a technique commonly used in ultra-cold atoms to infer the initial momentum distribution from the density distribution after a long time-of-flight~\cite{Pethick}.}. The aperture acts as a transverse momentum filter to suppress contributions from excited modes: From the zero-point energy in the harmonic trapping $\hbar\Omega$, we can estimate the momentum uncertainty of the ground mode $\Delta k_r=\sqrt{2m_\textrm{ph}\Omega/\hbar}\simeq 1.86\times 10^5~\si{m}^{-1}$, which is much smaller than the longitudinal wave vector component $k_z(0)=q\pi/D_0\simeq 1.6\times  10^7~\si{m}^{-1}$. Taking into account the quartz-air cavity interface ($\tilde n_{0,\textrm{\tiny Quartz}}\simeq 1.46$), the corresponding divergence angle $\Theta=0.95^\circ$ leads to a condensate diameter ${\sim}1.4~\si{\cm}$ at the momentum filter. Most of light in the first excited eigenstate ($2\hbar\Omega$, diameter ${\sim}2.0~\si{\cm}$) is expected to be blocked. After lifting the two-fold polarisation degeneracy of the photons, the transmitted light is equally split and directed onto two single-photon detectors (MPD PD5CTC, temporal resolution $\Delta t \simeq 50~\si{\ps}$, dead time $\tau_\textrm{\tiny PD}\simeq 79~\si{\ns}$), which are connected to an electronic correlation system (PicoQuant PicoHarp 300) that records and correlates photon detection events at times $t_{1,2}$ with a resolution $60~\si{\ps}$. To avoid errors during the coincidence measurement caused by the dead time of the system $\tau_\textrm{\tiny PicoHarp}\simeq 90~\si{\ns}$), the condensate light is sufficiently attenuated to provide photon count rates around ${\sim}0.5~\textrm{photons/pulse}$ ($2.5\times 10^6~\textrm{photons/s}$) at each detector. Evaluation of the time histograms yields the second-order correlation function for the BEC
\begin{equation}
g^{(2)}(t_1,t_2)=\frac{\langle n_0(t_1) n_0(t_2)\rangle}{\langle n_0(t_1) \rangle \langle n_0(t_2)\rangle},
\label{5_1}
\end{equation}
where $\langle ...\rangle$ denotes a temporal average, see Fig.~\ref{fig32}(b) for a typical data set. At $t_1=t_2$, we find significant photon bunching, $g^{(2)}(t_1,t_1)\approx 1.7$ (yellow diagonal), while for large time delays the photons are uncorrelated, $g^{(2)}(t_1,t_2)\approx 1.0$. Due to the nearly constant average photon number during the operation time (Fig.~\ref{fig32}(b), top), the second-order correlations depend only on the relative time delay $\tau=t_2-t_1$, and we hereafter only refer to the time-averaged correlation function $g^{(2)}(\tau)=\left\langle g^{(2)}(t_1,t_2) \right\rangle_{t_2-t_1=\tau}$. 

Moreover, we monitor the time evolution of the condensate intensity in the same optical path, see Fig.~\ref{fig32} (top), relying on a photomultiplier (Hamamatsu H9305-01, $\Delta t\simeq 1.4~\si{\ns}$, quantum efficiency ${\approx}10\%$) and a fast oscilloscope (Lecroy DDA 5005A, $5~\si{\GHz}$ bandwidth). This allows us to resolve the number fluctuations, which occur on time scales around $2~\si{\ns}$; examples are given in Fig.~\ref{fig32}(c). From the intensity traces $I_0(t)$ we can equally reconstruct the second-order correlation function
\begin{equation}
g^{(2)}(\tau)=\frac{\langle I_0(t+\tau)I_0(t)\rangle_t}{\langle I_0(t)\rangle_t \langle I_0(t+\tau)\rangle_t},
\label{5_2}
\end{equation}
where $\langle...\rangle_t =(T-\tau)^{-1} \int_{0}^{T-\tau}{(...)\textrm{d}t}$ denotes the temporal average of the pulse of duration $T$. We note, that despite consistent results for $g^{(2)}(0)$, the Hanbury Brown-Twiss interferometer is considered as the more reliable detection scheme for our purposes due to its high temporal resolution.

\subsection{Time-resolved photon correlations}
\label{Photonenkorrelationen}

\begin{figure}[t]
\centering\includegraphics{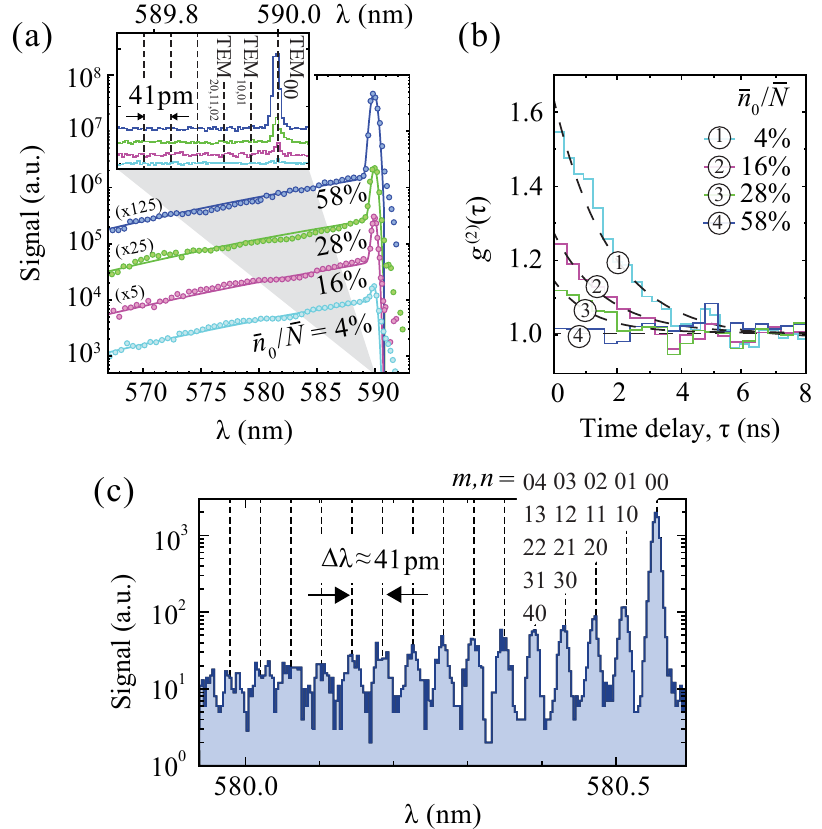}
\caption{(a) The measured spectra for increasing condensate fractions (circles) are well described by $300~\si{\kelvin}$ Bose-Einstein distributions (solid lines). \itshape Inset: \normalfont (Linear) high-resolution spectrum demonstrating the macroscopic occupation of the ground state only. All spectra have been vertically shifted for clarity. (b) Second-order correlation functions $g^{(2)}(\tau)$ (\ding{172}${-}$\ding{174}) exhibit photon bunching at short time delays $\tau$. ($\lambda_\textrm{\tiny c}=590~\si{\nano m}$, $\hbar\Delta=-6.7k_\textrm{\tiny B} T$, $\rho=1~\si{\milli \mole/\litre}$, Rhodamine 6G) (c) Mode-resolved spectrum of a Bose-Einstein condensed photon gas measured using a double monochromator. Reproduced with permission from \cite{Schmitt1}. Copyright 2014 by the American Physical Society.}
\label{fig34}
\end{figure}

In a first step, we study the number correlations of different-sized photon BECs coupled to a particle reservoir of constant size by fixing the dye concentration $\rho=1~\si{\milli \mole/\litre}$ (Rhodamine 6G) and the dye-cavity detuning $\hbar\Delta=hc(\lambda^{-1}_\textrm{\tiny c} - \lambda^{-1}_\textrm{\tiny zpl})=-6.7k_\textrm{\tiny B} T$ ($\lambda_\textrm{\tiny c}=590~\si{\nano m}$).

Figure~\ref{fig34}(a) shows spectral distributions in the Bose-Einstein condensed phase hallmarked by the macroscopically occupied ground mode and thermally populated excited states. All condensate fractions $\bar n_0/\bar N=\{4\%;16\%;28\%;58\%\}$ and reduced temperatures $T/T_\textrm{\tiny c}=\{0.98;0.92;0.85;0.65\}$, respectively, are obtained from fitting the data with $T=300~\si{\kelvin}$ Bose-Einstein distributions. This corresponds to absolute photon numbers $\bar n_0\simeq\{4;19;37;120\}\times 10^3$ and $\bar N\simeq\{100;119;132;207\}\times 10^3$. To confirm the single-mode property of the condensate, we show corresponding spectra (Fig.~\ref{fig34}(a), inset) with a $9~\si{\pico m}$-resolution which is below the transverse mode spacing $\Delta\lambda=41~\si{\pico m}$. By measuring the entire cavity emission the full periodic mode structure is revealed, see Fig.~\ref{fig34}(c).

The second-order correlation functions $g^{(2)}(\tau)$ shown in Fig.~\ref{fig34}(b) exhibit zero-delay autocorrelations $g^{(2)}(0)_\textrm{\ding{172}${-}$\ding{175}}=\{1.64(2);1.30(2);1.15(2);1.01(1)\}$ followed by an exponential decay to $g^{(2)}(\tau)\simeq 1$ at larger time delays (dashed lines). According to~(\ref{3_51}), we fit $g^{(2)}(\tau) = 1 + [g^{(2)}(0)-1] \exp(-{\tau}/{\tau^{(2)}_\textrm{c,exp}})$ to the data sets \ding{172}${-}$\ding{174} and obtain $\tau^{(2)}_\textrm{c,exp}\simeq\{1.75(5);1.56(8);1.18(3)\}~\si{\ns}$. For the largest condensate fraction \ding{175} the photon bunching vanishes, $g^{(2)}(0)\simeq 1$, such that we cannot determine the correlation time. Our observations reveal strikingly: above $N_\textrm{\tiny c}$, the number correlations do not rapidly drop to $g^{(2)}(0)=1$ as one would anticipate for a system with strictly conserved particle number~\cite{Klaers5,Glauber1}. Indeed, the observed behaviour provides a first evidence for grand-canonical particle exchange with an effective reservoir. The bunching amplitude $g^{(2)}(0)>1$, however, persists only up to a specific condensate fraction, where grand-canonical conditions cease to be applicable: the finite-size reservoir saturates and canonical ensemble conditions start to prevail in the system. According to ${\delta n_0}/{\bar n_0} = \sqrt{g^{(2)}(0) - 1}$,
the zero-delay autocorrelation $g^{(2)}(0)$ is directly associated to the relative condensate fluctuations. For the data shown in Fig.~\ref{fig34}(b), this gives $\delta n_0/\bar n_0=\{80(1);55(2);39(3);10(5)\}\%$.

\subsection{Grand-canonical condensate correlations}

\begin{figure}[t]
\centering\includegraphics{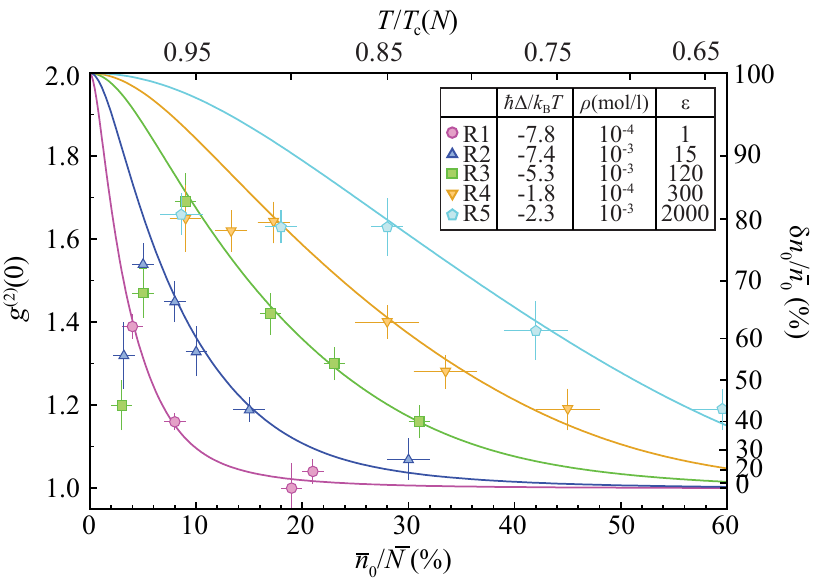}
\caption{Second-order autocorrelations $g^{(2)}(0)$ and relative condensate fluctuations $\delta n_0/\bar n_0$ as a function of the condensate fraction $\bar n_0/\bar N$ (reduced temperature $T/T_\textrm{\tiny c}$), for different reservoir sizes R1${-}$R5. The increasing effective reservoir size is quantified by $\varepsilon$. For large concentrations $\rho$ and reduced dye-cavity-detunings $\Delta$ (R5), the fluctuations of the ground mode populations persist deep into the condensate phase. The solid lines indicate the prediction from the theoretical model discussed in Section~\ref{photonenzahlstatistik}. Error bars give statistical uncertainties. ($\lambda_\textrm{\tiny c}=\{598;595;580;598;602\}\si{\nano m}$ für R1${-}$R5; $\rho=\{0.1;1.0;1.0\}~\si{\milli \mole/\litre}$ für R1${-}$R3 (Rhodamine 6G) und $\rho=\{0.1;1.0\}~\si{\milli \mole/\litre}$ für R4,R5 (PDI). For the theory curves, we use $M=\{5.5(22);20.5(71);16.0(57);2.1(4);10.8(37)\}\times 10^9$ for R1${-}$R5). Reproduced with permission from \cite{Schmitt1}. Copyright 2014 by the American Physical Society.}
\label{fig35}
\end{figure}

We systematically demonstrate the genuine grand-canonical nature of the dye-photon-system in the Bose-Einstein condensed phase by engineering different-sized reservoirs. According to~(\ref{3_53a}), the effective reservoir size is increased for high dye concentration and reduced dye-cavity-detunings.

Figure~\ref{fig35} shows zero-delay autocorrelations $g^{(2)}(0)$ and the fluctuation level, respectively, as a function of $\bar n_0/\bar N$ for five different combinations of dye concentration and detuning (R1-R3: Rhodamine 6G; R4-R5: PDI red). The main advantage of PDI red is the ability to implement small (absolute) dye-cavity-detunings $\hbar\Delta>-2.5k_\textrm{\tiny B} T$ with high reabsorption rates in a spectral region ($585{-}605~\si{\nano m}$), where the mirrors transmit a sufficient amount of light to be measured. In order to quantify the effective reservoir size (relative to R1), we introduce
\begin{equation}
\varepsilon = \frac{M_{\textrm{eff},R_i}}{M_{\textrm{eff},R1}} = \frac{\rho_{R_i}}{\rho_{R1}}\times \frac{1 + \cosh\left( \hbar\Delta_{R1}/k_\textrm{\tiny B} T  \right)}{1 + \cosh\left( \hbar\Delta_{R_i}/k_\textrm{\tiny B} T  \right)},
\label{5_6}
\end{equation}
see the table in Fig.~\ref{fig35}. For the case of the smallest reservoir (R1) the number fluctuations are quickly damped as the photons undergo BEC. Upon increasing the effective reservoir size (R1$\rightarrow$R5), we observe that the region with statistical fluctuations can be systematically extended to larger condensate fractions. For the largest implemented reservoir (R5), we find $g^{(2)}(0)\simeq 1.2$ at $\bar n_0/\bar N \simeq 60\%$. At this point, the photon condensate performs number fluctuations $\delta n_0/\bar n_0 = (g^{(2)}(0)-1)^{1/2}\simeq 45\%$, although its occupation $\bar n_0\approx 144\thinspace000$ is similar to the total number $\bar N\approx 240\thinspace000$. Our findings provide strong evidence for the photon statistics to be controlled by grand-canonical particle exchange~\cite{Holthaus1,Kocharovsky,Klaers5}.

The experimental results are recovered by our theoretical model (solid lines in Fig.~\ref{fig35}), except for condensate fractions below $5\%$. This is attributed to imperfect mode filtering that leads to an effective averaging of uncorrelated photons from a few equally populated transverse modes (at $\bar N\simeq N_\textrm{\tiny c}$) and suppresses the bunching amplitude. If the ground state contribution dominates ($\bar n_0 /\bar N\geq 5\%$), the effect becomes negligible. Furthermore, the largest detectable autocorrelation value is clamped at $g^{(2)}(0)\simeq 1.6{-}1.7$. Both issues can be resolved when the correlations are measured with a streak camera system~\cite{Oeztuerk}. To fit our data with the theory curves, the molecule number $M$ is treated as a free parameter and good agreement is obtained when we choose $10^9{-}10^{10}$ molecules, see the caption of Fig.~\ref{fig35}. The large $M$-values suggest that not only molecules located in the ground mode volume ($\approx 10^8$ for $\rho=1~\si{\milli \mole/\litre}$) contribute to the effective reservoir. A possible explanation is the residual overlap between the excited TEM$_{mn}$ modes and the TEM$_{00}$ ground mode that couples molecules in both volumes by absorption and emission of "secondary" photons, effectively increasing the reservoir size for the BEC. Alternatively, a modification of the autocorrelations could also be caused by photon-photon interactions~\cite{Wurff}. To this date, the role of interactions and the origin of photon nonlinearities in the optical condensate have not been fully resolved. Previous work has identified thermal lensing to cause effective (non-local) photon-photon-interactions associated with a dimensionless interaction parameter $\tilde g\simeq 10^{-5}{-}10^{-2}$\cite{Klaers1,Strinati,Marelic3,Dung,Alaeian,Greveling}. Promising candidates for the implementation of genuine quantum nonlinearities include e.g. polaritons of strongly interacting atomic Rydberg states~\cite{Gorshkov,Peyronel} or coupled cavity arrays \cite{Zhu}. In combination with these concepts photon BEC holds prospects for the realisation of strongly correlated many-body states of light.

\subsection{Intensity fluctuations \& photon statistics}

We have seen that the second-order correlation time ($\tau^{(2)}_\textrm{\tiny c}\simeq2~\si{\ns}$) of the Bose-Einstein condensed ground state is sufficiently slow to directly monitor the temporal number evolution with a fast photomultiplier. 

\begin{figure}[t]
\centering\includegraphics{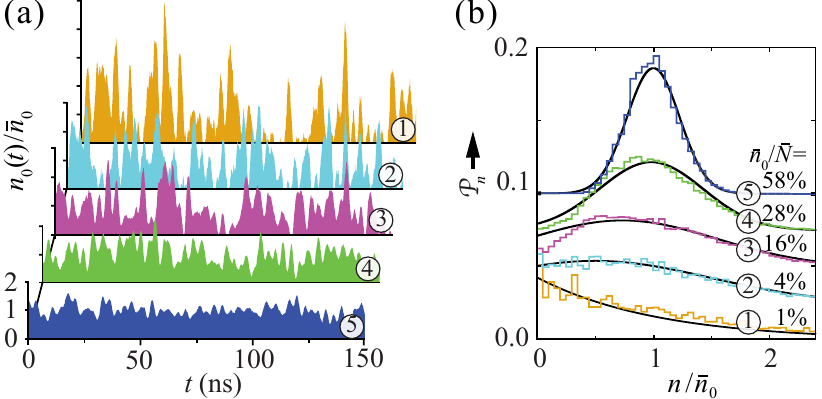}
\caption{(a) Temporal evolution of the normalised condensate population $n_0(t)/\bar n_0$ ($\simeq 1.4~\si{\ns}$ temporal resolution). For increasing condensate fractions (\ding{172}${\rightarrow}$\ding{176}) a damping of the fluctuations is observed. (b) The photon number distributions (vertically shifted) exhibit a crossover from Bose-Einstein-like to Poissonian statistics in agreement with theory (solid lines), see also Fig.~\ref{fig16}. (Parameters as in Fig.~\ref{fig34}). Reproduced with permission from \cite{Schmitt1}. Copyright 2014 by the American Physical Society.}
\label{fig36}
\end{figure}

Figure~\ref{fig36}(a) shows the time evolution of the (normalised) photon number $n_0(t)/\bar n_0(t)$ for a fixed reservoir size with parameters as in Fig.~\ref{fig35}. Close to the condensation threshold, the BEC exhibits large number fluctuations, which are gradually damped out as the condensate fraction is increased. By evaluating histograms of roughly 50 traces per condensate fraction, we reconstruct the underlying photon statistics $\mathcal{P}_n$, i.e. the probability to find $n$ photons in the condensate, see Fig.~\ref{fig36}(b). As the fluctuations are reduced, the distributions reveal a crossover from exponentially decaying Bose-Einstein towards Poissonian statistics, with a width $\delta n_0$ that measures the relative degree of fluctuations $\delta n_0/\bar n_0$. Our results are in excellent agreement with theory curves (solid lines) from~(\ref{3_18}), confirming the predicted crossover from grand-canonical to canonical statistical conditions (Section~\ref{photonenzahlstatistik}).

\section{Phase coherence of the condensate}
\label{phasenkohaerenzexp}
In the presence of large reservoirs, even strongly-occupied BECs that contain thousands of photons on average display a finite probability $\mathcal{P}_{0}>0$ to produce states without a single photon. Naturally, the question arises: how do such statistical (amplitude) fluctuations affect the temporal phase stability of the BEC? In comparison, the Poissonian statistics in the limit of small reservoirs causes the zero-photon probability to vanish $\mathcal{P}_{0}=0$, such that - despite residual phase diffusion~\cite{Imamoglu1,Lewenstein,Leeuw,Naraschewski} - a well-defined phase is expected. Similar observations with (micro)canonical atomic BECs prompt the emergence of phase coherence for the condensate wave function~\cite{Andrews,Ketterle1,Oettl,Saba}. In the last section of this Tutorial, we describe an experimental measurement of the temporal phase coherence for a BEC of light.

\subsection{Experimental scheme}
\begin{figure}[t]
\centering\includegraphics{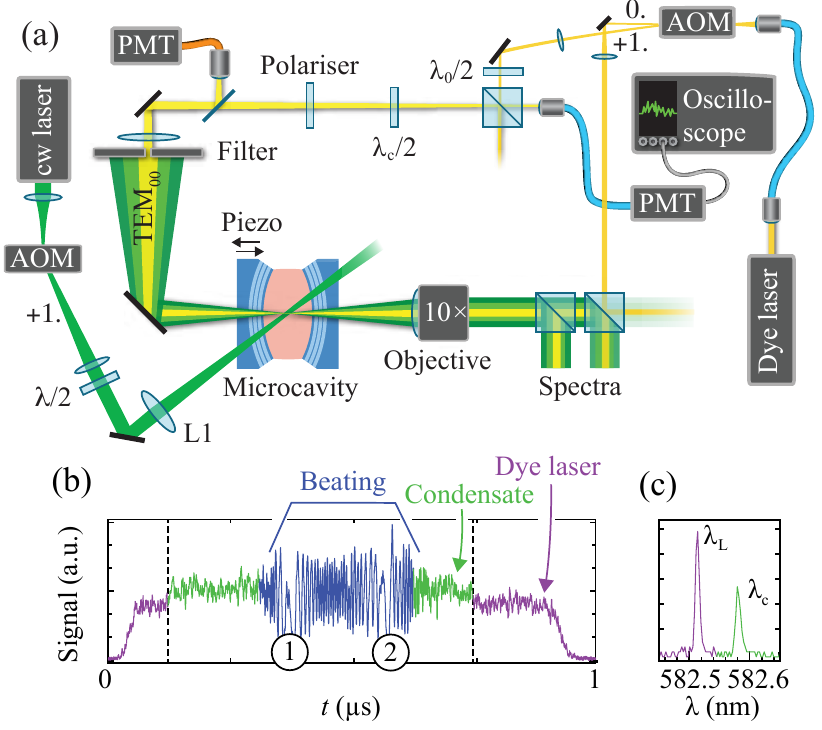}
\caption{Heterodyne interferometry to study the phase evolution of the photon condensate. The momentum-filtered ground mode emission (left cavity mirror) is superimposed with a dye laser in a single mode fibre. (b) A typical resulting interference signal recorded with a fast photomultiplier. (c) The emission transmitted through the right cavity mirror is used to monitor the wavelengths of BEC and dye laser, $\lambda_\textrm{\tiny c}$ and $\lambda_\textrm{\tiny L}$, in a double monochromator and measure the condensate fraction $\bar n_0/\bar N$.}
\label{fig37}
\end{figure}

To study the phase evolution, we rely on time-resolved heterodyne interference signals between the condensate emission superimposed with a dye laser acting as a phase reference, see Fig.~\ref{fig37}~\cite{Leeuw}. From a separate detection of the intensity of the condensate in the interferometer (blocked dye laser), we obtain the degree of second-order coherence $g^{(2)}(0)$ and the correlation time $\tau^{(2)}_\textrm{\tiny c}$. The experiments are performed for longitudinal wave number $q=7$ in the microcavity, which is filled with a Rhodamine 6G solution ($\rho=3~\si{\milli \mole/\litre}$). The microcavity is pumped with continuous laser light, which is here chopped into $600~\si{\ns}$ pulses at $40~\si{\hertz}$ repetition rate by an AOM.

As a local oscillator for the heterodyne interferometry we use a cw dye laser (Sirah Matisse), which offers a tuneable emission between $\lambda_\textrm{\tiny L}=560\textrm{-}605~\si{\nano m}$. Analog to the condensate operation cycle, the dye laser is acousto-optically chopped into $800~\si{\ns}$ pulses at the same repetition rate, while the zeroth diffraction order allows to measure $\lambda_\textrm{\tiny L}$ with a resolution of approximately $10~\si{\pico m}$, see Fig.~\ref{fig37}(c). The relatively long pulse duration is required in order to observe sufficiently long beatings between the condensate and dye laser emission, as will be elaborated in more detail later in this section. To obtain high-contrast interference signals, we use half-wave plates to project the polarisation axes of the momentum-filtered photon condensate and dye laser on top of each other, and combine both beams after passing a non-polarising beamsplitter (90\% transmission) in a single mode fiber (Thorlabs P1-488PM)\setcounter{footnote}{2}\footnote{The optical phase is commonly retrieved in a balanced heterodyne detection scheme~\cite{Carleton}, by subtracting the interference signals at both output ports of a symmetric (50:50) beamsplitter exploiting their $\pi$-phase difference. For low condensate powers, however, the usage of an asymmetric (90:10) beamsplitter turned out to enhance the signal-to-noise ratio of the observed beating signals.}. The temporal interference traces are detected by a fast photomultiplier tube (Hamamatsu H10721-20, $\Delta t\simeq 0.57~\si{\ns}$) and recorded with a digital oscilloscope (Tektronix DPO7000, $\Delta\nu\simeq 3.5~\si{\GHz}$). A typical time-resolved interference signal, where condensate and dye laser wavelength have been matched, is shown in Fig.~\ref{fig37}(b). The superposition of Bose-Einstein condensed light field, $\psi_\textrm{\tiny c}(t)=\sqrt{n_\textrm{\tiny c}(t)}\exp\{i[\omega_\textrm{\tiny c}(t) t+\varphi(t)]\}$, and dye laser field, $\psi_\textrm{\tiny L}(t)=\sqrt{n_\textrm{\tiny L}(t)}\exp(i\omega_\textrm{\tiny L}t)$, gives a beat signal
\begin{eqnarray}
\left|\psi_\textrm{\tiny c}(t)+\psi_\textrm{\tiny L}(t)\right|^2 = & n_\textrm{\tiny c}(t)+n_\textrm{\tiny L}(t)+ 2 \sqrt{n_\textrm{\tiny c}(t)n_\textrm{\tiny L}(t)} \nonumber\\
& \times \cos\left\{\left[\omega_\textrm{\tiny c}(t)-\omega_\textrm{\tiny L}\right] t+\varphi(t)\right\},
\label{6_0}
\end{eqnarray}
where $\varphi(t)$ denotes the time-dependent condensate phase. Notably, we here have explicitly maintained the time dependence of the condensate frequency $\omega_\textrm{\tiny c}(t)$, for reasons that will be discussed in the following.

The thermodynamic state of the photon gas is obtained from a spectroscopic measurement of the energy distributions with a $4f$-spectrometer (Section~\ref{secondordercorrelationsexp}). The spectra cover a wavelength (energy) range of $30\si{\nano m}$ ($\simeq 4k_\textrm{\tiny B} T$) and provide the ground state population $\bar n_0$ and total photon number $\bar N$, being calibrated with reference spectra at $N_\textrm{\tiny c}\simeq 79\thinspace000$. Moreover, a part of the transmitted cavity emission is injected into the high-resolution double monochromator together with the aforementioned dye laser to monitor the relative spectral position of condensate and dye laser wavelength, see Fig.~\ref{fig37}(c). At the smallest achievable cavity lengths $D_0\approx 1.4\si{\micro m}$, the curved mirrors are firmly pressed together, effectively reducing residual mechanical resonator drifts and vibrations. Under these conditions, minute piezo-tuning of the cavity length allows us to actively match the condensate with the dye laser wavelength with a spectral precision $\Delta\lambda\approx 10~\si{\pico m}$. At $\lambda=580~\si{\nano m}$, the mirror separation can thus be tuned with an accuracy of $\Delta D_0=D_0 \Delta\lambda/\lambda\approx 24.5~\si{\pico m}$.

\subsection{Modulation of the condensate frequency}
\label{modulation}
\begin{figure*}[t]
\centering\includegraphics{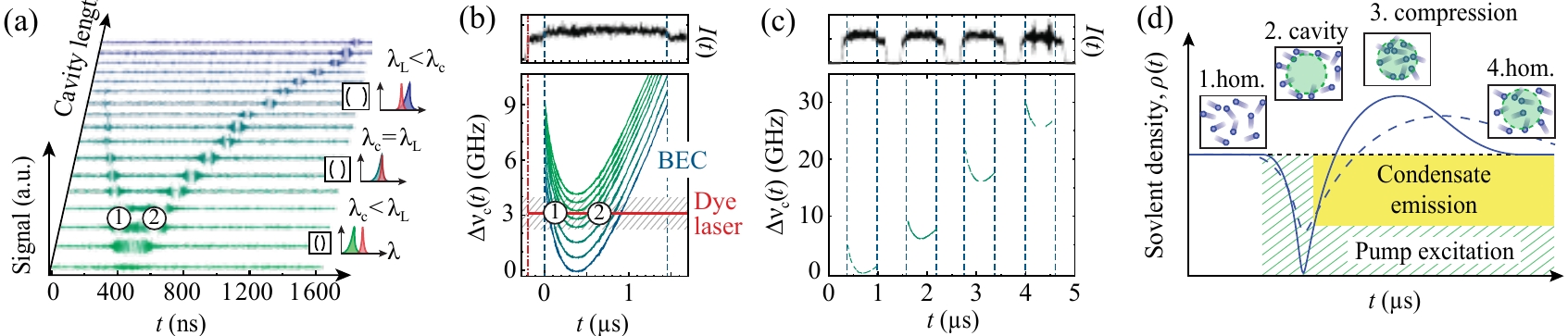}
\caption{(a) Interference signals upon tuning $\lambda_\textrm{\tiny c}$ at fixed dye laser wavelength $\lambda_\textrm{\tiny L}$ by increasing the cavity length. Both a frequency modulation and a relative temporal separation of two resonance crossings (\ding{172},\ding{173}) are observed. (b)~Non-resonant intensity trace (top) and relative drift of the condensate frequency $\Delta\nu_\textrm{\tiny c}(t)$, as reconstructed from (a) (bottom). Vertically shifted curves correspond to increased cavity lengths (from top to bottom), and the hatched area indicates the PMT detection bandwidth ${\simeq}1.75~\si{\GHz}$. (c) Combined intensity (top) and $\Delta\nu_\textrm{\tiny c}(t)$ (bottom) upon excitation with a sequence of 4 pulses showing a beating within the last pulse. We deduce a global slow linear frequency drift over several pulses and a parabolic frequency modulation is reproduced within each individual pulse, possibly caused by a (d) temporal density modulation of the dye medium following the pump excitation (hatched). A tight pump beam focus (solid line) leads to an enhanced amplitude and speed of the modulation during the BEC emission (yellow), compared to a weak focusing scenario (dashed line).}
\label{fig39}
\end{figure*}

Despite the mechanical stability of the microcavity, we have already seen in Fig.~\ref{fig37} that the measured intensity traces reveal a frequency modulation of the beating signal. Figure~\ref{fig39}(a) shows the observed temporal variation of the beat signal for different initial cavity lengths, the latter modifying $\lambda_\textrm{\tiny c}$. For an average condensate wavelength blue-detuned with respect to the dye laser ($\lambda_\textrm{\tiny c}<\lambda_\textrm{\tiny L}$, insets of Fig.~\ref{fig39}(a)), no beating signal is observed. For red-detuned light ($\lambda_\textrm{\tiny c}>\lambda_\textrm{\tiny L}$) however, the occurring beating signal shows two resonances that exhibit an increased temporal separation as the condensate is further detuned, with the beating frequency in between exceeding the detector bandwidth. The interference data allows us to reconstruct the frequency drift $\nu_\textrm{\tiny c}(t)$ of the condensate emission, which is shown in Fig.~\ref{fig39}(b) for various cavity lengths.

The frequency drift in Fig.~\ref{fig39}(b) has been recorded for temporally equidistant pump pulse excitation (pulse length $\Delta t= 1.5~\mu\textrm{s}$ every $T_\textrm{\tiny p}=25~\si{\milli\second}$), leading to beating signals that occur in every subsequent condensate pulse with nearly the same shape (due to mechanical shot-to-shot stability). However, when irradiating the dye-microcavity with a quick sequence of 4 pump pulses ($\Delta t=T_\textrm{\tiny p}=600~\si{\ns}$, followed by $100~\si{\milli\second}$ dark time), see Fig.~\ref{fig39}(c), only one of the four produced condensate pulses exhibits a beating with the laser, yet with the same characteristic, nearly parabolic "fast" frequency drift observed previously. We attribute the "slow" global frequency drift to modulation of the index of refraction of the dye medium that is caused by effectively heating the solution with the pump laser. The relaxation timescale is approximately $20~\si{\milli\second}$, similar to timescales of thermal lensing effects in our system~\cite{Klaers1,Klaers4,Dung}. The fast sloshing of the condensate eigenfrequency during a single pulse may be caused by the steep rising slope of the pump pulse itself, as it occurs in each pulse of the fast sequence scheme shown in Fig.~\ref{fig39}(c). Furthermore, the behaviour is observed only for longitudinal wave numbers $q\leq10$ when the dye film in between the cavity mirrors becomes kinematically 2D. Both observations give reason to conclude that the parabolic contribution to the frequency modulation $\nu_\textrm{\tiny c}(t)$ is based on a refractive index change, which originates from an overdamped density oscillation in the dye film. Due to the cavity length stability within a single pulse,
\begin{equation}
D_0 = q \frac{\lambda_\textrm{\tiny c}(t)}{2\tilde n_0(t)} = q \frac{c}{2\tilde n_0(t)\nu_\textrm{\tiny c}(t)} \stackrel{!}{=} \textrm{const.}
\label{6_1}
\end{equation}
Therefore, the density of the dye solution in the ground mode volume reads
\begin{equation} 
\rho(t) \propto \tilde n_0(t)= q\frac{c}{2D_0 \nu_\textrm{\tiny c}(t)}=q\frac{1}{\tau_\textrm{\tiny rt}\nu_\textrm{\tiny c}(t)},
\label{6_2}
\end{equation}
where the (vacuum) resonator round trip time $\tau_\textrm{\tiny rt}=2D_0/c$ has been inserted. According to~(\ref{6_2}), the density scales inversely with the frequency drift from Fig.~\ref{fig39}(b), corresponding to a compression of the solvent in the area of the pump beam. Presumably, this could be caused by an initial localised heating and dilution of the medium due to the pump pulse, see the illustrated sequence in Fig.~\ref{fig39}(d). The resulting density hole leads to a reflow and densification of ethylene glycol molecules until the medium is finally homogenised. The observed time scale of the overdamped density modulation is consistent with an estimate based on the propagation time of a sound wave through the ground mode area of diameter $d_0\approx15~\si{\micro m}$, $t_\textrm{s}=d_0/v_\textrm{s}\approx 10^{-8}~\si{\second}$, where $v_\textrm{s}=1688~\si{\meter\per\second}$ is the speed of sound in ethylene glycol at $300~\si{\kelvin}$~\cite{Gopalakrishna}, and it occurs on a  considerably shorter time scale than thermal lensing ($10^{-3}\textrm{s}$). Moreover, our interpretation is affirmed by the notion that the pump beam geometry affects the condensate frequency modulation: for a larger pump beam waist, the dynamics becomes slower and the maximum of the compression is postponed to later times, see Fig.~\ref{fig39}(d)\setcounter{footnote}{2}\footnote{In principle, defocusing allows one to observe temporally extended beating signals. However, this is limited by the required increased length of the pump pulses, which inevitably leads to a breakdown of the condensate operation due to photodegradation and triplet-state pumping of the dye.}. For all subsequently discussed measurements, we use a fixed pumping geometry with a beam diameter $2w_{01}=2\lambda f_\textrm{\tiny L1}/\pi w_0\simeq 140~\si{\micro m}$ (beam waist $w_0=1~\si{\milli m}$ and $f_\textrm{\tiny L1}=40~\si{\cm}$, see Fig.~\ref{fig37}).

\begin{figure}[t]
\centering\includegraphics{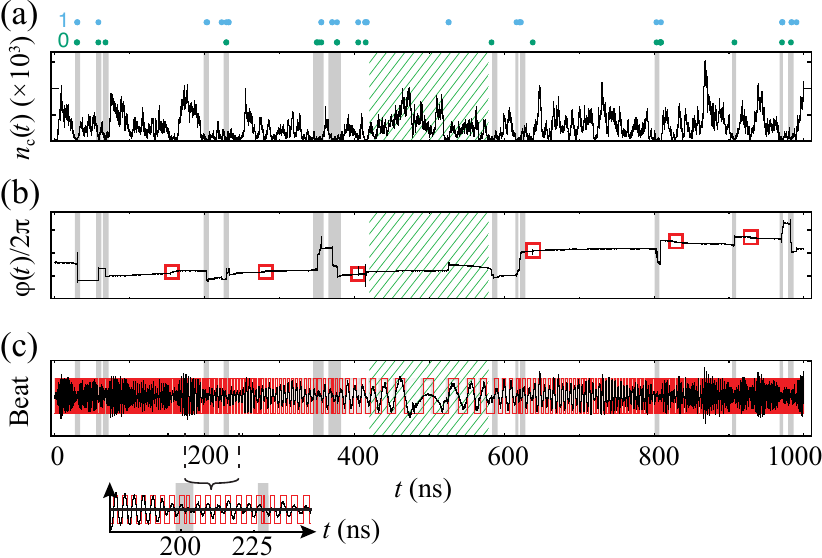}
\caption{Phase jump detection benchmark with simulated interference signals. (a) Number evolution of a fluctuating BEC and points in time with zero or one photon in the ground mode (dots, top). (b) Corresponding phase evolution (units of $2\pi$) with a phase jump detection resolution $\Delta\varphi>0.2\pi$ (red boxes indicate missed events). (c) Beating between photon condensate and reference laser and digitalised signal (red). Irregularities in the periodicity of subsequent square signals selected by our algorithm (grey shaded, see also zoomed time trace) coincide with actual phase jumps in (b). The resonance is excluded from the detection (hatched). ($M=10^6$, $\bar n_\textrm{\tiny c} = 280$ with $g_\textrm{\tiny c}^{(2)}(0)=1.67$, $\hbar\Delta/k_\textrm{\tiny B} T=0$, wavelength drift $\lambda_\textrm{\tiny c}(t)=\lambda_\textrm{\tiny L} [1-(0.5- t/T)/20]$ with $T=1~\mu\textrm{s}$. Reference laser: $\bar n_\textrm{\tiny L} = 1000$ with $g_\textrm{\tiny L}^{(2)}(0)=1.0$)}
\label{fig40}
\end{figure}

\subsection{Phase jump detection algorithm}
The microcavity frequency drift prohibits a temporally stable resonance condition to be fulfilled between photon condensate and dye laser, making a direct observation of the BEC phase evolution difficult. However, discrete phase jumps of the condensate can be easily unveiled if the recorded chirped interference signals are examined for irregularities in their oscillatory behaviour. For an automated analysis, we develop a phase jump detection algorithm that we benchmark with Monte-Carlo-simulated data (Section~\ref{phasenkohaerenz}).

Figure~\ref{fig40} shows the simulated (a) intensity and (b) phase evolution of a BEC under grand-canonical statistics, and (c) depicts the corresponding simulated beating signal between the photon BEC and a dye laser. In the first step of the analysis, the analog interference signal is digitalised (red). Subsequently, the procedure evaluates the digital square-signal for irregularities in the (i) width and (ii) central position of adjacent high- or low-valued segments. If the irregularities exceed predefined limits, the algorithm flags these points in time (grey shaded). The low-frequency region near the resonance (hatched) is excluded from the detection. The simulated data confirms the operability of the algorithm, as demonstrated by coincidences of grey regions with zero- or one-photon-states in the ground state (Fig.~\ref{fig40}(a), top, dots). It enables the detection of discrete phase rotations between $[0.2\pi,1.8\pi]$. As the analysis is based on the detection of relative irregularities, the temporal resolution is limited by the beating oscillation period.

\begin{figure}[htbp]
\centering\includegraphics{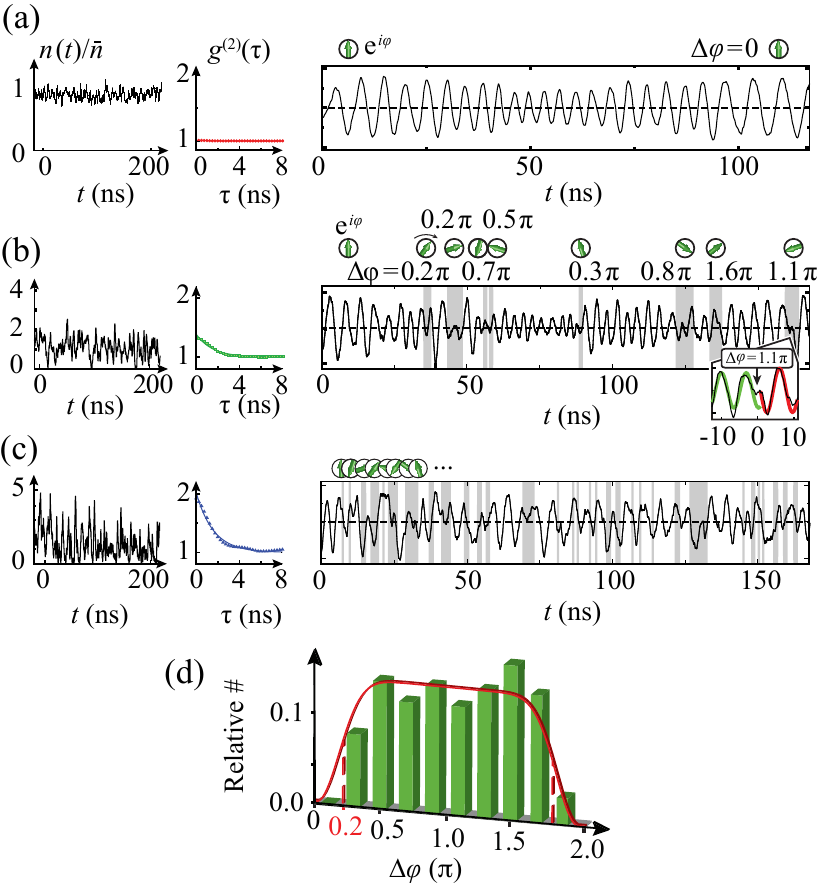}
\caption{Interference between photon BEC and dye laser (right) for average photon numbers (a) $\bar n = 114\thinspace000$ ($\bar n/ \bar N=57\%$) (b) $8\thinspace300$ ($10\%$) and (c) $3\thinspace700$ ($5\%$), which realises different levels of statistical number fluctuations, as visible in the normalised photon number evolution (left) and the autocorrelation (middle). Regions that have been identified by our detection algorithm (grey shaded) indicate phase jumps at increasing rates from (a) to (c). The magnitude of the phase rotations is obtained from a fit as shown in the inset in (b). (d) Histogram of the phase rotation angles $\Delta\varphi$ for $108$ fitted phase jumps in signals as in (b). Within the detection window $[0.2\pi,1.8\pi]$ (red line), the random distribution reflects the $U(1)$ symmetry of the ground state, which is broken upon condensation. (Rhodamine 6G, $\rho=3~\si{\milli \mole/\litre}$, $\lambda_\textrm{\tiny c}=582~\si{\nano m}$).  Reproduced with permission from \cite{Schmitt4}. Copyright 2016 by the American Physical Society.}
\label{fig41}
\end{figure}

\subsection{Phase evolution of the photon condensate}

Figure~\ref{fig41} shows the time evolution of the interference between photon BEC and dye laser for three different cases of photon statistics at a fixed reservoir size, starting from a strongly occupied second-order coherent condensate in Fig.~\ref{fig41}(a) towards a strongly fluctuating population in (c). The left column gives the time of the (normalised) condensate number $n(t)/\bar n$, which is recorded after each interference measurement by blocking the dye laser. From this, the autocorrelation function $g^{(2)}(\tau)$ is computed (middle), implicitly providing a measure of the fluctuation level $\delta n =\bar n\sqrt{g^{(2)}(0)-1}$ and the second-order correlation time $\tau^{(2)}_\textrm{\tiny c}$. In all measurements with a significant bunching amplitude ($g^{(2)}(0)>1$), an exponential fit to the autocorrelation data yields $\tau^{(2)}_\textrm{\tiny c}\approx 2~\si{\nano s}$.

For canonical ensemble conditions with Poissonian number statistics, see Fig.~\ref{fig41}(a) with $g^{(2)}(0)=1.01(2)$, the beating oscillates regularly, which demonstrates the temporal coherence of the BEC throughout $120~\si{\ns}$\setcounter{footnote}{2}\footnote{For large waists of the pump beam, the longest recorded time span without phase jumps was $1~\mu\textrm{s}$ ($300~\si{m}$ coherence length).}. As the condensate fraction is reduced, the reservoir size becomes sufficiently large to realise grand-canonical statistical conditions, which is hallmarked by the occurrence of intensity fluctuations in Figs.~\ref{fig41}(b) with $g^{(2)}(0)=1.33(4)$ and (c) with $g^{(2)}(0)=1.93(13)$, respectively. This is accompanied by a discontinuous phase behaviour manifested in the beating signals, which for increased fluctuations shows a reduction of the time separation between adjacent phase jumps $\Gamma_\textrm{\tiny PJ}^{-1} \approx 21.3~\si{\ns}$ in (b) and $\Gamma_\textrm{\tiny PJ}^{-1} \approx 5.3~\si{\ns}$ in (c). In the vicinity of the detected phase jumps (grey shaded) a fit yields the magnitude of the imparted phase shift, see the inset of Fig.~\ref{fig41}(b). To good approximation, the phase rotation angles are evenly distributed within the detection range $[0.2\pi,1.8\pi]$, as indicated by the histogram in Fig.~\ref{fig41}(d). The random distribution gives evidence for the $U(1)$ symmetry of the infinitely phase-degenerate ground state. Physically, this equipartition is attributed to the intrinsic randomness of a spontaneous emission event, which is expected to trigger the emergence of a condensate after a previous fluctuation to low photon numbers.

\begin{figure}[t]
\centering\includegraphics{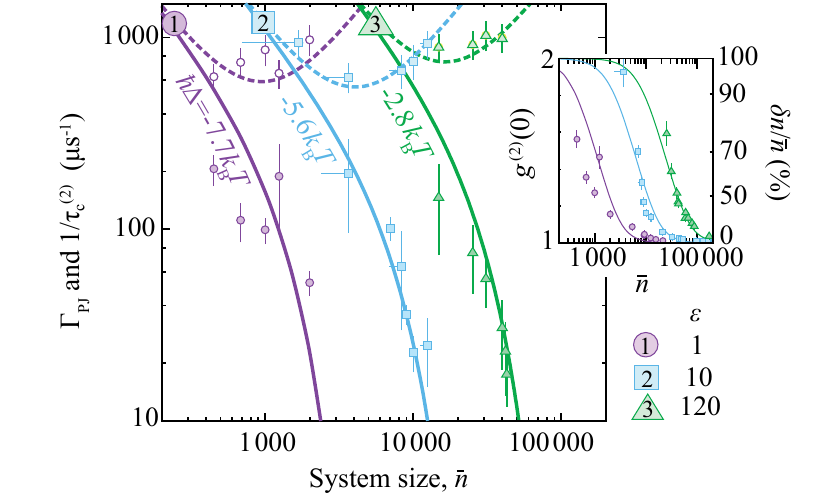}
\caption{Inverse second-order correlation time $1/\tau_\textrm{\tiny c}^{(2)}$ (open symbols) and phase jump rates $\Gamma_\textrm{\tiny PJ}$ (filled symbols) versus condensate number for three different-sized reservoirs, realised by $\hbar\Delta=-\{7.7;5.6;2.8\}k_\textrm{\tiny B}T$ and quantified by $\varepsilon$, see~(\ref{5_6}). Solid lines give theory curves for $\Gamma_\textrm{\tiny PJ}^0$ with $M=\{2.0;4.5;5.0\}\times 10^9$, $\hat B_{12}=\{140;250;1\thinspace300\}\textrm{s}^{-1}$. In the region accessed by our experiments, the phase jump rates are significantly smaller that the inverse second-order correlation times. The inset gives the corresponding zero-delay autocorrelation $g^{(2)}(0)$ along with numerical calculations (solid lines). ($\lambda_\textrm{\tiny c}=\{597;582;563\}\si{\nano m}$, $\rho=3~\si{\milli \mole/\litre}$, $\bar n = \sqrt{M_\textrm{\tiny eff}}\simeq\{1.0;4.1;16.5\}\times 10^3$, error bars are statistical uncertainties). Reproduced with permission from \cite{Schmitt4}. Copyright 2016 by the American Physical Society.}
\label{fig43}
\end{figure}

\subsection{First- and second-order coherence times}
\label{korrelationszeitenexp}
As previously discussed, Fig.~\ref{fig41} indicates a separation of the dynamics for number and phase fluctuations: while $\tau^{(2)}_\textrm{\tiny c}$ remains nearly constant, the measured values for $\Gamma_{\textrm{\tiny PJ}}^{-1}$ change by 2 orders of magnitude and seem to depend on the choice of the statistical ensemble and its associated zero-photon-probability $\mathcal{P}_0$.

For a quantitative analysis of the time scale separation, Fig.~\ref{fig43} summarises experimental results of the phase jump rates $\Gamma_\textrm{\tiny PJ}$ and inverse second-order correlation times $1/\tau_\textrm{\tiny c}^{(2)}$ as a function of the average photon number in the condensate for three different-sized particle reservoirs. The phase jump rates (filled symbols) increase strongly for both growing reservoir size as well as decreasing condensate photon number ("system size") based on the here enhanced probability to have a low photon number given the increased fluctuation level (inset), which reduces the phase stability. The rates deduced from the zero-photon-probabilities $\Gamma_\textrm{\tiny PJ}^0= \hat B_{12} M \mathcal{P}_0$ (solid lines) show an excellent agreement with the experimental data. This suggests that a drop of the condensate population to zero followed by a spontaneous emission process is physically responsible for the observed phase jumps. Similarly, the inverse second-order correlation times $1/\tau_\textrm{\tiny c}^{(2)}$ (open symbols) present a good agreement with theory curves (dashed lines) based on~(\ref{3_52}). 

For all three configurations, a separation of the time scales for first- and second-order coherence is visible in the statistics crossover region, i.e. near $\bar n=\sqrt{M_\textrm{\tiny eff}}$. What is its physical origin? On the one hand, spontaneous emission events can cause arbitrary phase fluctuations. However, these matter only when a few photons occupy the ground state with a likelihood given by the photon statistics, which therefore dominates the first-order phase jump dynamics, see~(\ref{3_58}). On the other hand, the dynamics of particle number fluctuations is subject to absorption and emission rates of photons by the dye medium, according to~(\ref{3_46}). Although for increased condensate populations (at a fixed reservoir) the relative fluctuations $\delta n/\bar n$ are reduced, the fluctuation time scale is still controlled by the Einstein coefficients. In fact, even larger condensate populations lead to a reduction of the second-order correlation time, in stark contrast to the increased first-order correlation time. Although our analysis does not account for diffusive contributions to the temporal phase coherence~\cite{Leeuw}, it conveys the unusual properties of Bose-Einstein condensed light: a light source comprised of a single macroscopically occupied emitter that exhibits statistical intensity fluctuations as large as in a thermal source. The relation between first and second-order coherence for thermal emitters, $g^{(2)}(\tau)=1+|g^{(1)}(\tau)|^2$, is however expected to hold only in the extreme grand-canonical regime with $\bar n\geq\sqrt{M_\textrm{\tiny eff}}$~\cite{Loudon,Mandel}.

\subsection{Extrapolation to the thermodynamic limit}
Finally, we discuss the physical significance of statistical number fluctuations and phase coherence for a photon BEC in the thermodynamic limit. For this, we study the phase jump rate for enlarged system sizes. Importantly, we ensure to increase the sizes of both condensate $\bar n$ and effective particle reservoir $M_\textrm{\tiny eff}$ in a way that conserves the statistical ensemble conditions.

\begin{figure}[t]
\centering\includegraphics{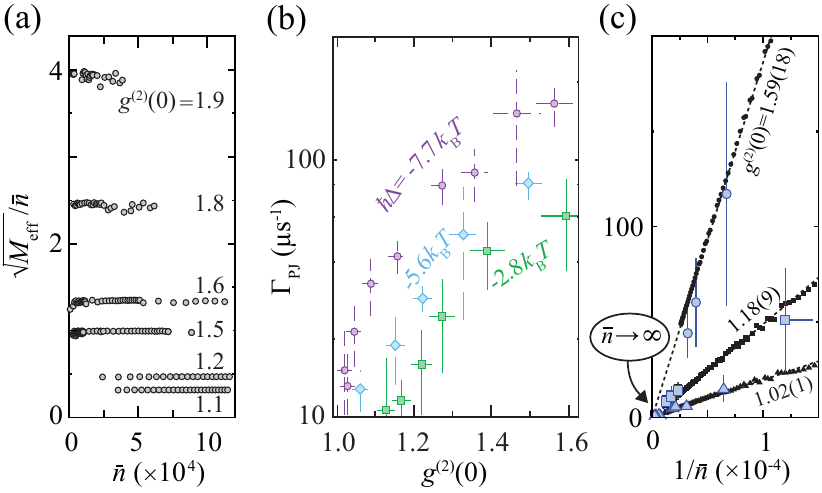}
\caption{(a) Numerical scaling of effective reservoir and condensate size versus $\bar n$ for various fluctuation levels $g^{(2)}(0)$, demonstrating the conservation of $\sqrt{M_\textrm{\tiny eff}}/\bar n$ as the autocorrelation remains fixed. ($\hbar\Delta=0$, $M=0.5\textrm{-}2.0\times 10^{10}$). (b) Measured $\Gamma_\textrm{\tiny PJ}$ versus autocorrelation for three effective reservoirs. For each fluctuation level, the larger system exhibits an enhanced phase stability. (c) Extrapolation of phase coherence for large photon-reservoir-systems at three fixed degrees of fluctuations $g^{(2)}(0)$ (filled symbols). By extrapolating $\bar n \rightarrow\infty$ (dashed), all curves indicate a full suppression of phase jumps in the thermodynamic limit. Black symbols are numerical results for $\Gamma_\textrm{\tiny PJ}^0=\hat B_{12} M \mathcal{P}_0$ at corresponding $g^{(2)}(0)$ and $M=4\times 10^9$, $\hat B_{12}=1000~\si{\second}^{-1}$. (Parameters as in Fig.~\ref{fig43}). Reproduced with permission from \cite{Schmitt4}. Copyright 2016 by the American Physical Society.}
\label{fig45}
\end{figure}

Figure~\ref{fig45}(a) shows the reservoir-system-ratio $\sqrt{M_\textrm{\tiny eff}}/\bar n$ as a function of $\bar n$ for different values of $g^{(2)}(0)$ obtained from numerical calculations. For a given photon number $\bar n$, the reservoir size $M_\textrm{\tiny eff}$ is adjusted iteratively until the corresponding photon number distribution $\mathcal{P}_n$ reproduces one of the target values $g^{(2)}(0)=\{1.10;...;1.90\}$. Subsequently, the procedure is repeated for larger condensate populations to yield further data points at the same fluctuation level. Our numerical results indicate that conserving the the statistical ensemble conditions, i.e. $\sqrt{M_\textrm{\tiny eff}}/\bar n$, is equivalent to a constant zero-delay autocorrelation. This suggests that the phase coherence may be extrapolated towards the thermodynamic limit ($\bar n, \bar N, M\rightarrow \infty$, $\sqrt{M_\textrm{\tiny eff}}/\bar n=\textrm{const.}$), provided that one does maintain the fluctuation level $g^{(2)}(0)$. Strictly speaking, an extrapolation also requires the critical temperature $T_\textrm{\tiny c}(\bar N)\propto \sqrt{\bar N/R}$ to be constant. This could be achieved by increasing the radius of curvature of the cavity mirrors $R\rightarrow \infty$ proportional to $\bar N$. Experimentally, this compensation is unfeasible with the described setup, such that we only comply with the requirements for a fixed statistical ensemble\setcounter{footnote}{2}\footnote{Recently realised photon gases in variable micropotentials might however render a conservation of $T_\textrm{\tiny c}$ tractable~\cite{Dung,Walker}.}.

Figure~\ref{fig45}(b) gives the phase jump rate versus autocorrelation for three reservoirs. For all values of $g^{(2)}(0)$, we observe that photon condensates coupled to the smallest reservoir ($\hbar\Delta=-7.7 k_\textrm{\tiny B} T$) exhibit shorter coherence times than condensates coupled to the medium-sized ($-5.6 k_\textrm{\tiny B} T$) and largest ($-2.3 k_\textrm{\tiny B} T$) reservoir. This meets our expectations: for the same level of fluctuations, i.e. the same statistical ensemble, an increased condensate population should reduce the zero-photon-probability $\mathcal{P}_0$, see~\ref{3_60}.

From our data, we extract three sets of phase jump rates for selected zero-delay autocorrelations $g^{(2)}(0)=1.59(18)$, $1.18(9)$ and $1.02(1)$, which are shown in Fig.~\ref{fig45}(c) versus the inverse condensate population $1/\bar n$. All data sets lie in the range $\bar n\geq \sqrt{M_\textrm{\tiny eff}}$, for which a separation of $\Gamma_\textrm{\tiny PJ}$ and $1/\tau_\textrm{\tiny c}^{(2)}$ has been observed. A linear extrapolation of the data towards an infinitely large condensate ($1/\bar n\rightarrow 0$) is consistent with a full suppression of discrete phase jumps in the thermodynamic limit, in spite of the absence of second-order coherence. Numerical calculations (black symbols) for $g^{(2)}(0)=1.50,1.18$ und $1.05$ support this conclusion. The largest realised fluctuation level comes close to the photon statistics crossover, $g^{(2)}(0)=\pi/2$, with a zero-photon-probability $\mathcal{P}_{0}\simeq 0.64\thinspace \bar n^{-1}$, see~(\ref{3_41}). Under the assumption that phase jumps occur due to vanishing photon numbers, a fit to the data in Fig.~\ref{fig45}(c) yields $\Gamma_\textrm{\tiny PJ}/ \hat B_{12} M =0.51(14) \bar n^{-1}$ reproducing the expected slope within the quoted uncertainty. For lower fluctuation levels, the exact scaling of $\mathcal{P}_0$ with $\bar n$ remains elusive and we therefore compare our data only with numerical results, which similarly demonstrate a linear scaling of $\Gamma$ with the inverse photon number ($\mathcal{P}^{1.18}_0=0.13(5)\bar n^{-1}$ and $\mathcal{P}^{1.02}_0=0.06(2)\bar n^{-1}$). Although the presence of amplitude fluctuations of the condensate wave function $\sqrt{n(t)}\exp(i\phi(t))$ reduces the degree of first-order coherence, we expect that in the investigated parameter regime discrete phase jumps will be fully suppressed in the thermodynamic limit.

\section{Conclusions and Outlook}
\label{conclusion}

This Tutorial has presented a study of the thermalisation dynamics and temporal coherence properties of a Bose-Einstein condensed photon gas in the grand-canonical statistical ensemble. Key evidences are provided by measurements of {\itshape (i)} the spectral photon dynamics, which demonstrates the thermalisation of the photons due to reabsorptive coupling to a dye heat bath, {\itshape (ii)} the large (grand-canonical) statistical number fluctuations at significant condensate fractions, and {\itshape (iii)} the observed variation of the temporal phase coherence of the condensate wave function. An extrapolation to the thermodynamic limit gives BECs with super-Poissonian number statistics despite suppressed phase jumps.

The realisation of BEC in the grand-canonical ensemble has for the first time shed light on the long-discussed {\itshape grand-canonical fluctuation catastrophe}~\cite{Haar,Holthaus1,Ziff,Fujiwara,Fierz,Kocharovsky,Navez,Weiss}. The observation of extremely large, statistically fluctuating condensate populations demonstrates the physical significance of the grand-canonical ensemble for the Bose-Einstein condensed phase. Moreover, the results provide the fundamental insight that BEC does not strictly imply first- or second-order coherence.

For the future, it will be exciting to study phase diffusive contributions to the condensate linewidth, as has been theoretically predicted but remains elusive in any Bose-condensed system to date~\cite{Leeuw,Imamoglu1,Naraschewski,Lewenstein}. A major experimental challenge here depicts the required frequency stability of the photon BEC to observe minute phase drifts over long measurement durations. Moreover, it is expected that in-depth studies of the thermal character of the grand-canonical statistical fluctuations may reveal unusual fluctuation-dissipation-relations in the ideal Bose gas, associated with macroscopic thermodynamic quantities as e.g. a generalised statistical compressibility imposed by the particle reservoir. From a technical point of view, macroscopically occupied, but incoherent photon condensates under grand-canonical conditions could pose interesting novel light sources for speckle-free imaging applications due to their high directional brilliance and (tuneable) low degree of coherence.

Further exciting research directions for grand-canonical BECs might be pursued in conjunction with variable potentials for thermalised light and coupled condensates, as has been demonstrated in microstructured optical cavities~\cite{Dung,Walker}. Phase-stable, macroscopically occupied condensates arranged in a lattice are expected to constitute a realisation of the XY model of 2D interacting spins, that could provide a fruitful platform to address complex optimisation problems~\cite{Cuevas,Berloff}. In this regard, the phase jumps associated with grand-canonical statistical fluctuations could mimic spin fluctuations at an effective temperature: at sufficiently low "temperatures" one expects the emergence of the BKT phase associated with algebraic long-range spin order~\cite{Berezinski,Kosterlitz,Hadzibabic,Hadzibabic3}.

\section*{Acknowledgements}
I thank F. \"Ozt\"urk, E. Busley, J. Klaers and M. Weitz for their critical reading of the manuscript. Many insightful discussions on photon condensation, in particular with R. Nyman and P. Kirton, have been fruitful for the writing of this manuscript. Support by the Bonn-Cologne Graduate School of Physics and Astronomy is acknowledged.

\printbibliography

\end{document}